\documentclass[%
twocolumn, 
notitlepage, 
aps,
prd, 
longbibliography,
10pt, 
superscriptaddress,
floatfix,
letterpaper,
reprint,
bibnotes,
amsmath,amssymb,
]{revtex4-2}

\usepackage{hyperref}
\usepackage{graphicx}
\usepackage{dcolumn}
\usepackage{bm}
\usepackage[compat=1.1.0]{tikz-feynman}
\usepackage{physics}
\usepackage{lipsum}

\definecolor{linkcolor}{HTML}{399B03}
\definecolor{urlcolor}{HTML}{399B03}
\hypersetup{pdfstartview=FitH, linkcolor=linkcolor,urlcolor=urlcolor,colorlinks=true}
\hypersetup{
    unicode=false,          
    pdftoolbar=true,        
    pdfmenubar=true,        
    pdffitwindow=false,     
    pdfstartview={FitH},    
    pdfauthor={},         
    colorlinks=true,        
    linkcolor=blue,         
    citecolor=red,          
}

\begin{document}

\title{Relativistic Self-Consistent $GW$: Exact Two-Component Formalism with
One-Electron Approximation for Solids}

\author{Chia-Nan Yeh}%
\affiliation{%
Department of Physics, University of Michigan, Ann Arbor, Michigan 48109, USA
}%

\author{Avijit Shee}%
\affiliation{%
Department of Chemistry, University of Michigan, Ann Arbor, Michigan 48109, USA
}%

\author{Qiming Sun}%
\affiliation{
AxiomQuant Investment Management LLC, Shanghai 200120, China
}

\author{Emanuel Gull}%
\affiliation{%
Department of Physics, University of Michigan, Ann Arbor, Michigan 48109, USA
}%

\author{Dominika Zgid}%
\affiliation{%
Department of Chemistry, University of Michigan, Ann Arbor, Michigan 48109, USA
}%
\affiliation{%
Department of Physics, University of Michigan, Ann Arbor, Michigan 48109, USA
}%

\date{\today}

\begin{abstract}
We present a formulation of relativistic self-consistent $GW$ for solids based on the exact two-component formalism with one-electron approximation (X2C1e) and non-relativistic Coulomb interactions. Our theory allows us to study scalar relativistic effects, spin-orbit coupling, and the interplay of relativistic effects with electron correlation without adjustable parameters. Our all-electron implementation is fully \emph{ab initio} and does not require a pseudopotential constructed from atomic calculations. 
We examine the effect of the X2C1e approximation by comparison to the established four-component formalism and reach excellent agreement. 
The simplicity of X2C1e enables the construction of higher order theories, such as embedding theories, on top of perturbative calculations.
\end{abstract}

\maketitle

\section{Introduction}
Relativistic effects, such as spin-orbit coupling (SOC), are essential for understanding the physics of quantum materials including 
correlated topological insulators~\cite{Rachel_2018}, topological superconductors~\cite{Sato_2017}, quantum spin liquids~\cite{Zhou_RMP_2017}, and topological semimetals~\cite{Lv_RMP_2021}.

SOC effects are particularly important in materials with heavy elements, such as those with partially occupied $d-$ and $f-$electron shells. They include several new $5d$ transition metal oxides (iridates, osmates)~\cite{Kim_iridate_PRL_2008,DMFT_BaOaO3_2021}, multiferroic materials~\cite{Fiebig2016}, and heterostructures of transition metal systems~\cite{PhysRevB.92.075439}, where the interplay of relativistic effects and electron correlation may lead to magnetism and electron localization. 
Analyzing SOC effects in these systems is crucial for understanding the nature of electronic states. 
Harnessing and controlling SOC effects may lead to novel designs for applications and devices.

The computational description of relativistic effects in molecular and periodic systems has a long history. 
Relativistic quantum effects are described by the Dirac, rather than the Schr\"{o}dinger, equation \cite{Dirac_eqn_1928}. A solution of the Dirac equation employs the so-called four-component formalism, where the problem is expanded into Dirac bispinors, which describe spin as well as electrons and positrons. 
In molecular systems, the Dirac equation for Gaussian-type orbitals (GTOs)
 has been studied extensively in the context of mean-field and density functional theory (DFT) yielding numerous mature implementations~\cite{Dirac_2020,BDF_2020,PySCF_2020,ReSpect_Repisky2020}. 
Extensions of the four-component theory to configuration interaction (CI) and coupled-cluster (CC) theory~\cite{review_Visscher2002,review_relativistic_correlated_method_Fleig2012} are active fields of research.

Two-component relativistic Hamiltonians, where the positronic degrees of freedom are eliminated, result in
a useful compromise in terms of computational cost between the scalar non-relativistic and the four-component relativistic one-electron Hamiltonians. They typically increase the computational cost  by one order of magnitude in comparison to the scalar relativistic approaches, due to the transition from real to complex quantities and the inclusion of two-component matrices. 

In molecular chemistry, the two-component formalism resulted in numerous interesting applications, see Ref.~\cite{Liu_relativistic_review_2010,Saue_Review} for reviews. In general, two-component Hamiltonians can be divided into two broad classes. {\em Inexact two-component Hamiltonians} such as Pauli~\cite{Liu_relativistic_review_2010,Saue_Review}, Douglas-Kroll-Hess (DKH)~\cite{DKH_2004}, and ZORA~\cite{ZORA_1986,ZORA_1993} Hamiltonians are considered {\em inexact} due to the approximate decoupling schemes used to transform the four-component to the two-component theory. In contrast, {\em exact two-component} (X2C) Hamiltonians reproduce  the positive-energy spectrum of the parent four-component Hamiltonian exactly~\cite{Liu_relativistic_review_2010,Saue_Review,NESC_Review_Cremer2014}. The  formulation of the X2C theories generated a lot of excitement in molecular electronic structure theory due to its transparent nature, lack of ad hoc approximations, and computational efficiency.

Numerous applications of the relativistic formalism to periodic systems have been performed. While  the choice of GTOs as one-particle basis functions is overwhelmingly common for molecular systems, for periodic systems relativistic calculations were performed for several choices of one-particle basis functions including as plane waves~\cite{2cG0W0_Scherpelz_2016}, augmented plane waves (APW)~\cite{relativistic_APW_1965}, linear-APWs (LAPW)~\cite{LAPW_MacDonald_1980,LAPW_Wimmer_1981}, linear muffin-tin
orbitals (LMTO)~\cite{relativistic_LMTO_1988,relativistic_LMTO_second_1988},  projector augmented waves (PAW)~\cite{relativistic_PAW_2010}, analytic Slater-type orbitals (STOs)~\cite{ZORA_BAND_1997,Rundong_JCP2016}, and Gaussian-type orbitals (GTOs)~\cite{Kadek_PRB2019}. For a discussion of these developments see Ref.~\onlinecite{Kadek_PRB2019}. Note that, while many of these applications involved inexact two-components Hamiltonians, the  application of the full four-component formalism in the density functional theory (DFT) framework to periodic systems employing GTOs  was only performed in 2019 by \textcite{Kadek_PRB2019}.

While DFT, due to its affordable computational scaling, can be applied to many of the one-particle orbital bases, the situation is more complicated for correlated \emph{ab initio}  methods with a higher computational scaling. 
For those, one would ideally want to employ a compact one-particle basis such as GTOs and retain the possibility of describing both core and valence electrons by the same type of basis function.  Moreover, due to their computational demand, it is advantageous to avoid the expensive four-component formalism in favor of a more manageable two-component formulation. 

Motivated by these considerations, we describe here the application of an exact two-component theory in the one-electron approximation (X2C1e) to fully self-consistent $GW$ (sc$GW$) for periodic problems in the  one-particle GTO basis. We call the method X2C1e-sc$GW$.
The exact two-component methods (X2C) generate an electron-only two-component Hamiltonian that exactly reproduces the one-electron energies of the original four-component Dirac Hamiltonian while approximating some of the relativistic two-body integrals, which are expected to be small for atoms that are not extremely heavy \cite{intro_relativistic_QC_Dyall_2007,Liu_relativistic_review_2010,Saue_Review,NESC_Review_Cremer2014}. 

Two-component methods are particularly appealing for a numerical implementation in solids for two reasons. First, the restriction to two components, rather than four, substantially reduces  computation and memory demands. Second, because of the particularly simple form of the two-body integrals (which are just the regular non-relativistic two-body Coulomb integrals), two-component methods open a direct route towards parameter-free embedding calculations with self-energy embedding (SEET) \cite{Alexie_SEET_PRB2015,Lan_Generalized_SEET_2017,Dominika_SEET_2017,Zgid_periodic_seet2019,SEET_Sergei20} or   dynamical mean field theory (DMFT) \cite{Kotliar06}.
For instance, at present combinations of DFT with DMFT (DFT+DMFT) for relativistic compounds rely on adding a phenomenological $L\cdot S$ spin-orbit coupling term to the DMFT impurity Hamiltonian, the parameters of which are unknown and need to be adjusted on a case-by-case basis~\cite{LDA_DMFT_Sr2IrO4_2011,TRIQS_Aichhorn2016,DMFT_Sr2RuO4_2018,Expt_DMFT_Sr2RuO4_2019,DMFT_BaOaO3_2021}. Exact two-component theories can be used to remove this phenomenological parametrization from DFT+DMFT.
In addition, the two-component theory eliminates the need for simultaneous optimization of positive and negative energy solutions.

For periodic systems, the introduction of relativistic treatment into the $GW$ approach has a long history. In Ref.~\onlinecite{4cG0W0_Sakuma_2011}, within the full-potential linearized augmented-plane-wave (FLAPW) method, a fully spin-dependent formulation of the quasiparticle $GW$ approximation was presented, which described many-body renormalization effects arising from spin-orbit coupling. 
This approach took into account the spin off-diagonal elements of the Green's function and the self-energy. 
The core, valence, and conduction states of the reference one-particle system were
treated fully relativistically as four-component spinor wave functions.
In Ref.~\onlinecite{4c_scGW_Kutepov_2012}, spin-orbit interactions were included in $GW$ by using Dirac's form of the kinetic energy operator and full self-consistency was performed.
Recently, Ref.~\onlinecite{2cG0W0_Scherpelz_2016} reported the inclusion of SOC in a $GW$ code, WEST, with calculations at the $G_0W_0$ level. In this $G_0W_0$ calculation, both $G$ and $W$ were computed at a fully relativistic level without the use of empty states. 

In this paper, we discuss the  exact two-component theory in the one-electron approximation (X2C1e) for periodic systems described by a GTO one-particle basis and demonstrate results from its implementation into a fully self-consistent $GW$ (sc$GW$) method. We call this method X2C1e-sc$GW$. 
The methodology is designed to preserve the computational advantages of the two-component formalism as well as compactness of the GTO basis when treating periodic systems.
As an example of the X2C1e-sc$GW$ methodology, we discuss the series of silver halides ($\mathrm{AgCl}$, $\mathrm{AgBr}$, $\mathrm{AgI}$) in which scalar relativistic effects and SOC becomes gradually more important as the halogen is changed from $\mathrm{Cl}$ to $\mathrm{I}$. We show that, in these systems, X2C1e-sc$GW$ recovers all of the relativistic effects identified in the four-component DFT while yielding better experimental agreement than four-component DFT.

The remainder of this paper proceeds as follows. In Sec.~\ref{sec:relth}, we introduce the relativistic theory. Sec.~\ref{sec:compdet} focuses on computational details while Sec.~\ref{sec:res} contains results for the silver halides. Our conclusions are presented in Sec.~\ref{sec:conc}.

\section{Relativistic Theory} \label{sec:relth}
This section discusses the X2C1e approximation in solids and the diagrammatic perturbation theory applied to the relativistic two-component Hamiltonian. 
Starting form the non-interacting Dirac Hamiltonian $\mathcal{\hat{H}}_{0}$~\cite{Dirac_eqn_1928,minial_coupling_Gell-Mann1956,Liu_relativistic_review_2010,Saue_Review} presented in Sec.~\ref{subsec:Dirac} and the kinetic balance Gaussian type orbitals (KB-GTO)~\cite{RKB_JCP,RKB_ISHIKAWA1985,RKB_DYALL1990} presented in Sec.~\ref{subsec:KB-GTO}, we show in Sec.~\ref{subsec:modified_Dirac} how expanding $\mathcal{\hat{H}}_{0}$ using the KB-GTO basis will lead to the modified Dirac equation~\cite{modified_Dirac_Kutzelnigg1984,modified_Dirac_Dyall1994}. The  non-interacting X2C1e Hamiltonian~\cite{x2c1e_Dyall_2001,x2c1e_Dyall_2002} can then be constructed via the normalized elimination of the small component (NESC) of the modified Dirac Hamiltonian~\cite{FW_transformation_1950,NESC_Dyall_1997,FW_transformation_2_1999,x2c1e_Dyall_2001,x2c1e_Dyall_2002,NESC_Review_Cremer2014}, as shown in Sec.~\ref{subsec:x2c1e}. 
In Sec.~\ref{subsec:x2c1e-Coulomb}, we define the X2C1e-Coulomb Hamiltonian as a combination of the non-interacting X2C1e Hamiltonian with the non-relativistic Coulomb interactions. 
The formulation of diagrammatic perturbation theory such as the sc$GW$ approximation using the X2C1e-Coulomb  Hamiltonian is described in Sec.~\ref{subsec:2c_GW}. 

\subsection{Non-interacting Dirac Hamiltonian\label{subsec:Dirac}}
In the absence of electron-electron interactions and external magnetic fields, and within the Born-Oppenheimer approximation, the Dirac equation with minimal coupling to the attractive nuclear Coulomb potential $V(\bold{r})$~\cite{Dirac_eqn_1928,minial_coupling_Gell-Mann1956} can be recast as an eigenvalue problem, $ \hat{\mathcal{H}}_{0} \Psi=E\Psi$~\cite{Liu_relativistic_review_2010,Saue_Review}, where $\Psi=(\Psi^L,  \Psi^S)^T$ denotes a four-component spinor written in terms of two `large' and `small'-component spinors, and $\hat{\mathcal{H}}_{0}$ denotes the $4\times4$ Hamiltonian matrix
\begin{align}
\hat{\mathcal{H}}_{0} = 
\begin{pmatrix}
V(\bold{r}) & c\sigma\cdot\hat{\bold{p}}\\
c\sigma\cdot\hat{\bold{p}} & V(\bold{r})-2c^{2}
\end{pmatrix}.
\label{Eq:dirac_operator} 
\end{align}
Here, $c$ is the speed of light, $\bold{\sigma}$ are Pauli matrices, and $\hat{\bold{p}} = -i\nabla$ is the momentum operator. In order to discuss the exact two-component formalism, we will first discuss the solution of this non-interacting Hamiltonian.

\subsection{Kinetic balance Gaussian type orbital\label{subsec:KB-GTO}}
In practical calculations, Hamiltonians are expanded into a finite basis set. We will limit our discussion here to Bloch waves constructed from a periodic arrangement of Gaussian orbitals, which are one possible choice of basis sets for solids. 

In the non-relativistic case, the non-relativistic Hamiltonian 
is expanded into scalar Gaussian Bloch orbitals $g^{\bold{k}}_{i}(\bold{r})$ constructed from Gaussian atomic basis functions $g^{\bold{R}}_{i}(\bold{r})$ as 
\begin{align}
g^{\bold{k}}_{i}(\bold{r}) = \sum_{\bold{R}} g^{\bold{R}}_{i}(\bold{r})e^{i\bold{k}\cdot\bold{R}},
\end{align}
where $\bold{k}$ is a wave vector in the first Brillouin zone of the reciprocal space, and $g^{\bold{R}}_{i}(\bold{r})$ is the $i$-th Gaussian atomic orbital centered in unit cell $\bold{R}$~\cite{Boys_Gaussian_basis_1950}. The summation over $\bold{R}$ extends over the whole lattice. 
The overlap matrix 
\begin{align}
S^{\bold{k}}_{ij} = \int_{\Omega} d\bold{r} g^{\bold{k}*}_{i}(\bold{r})g^{\bold{k}}_{j}(\bold{r})\delta_{\bold{k}\bold{k}'}
\label{Eq:scalar_S}
\end{align}
is diagonal in reciprocal space indices due to the translational invariance of the lattice but generally non-diagonal in the orbital space indices ($\Omega$ denotes the unit cell).

In the relativistic case, in order to expand the four-component relativistic operator of Eq.~\ref{Eq:dirac_operator}, we define a four-component Bloch bispinor basis 
\begin{align}
\chi^{\bold{k}}_{i}(\bold{r}) = 
\begin{pmatrix}
\chi^{\bold{k},L}_{i}(\bold{r}) \\
\chi^{\bold{k},S}_{i}(\bold{r})
\end{pmatrix}
\label{Eq:4c_bispinor}
\end{align}
where $\chi^{\bold{k},L}_{i}(\bold{r})$ and $\chi^{\bold{k},S}_{i}(\bold{r})$ denote a large ($L$) and small ($S$) component spinor. 
In the present work, in analogy to the non-relativistic case, the large component spinor basis is defined in terms of a scalar Gaussian Bloch orbital 
\begin{align}
\chi^{\bold{k},L}_{i}(\bold{r}) =
\begin{pmatrix}
\chi^{\bold{k},L}_{i,\uparrow}(\bold{r}) \\
\chi^{\bold{k},L}_{i,\downarrow}(\bold{r})
\end{pmatrix}
\label{Eq:L_spinor}
\end{align}
where $\chi^{\bold{k},L}_{i\uparrow}(\bold{r})$ and the $\chi^{\bold{k},L}_{i\downarrow}(\bold{r})$ are the spin-up and the spin-down components of the large component spinor basis which is expressed in terms of the scalar Gaussian Bloch orbitals $g^{\bold{k}}_{i}(\bold{r})$. 
Rather than using this basis also for the small component, we define a relativistic small component  basis through  the restricted kinetic balance (RKB) condition~\cite{RKB_JCP,RKB_ISHIKAWA1985,RKB_DYALL1990} as 
\begin{align}
\chi^{\bold{k},S}_{i}(\bold{r}) &= \frac{1}{2c}(\bold{\sigma}\cdot \hat{\bold{p}})\chi^{\bold{k},L}_{i}(\bold{r}).
\label{Eq:S_spinor}
\end{align}
The RKB condition enforces the exact coupling of large and small components in the non-relativistic limit~\cite{Liu_relativistic_review_2010} and is essential to achieve variationally stable four-component solutions in a finite basis set~\cite{RKB_JCP,variational_collapse_1984,RKB_ISHIKAWA1985,RKB_DYALL1990,Liu_relativistic_review_2010,Saue_Review}. 
For a physical single-particle state, the expansion coefficients for large and small components are allowed to be different. The same holds for the spin-up and spin-down parts in Eq.~\ref{Eq:L_spinor} and Eq.~\ref{Eq:S_spinor}. 
In the following, we will refer to the basis of Eq.~\ref{Eq:4c_bispinor} as `kinetic balance Gaussian-type orbitals' (KB-GTO). 

\subsection{Modified Dirac Hamiltonian\label{subsec:modified_Dirac}}
Expanding $\hat{\mathcal{H}}_{0}$ into $N$ basis functions of the KB-GTO basis per unit cell, we arrive at the modified non-interacting Dirac Hamiltonian~\cite{modified_Dirac_Kutzelnigg1984,modified_Dirac_Dyall1994} 
\begin{align}
\mathcal{H}_{0}^{\bold{k}} &= 
\begin{pmatrix}
\bold{V}^{\bold{k}} & \bold{T}^{\bold{k}} \\
\bold{T}^{\bold{k}} & \bold{W}^{\bold{k}}-\bold{T}^{\bold{k}}
\end{pmatrix}.
\label{Eq:modified_dirac_matrix}
\end{align}
The overlap matrix of the bispinor basis is defined as
\begin{align}
\mathcal{S}^{\bold{k}} = 
\begin{pmatrix}
\bold{S}^{\bold{k}} & 0_{2N} \\ 
0_{2N} & \bold{T}^{\bold{k}}/2c^2
\end{pmatrix}.
\label{Eq:dirac_overlap_matrix}
\end{align}
$\bold{V}^{\bold{k}}$, $\bold{T}^{\bold{k}}$, $\bold{S}^{\bold{k}}$, and $\bold{W}^{\bold{k}}$ are matrices of size  $2N\times2N$ defined as  
\begin{align}
\bold{V}^{\bold{k}} = I_{2}\otimes V^{\bold{k}} =
\begin{pmatrix}
V^{\bold{k}} & 0_{N} \\
0_{N} & V^{\bold{k}}
\end{pmatrix}, \\
\bold{T}^{\bold{k}} = I_{2}\otimes T^{\bold{k}} =
\begin{pmatrix}
T^{\bold{k}} & 0_{N} \\
0_{N} & T^{\bold{k}}
\end{pmatrix}, \\
\bold{S}^{\bold{k}} = I_{2}\otimes S^{\bold{k}} =
\begin{pmatrix}
S^{\bold{k}} & 0_{N} \\
0_{N} & S^{\bold{k}}
\end{pmatrix}.
\label{eqn:S_L}
\end{align}
Here $V^{\bold{k}}$ is a matrix of size $N\times N$ and contains the contributions of the external potential, $T^{\bold{k}}$ is the kinetic energy matrix, and $S^{\bold{k}}$ is the scalar overlap matrix defined in Eq.~\ref{Eq:scalar_S}. 
$\bold{W}^{\bold{k}}$ defines the matrix for the potential of the small component. Via the Dirac identity $(\bold{\sigma}\cdot\hat{\bold{p}})\hat{V}(\bold{\sigma}\cdot\hat{\bold{p}}) = (\hat{\bold{p}}\hat{V}\cdot\hat{\bold{p}})I_{2} + i\bold{\sigma}\cdot(\hat{\bold{p}}\hat{V}\times\hat{\bold{p}})$ it  can be separated into a spin-free $\bold{W}^{\bold{k}}_{\mathrm{SR}}$ and a spin-dependent part~\cite{modified_Dirac_Dyall1994} $\bold{W}^{\bold{k}}_{\mathrm{SOC}}$ ,
\begin{align}
&\bold{W}^{\bold{k}} = \bold{W}^{\bold{k}}_{\mathrm{SR}} + \bold{W}^{\bold{k}}_{\mathrm{SOC}}, \label{Eq:W1}\\
&\bold{W}^{\bold{k}}_{\mathrm{SR}} = 
\begin{pmatrix}
(W_{\mathrm{SR}})^{\bold{k}} & 0 \\
0 & (W_{\mathrm{SR}})^{\bold{k}}
\end{pmatrix}, \label{Eq:W2}\\
&\bold{W}^{\bold{k}}_{\mathrm{SOC}} = 
\sum_{\mu = x,y,z}
\begin{pmatrix}
(W_{\mathrm{SOC}})^{\bold{k},\mu} & 0 \\
0 & (W_{\mathrm{SOC}})^{\bold{k},\mu}
\end{pmatrix}
\tilde{\sigma}_{\mu}, \label{Eq:W3}
\end{align}
where $\tilde{\sigma}_{\mu} = I_{N} \otimes \sigma_{\mu}$, $\sigma_{\mu}$ are Pauli matrices and 
\begin{align}
&(W_{\mathrm{SR}})^{\bold{k}}_{ij}= \int_{\Omega} \frac{1}{4c^2}g^{\bold{k}}_{i}(\bold{r})^{*}\big[\hat{\bold{p}}V(\bold{r})\cdot\hat{\bold{p}}\big]g^{\bold{k}}_{j}(\bold{r})d^{3}\bold{r}, \label{Eq:W_SR}\\
&(W_{\mathrm{SOC}})_{ij}^{\bold{k},\mu}=\int_{\Omega} \frac{1}{4c^2}g^{\bold{k}}_{i}(\bold{r})^{*}\big[i(\hat{\bold{p}}V(\bold{r})\times \hat{\bold{p}})_{\mu}\big]g^{\bold{k}}_{j}(\bold{r})d^{3}\bold{r}. \label{Eq:W_SOC}
\end{align}

For the case where, $V(\bold{r})$ corresponds to the nuclear potential $Z/r$, the spin-dependent part $\bold{W}^{\bold{k}}_{\mathrm{SOC}}$ can be re-expressed as the SOC of the electron spin with the magnetic field induced by the nucleus of charge $Z$ at the origin~\cite{intro_relativistic_QC_Dyall_2007}. 
The spin-free part $\bold{W}^{\bold{k}}_{\mathrm{SR}}$ is referred to as the scalar relativistic potential and contributes to the relativistic mass enhancement. 

Based on Eqs.~\ref{Eq:modified_dirac_matrix} and ~\ref{Eq:dirac_overlap_matrix}, the non-interacting Dirac equation $\hat{\mathcal{H}}_{0} \Psi=E\Psi$ can then be recast into a generalized eigenvalue problem 
\begin{align}
\mathcal{H}_{0}^{\bold{k}}\mathcal{C}^{\bold{k}} = \mathcal{S}^{\bold{k}}\mathcal{C}^{\bold{k}}\epsilon^{\bold{k}},
\label{Eq:modified_dirac_eqn}
\end{align}
where $\mathcal{C}^{\bold{k}}$ and $\epsilon^{\bold{k}}$ are the coefficient matrix for the corresponding one-particle states and the diagonal matrix for the one-particle energies. 
Due to the presence of large and small-component spinors as well as the electronic and positronic degrees of freedom, both $\mathcal{C}^{\bold{k}}$ and $\epsilon^{\bold{k}}$ are $4N\times 4N$ matrices for any given $k$-point. 
The resulting electronic and positronic one-particle states will be separated by an energy gap of $2c^{2}$~\cite{intro_relativistic_QC_Dyall_2007}. 
In the following sections, we will use subscripts $+$ and $-$ to denote electronic and positronic states, respectively. 
If one-particle states are organized in descending order of orbital energies, i.e.  
\begin{align}
\epsilon^{\bold{k}} = 
\begin{pmatrix}
\bold{\epsilon}^{\bold{k}}_{+} & 0_{2N} \\
0_{2N} & \bold{\epsilon}^{\bold{k}}_{-}
\end{pmatrix},
\label{Eq:modified_dirac_spectrum}
\end{align}
the coefficient matrix $\mathcal{C}^{\bold{k}}$ can be expressed as  
\begin{align}
\mathcal{C}^{\bold{k}} = 
\begin{pmatrix}
\bold{A}^{\bold{k}}_{+} & \bold{A}^{\bold{k}}_{-} \\
\bold{B}^{\bold{k}}_{+} & \bold{B}^{\bold{k}}_{-} \\
\end{pmatrix}
\label{Eq:modified_dirac_states}
\end{align}
where $\bold{A}^{\bold{k}}_{+}$ and $\bold{B}^{\bold{k}}_{+}$ are $2N\times 2N$ coefficient matrices of the large and small-component spinors for electronic states. 
The large and small-component positronic states are expressed in terms of $\bold{A}^{\bold{k}}_{-}$ and $\bold{B}^{\bold{k}}_{-}$. 

\subsection{Exact two-component theory with one-electron approximation\label{subsec:x2c1e}}
The exact two-component (X2C) theory aims to construct a two-component Hamiltonian that reproduces the electronic spectrum ($\bold{\epsilon}^{\bold{k}}_{+}$) of the parent four-component Hamiltonian $\mathcal{\hat{H}}_{0}$~\cite{Liu_relativistic_review_2010,Saue_Review,NESC_Review_Cremer2014}.
Different choices of $\mathcal{\hat{H}}_{0}$ lead to different variants of X2C~\cite{Liu_relativistic_review_2010}. 
Common choices include the free-particle Dirac Hamiltonian, the non-interacting Dirac Hamiltonian in the presence of nuclear Coulomb potential (Eq.~\ref{Eq:dirac_operator}), the Dirac Hartree-Fock (DHF) Hamiltonian, and the Dirac Kohn-Sham (DKS) Hamiltonian~\cite{Liu_relativistic_review_2010}. 
In this work, we will use the non-interacting Dirac Hamiltonian of Eq.~\ref{Eq:dirac_operator} as our $\mathcal{\hat{H}}_{0}$ and refer to this formulation as the X2C with the one-electron approximation (X2C1e)~\cite{x2c1e_Dyall_2001,x2c1e_Dyall_2002}. 

The effective two-component Hamiltonian is obtained via the normalized elimination of the small component (NESC)~\cite{NESC_Dyall_1997}. 
Defining the coupling matrix $\bold{X}^{\bold{k}}$ between the large ($\bold{A}^{\bold{k}}_{+}$) and the small component ($\bold{B}^{\bold{k}}_{+}$) coefficients for the electronic solutions as 
\begin{align}
\bold{B}^{\bold{k}}_{+} = \bold{X}^{\bold{k}}\bold{A}^{\bold{k}}_{+} 
\label{Eq:X_definition}
\end{align}
and inserting Eq.~\ref{Eq:X_definition} into the electronic part of Eq.~\ref{Eq:modified_dirac_eqn}, we obtain \begin{align}
&\bold{V}^{\bold{k}}\bold{A}^{\bold{k}}_{+} + \bold{T}^{\bold{k}}\bold{X}^{\bold{k}}\bold{A}^{\bold{k}}_{+} = \bold{S}^{\bold{k}}\bold{A}^{\bold{k}}_{+}\epsilon^{\bold{k}}_{+}\label{Eq:NESC_1},\\
&\bold{T}^{\bold{k}}\bold{A}^{\bold{k}}_{+} + (\bold{W}^{\bold{k}}-\bold{T}^{\bold{k}})\bold{X}^{\bold{k}}\bold{A}^{\bold{k}}_{+} = \frac{1}{2c^{2}}\bold{T}^{\bold{k}}\bold{X}^{\bold{k}}\bold{A}^{\bold{k}}_{+}\epsilon^{\bold{k}}_{+}.\label{Eq:NESC_2}
\end{align}
By multiplying Eq.~\ref{Eq:NESC_2} on the left by $(\bold{X}^{\bold{k}})^{\dag}$ and adding it to Eq.~\ref{Eq:NESC_1}, we obtain an un-normalized effective two-component equation for the positive-energy solution ($\bold{A}^{\bold{k}}_{+}$ and $\epsilon^{\bold{k}}_{+}$), 
\begin{align}
\tilde{\bold{L}}^{\bold{k}}_{+} \bold{A}^{\bold{k}}_{+} = \tilde{\bold{S}}^{\bold{k}}\bold{A}^{\bold{k}}_{+}\epsilon^{\bold{k}}_{+},
\label{Eq:unnormalized_NESC}
\end{align}
where 
\begin{align}
&\tilde{\bold{L}}^{\bold{k}}_{+} = \bold{V}^{\bold{k}} + (\bold{X}^{\bold{k}})^{\dag}\bold{T}^{\bold{k}} + \bold{T}^{\bold{k}}\bold{X}^{\bold{k}} + (\bold{X}^{\bold{k}})^{\dag}(\bold{W}^{\bold{k}}-\bold{T}^{\bold{k}})\bold{X}^{\bold{k}}, \\
&\tilde{\bold{S}}^{\bold{k}} = \bold{S}^{\bold{k}} + \frac{1}{2c^2}(\bold{X}^{\bold{k}})^{\dag}\bold{T}^{\bold{k}}\bold{X}^{\bold{k}}.
\end{align}
$\tilde{\bold{L}}_{+}^{\bold{k}}$ is the un-normalized electronic two-component Hamiltonian with the effective relativistic metric $\tilde{\bold{S}}^{\bold{k}}$.
In order to later combine this expression with the non-relativistic two-body integrals, we aim to rescale $\tilde{\bold{L}}_{+}^{\bold{k}}$ with respect to the non-relativistic metric $\bold{S}^{\bold{k}}$ (which is just the overlap matrix within the primitive basis) with the help of the matrix $\bold{R}^{\bold{k}}_{+}$ derived by Liu and Peng~\cite{X2C_revisit_Liu_2009}, 
\begin{align}
\bold{R}^{\bold{k}}_{+} = (\bold{S}^{\bold{k}})^{-\frac{1}{2}}[(\bold{S}^{\bold{k}})^{-\frac{1}{2}}\tilde{\bold{S}}^{\bold{k}}(\bold{S}^{\bold{k}})^{-\frac{1}{2}}]^{-\frac{1}{2}}(\bold{S}^{\bold{k}})^{\frac{1}{2}}.
\end{align}
Multiplying Eq.~\ref{Eq:unnormalized_NESC} on the left by $(\bold{R}^{\bold{k}}_{+})^{\dag}$, we arrive at a two-component equation expressed in terms of the non-relativistic metric $\bold{S}^{\bold{k}}$, 
\begin{align}
(\bold{H}^{\mathrm{X2C1e}}_{+})^{\bold{k}}\bold{C}^{\bold{k}}_{2c} = \bold{S}^{\bold{k}}\bold{C}^{\bold{k}}_{2c} \epsilon^{\bold{k}}_{+},
\label{Eq:X2C_solution}
\end{align}
where 
\begin{align}
&(\bold{H}^{\mathrm{X2C1e}}_{+})^{\bold{k}} = (\bold{R}^{\bold{k}}_{+})^{\dag}\tilde{\bold{L}}^{\bold{k}}_{+} \bold{R}^{\bold{k}}_{+}, \label{Eq:HX2C1e}\\
&\bold{S}^{\bold{k}}=(\bold{R}^{\bold{k}}_{+})^{\dag}\tilde{\bold{S}}^{\bold{k}}\bold{R}^{\bold{k}}_{+},\label{Eq:R_def1}\\
&\bold{C}^{\bold{k}}_{2c} = (\bold{R}^{\bold{k}}_{+})^{-1}\bold{A}^{\bold{k}}_{+}.
\end{align}

Due to the presence of the spin-dependent $\bold{W}_{\mathrm{SOC}}^{\bold{k}}$, $(\bold{H}^{\mathrm{X2C1e}}_{+})^{\bold{k}}$ contains non-zero off-diagonal spin components. 
While the presence of these terms incorporates the full SOC effect \emph{ab initio} at the one-electron level, it also introduces an extra computational cost compared to spin-free (scalar) theories such as the ones containing a non-relativistic Hamiltonian. 
In cases where SOC is negligibly weak, an additional approximation can be made, which we will refer to as the spin-free X2C1e (sfX2C1e)~\cite{x2c1e_Dyall_2001}, which consists of approximating $\bold{W}^{\bold{k}} \approx \bold{W}^{\bold{k}}_{\mathrm{SR}}$ such that $(\bold{H}_{+}^{\mathrm{sfX2C1e}})^{\bold{k}}$ becomes diagonal in spin space. 

Note that the $\bold{X}^{\bold{k}}$ matrix in the X2C1e formalism is constructed only once by solving the non-interacting Dirac equation, Eq.~\ref{Eq:modified_dirac_eqn}, since no self-consistent loop is required for a non-interacting Dirac solution. 
On the other hand, if the Dirac Hartree-Fock (DHF) Hamiltonian or the Dirac Kohn-Sham (DKS) Hamiltonian are taken as the $\mathcal{H}_{0}$, a self-consistent loop between the $\bold{X}^{\bold{k}}$ matrix and the single-particle solutions is required.

\subsection{X2C1e-Coulomb Hamiltonian\label{subsec:x2c1e-Coulomb}}
So far, we have not discussed the two-body electron-electron interactions.
Effectively, the NESC procedure can be rewritten as a block diagonalization using a unitary transformation matrix $U^{\bold{k}}$~\cite{Liu_relativistic_review_2010,NESC_Review_Cremer2014}. 
Similarly, the same transformation $U^{\bold{k}}$ has to be applied to electron-electron interactions as well~\cite{intro_relativistic_QC_Dyall_2007,Liu_relativistic_review_2010,Saue_Review}. 
However, this procedure would involve the evaluation of two-body electron-electron integrals at the four-component level and explicit transformation of different Hamiltonian blocks, which is expensive. 
Instead, one may perform an approximation and use the un-transformed electron-electron Coulomb integrals~\cite{intro_relativistic_QC_Dyall_2007,Saue_Review}. 
The resulting Hamiltonian consists of  $(\bold{H}^{\mathrm{X2C1e}}_{+})^{\bold{k}}$ (Eq.~\ref{Eq:HX2C1e}) along with the non-relativistic two-electron Coulomb integrals $U^{\bold{k}_{1}\bold{k}_{2}\bold{k}_{3}\bold{k}_{4}}_{\ i\ \ j\ k\ \ l}$,
\begin{align}
&H = \sum_{\bold{k}}\sum_{ij}\sum_{\sigma,\sigma'}(H^{\mathrm{X2C1e}}_{+})^{\bold{k}}_{i\sigma,j\sigma'}c^{\bold{k},\dag}_{i\sigma}c^{\bold{k}}_{j\sigma'} \nonumber\\
&+ \frac{1}{2N_{k}}\sum_{ijkl}\sum_{\bold{k}\bold{k}'\bold{q}}\sum_{\sigma\sigma'}U^{\bold{k},\bold{k}-\bold{q}, \bold{k}',\bold{k}'+\bold{q}}_{i\ \ \ j\ \ \ k\ \ \ l}c^{\bold{k}\dag}_{i\sigma}c^{\bold{k}'\dag}_{k\sigma'}c^{\bold{k}'+\bold{q}}_{l\sigma'}c^{\bold{k}-\boldsymbol{q}}_{j\sigma},
\label{Eq:X2C-Coulomb}
\end{align}
where $c^{\bold{k}\dag}_{i\sigma}$ ($c^{\bold{k}}_{i\sigma}$) are the creation (annihilation) operators for the single-particle spin-orbital state with crystal momentum $\bold{k}$, spin $\sigma$, and scalar Gaussian orbital $i$. 
The two-electron Coulomb integrals are defined as 
\begin{align}
&U^{\bold{k}_{1}\bold{k}_{2}\bold{k}_{3}\bold{k}_{4}}_{\ i\ \ j\ k\ \ l} = \label{Eq:V_ijkl}\\
&\int_{\mathbb{R}^{3}}d\bold{r}_{1}\int_{\mathbb{R}^{3}}d\bold{r}_{2}g^{\bold{k}_{1}*}_{i}(\bold{r}_{1})g^{\bold{k}_{2}}_{j}(\bold{r}_{1})\frac{1}{|\bold{r}_{1}-\bold{r}_{2}|}g^{\bold{k}_{3}*}_{k}(\bold{r}_{2})g^{\bold{k}_{4}}_{l}(\bold{r}_{2}). \nonumber
\end{align}
Note that translational invariance guarantees $\bold{k}_1+\bold{k}_3=\bold{k}_2+\bold{k}_4$. The singularity at $\bold{q}=\bold{k}_{1}-\bold{k}_{2}=\bold{k}_{4}-\bold{k}_{3}=\bold{0}$ is excluded manually. 

The Hamiltonian of Eq.~\ref{Eq:X2C-Coulomb} is referred to as the X2C1e-Coulomb Hamiltonian. In the case where the sfX2C1e Hamiltonian is taken as the one-electron part in Eq.~\ref{Eq:X2C-Coulomb}, we refer to it as the sfX2C1e-Coulomb Hamiltonian. 
Due to the decoupling in the spin space, existing non-relativistic many-body methods can be directly applied to the sfX2C1e-Coulomb Hamiltonian without further modification. 

Due to the use of un-transformed Coulomb interactions, the resulting many-body Hamiltonian will suffer from the so-called `picture-change' error~\cite{intro_relativistic_QC_Dyall_2007,Saue_Review,NESC_Review_Cremer2014}. 
In such a case, the missing contribution is the small-component Coulomb interaction which is responsible for the spin-same-orbit interactions between electrons~\cite{intro_relativistic_QC_Dyall_2007}. 
Contributions from the transformed Coulomb term are important for properties of electrons close to the nucleus but are typically small for valence electrons~\cite{Saue_Review,NESC_Review_Cremer2014}. 
Note that all X2C methods, including X2C1e, combined with un-transformed Coulomb interactions have this issue, and it is not related to the one-electron approximation or the exclusion of SOC. 

The approximation of neglecting the relativistic corrections in the electron-electron interaction can also be understood in terms of perturbation theory. Beyond the non-relativistic Coulomb potential, the first-order relativistic correction to the two-electron interaction is referred to as the Breit term and is on the order of $\alpha^{2}$, where $\alpha=e^2/(\hbar c)\sim 1/137$ is the fine structure constant~\cite{Coulomb_Breit_1929,intro_relativistic_QC_Dyall_2007}. 
Physically, neglecting this first order correction corresponds to not including spin-other-orbit and spin-spin interactions between electrons~\cite{intro_relativistic_QC_Dyall_2007}.

\subsection{Diagrammatic perturbation theory for the X2C1e-Coulomb Hamiltonian\label{subsec:2c_GW}}
All relativistic contributions in the X2C1e-Coulomb Hamiltonian are contained in the one-electron term, while the two-electron integrals stay non-relativistic. This implies that when diagrammatic perturbation theories are extended to treat relativistics and use the X2C1e-Coulomb Hamiltonian, they simply acquire a changed non-interacting Green's function. The only complication consists of the spin-orbit coupling term, which mixes the two spin species in the one-body term.

Here, we discuss the self-consistent $GW$ (sc$GW$) theory \cite{Hedin65} in more detail, following the description of Ref.~\onlinecite{SEET_Sergei20} with the addition of off-diagonal spin components for the one-electron quantities.  
The formulation of sc$GW$ based on the X2C1e-Coulomb Hamiltonian
can be understood as a simplified version of the fully spin-dependent $GW$ approximation~\cite{Aryasetiawan_Generalized_Hedin_2008,Aryasetiawan_Generalized_GW_2009} 
where the Coulomb interactions remain spin-independent and the positronic degrees of freedom are frozen at the non-interacting level. We emphasize that the generalization to other diagrammatic methods, such as self-consistent second order perturbation theory (GF2)~\cite{Rusakov16,Phillips14,Iskakov19}, and embedding theories, such as SEET~\cite{Alexie_SEET_PRB2015,Dominika_SEET_2017,Lan_Generalized_SEET_2017} or DMFT~\cite{Antoine_DMFT_RevModPhys_1996,Kotliar06}, is straightforward when the X2C1e-Coulomb Hamiltonian is employed.

The computational cost of relativistic self-consistent  $GW$ with \emph{ab initio} Coulomb interactions has so far been substantial. 
Existing relativistic implementations include four-component one-shot $G_{0}W_{0}$~\cite{4cG0W0_Sakuma_2011} and sc$GW$~\cite{4c_scGW_Kutepov_2012} based on the no-pair approximation for the Dirac-Coulomb Hamiltonian, and a two-component one-shot $G_{0}W_{0}$ implementation constructed from a pseudopotential~\cite{2cG0W0_Scherpelz_2016}. Fully self-consistent $GW$ in an \emph{ab initio} two-component theory has not yet been explored. The full self-consistency guarantees that the method is thermodynamically consistent and conserving. Both properties are essential to prevent ambiguities in embedding theories such as DMFT or SEET. 

We express the general spin-dependent single-particle Green's function $\bold{G}$ for the X2C1e-Coulomb Hamiltonian via the Dyson equation on the Matsubara frequency axis as
\begin{align}
\bold{G}^{\bold{k}}&(i\omega_{n}) 
= \begin{pmatrix}
G^{\bold{k}}_{\uparrow\uparrow}(i\omega_{n}) & G^{\bold{k}}_{\uparrow\downarrow}(i\omega_{n}) \\
G^{\bold{k}}_{\downarrow\uparrow}(i\omega_{n}) & G^{\bold{k}}_{\downarrow\downarrow}(i\omega_{n})
\end{pmatrix} \label{Eq:G_blocks}\\
&= \big [ (i\omega_{n}+\mu)\bold{S}^{\bold{k}} - (\bold{H}^{\mathrm{X2C1e}}_{+})^{\bold{k}} - \bold{\Sigma}^{\bold{k}}(i\omega_{n})\big ]^{-1}\label{Eq:Dyson}
\end{align}
where $\mu$ is the chemical potential, $\omega_{n} = (2n+1)/\beta$ the fermionic Matsubara frequency, $\beta$ the inverse temperature, and $\bold{\Sigma}^{\bold{k}}$ the self-energy. 
All quantities in bold fonts are expressed in the large-component spinor basis (Eq.~\ref{Eq:L_spinor}) with the overlap matrix $\bold{S}^{\bold{k}}$ (Eq,~\ref{eqn:S_L}).
In Eq.~\ref{Eq:G_blocks}, each spin block of the Green's function is an $N\times N$ matrix in which all occupied and virtual states are included. 

In sc$GW$, the spin-dependent self-energy $(\bold{\Sigma}^{GW})^{\bold{k}}[\bold{G}] (i\omega_n)$ is a functional of the interacting single-particle Green's function $\bold{G}$ and can be separated into a static and a dynamic part as
\begin{align}
(\bold{\Sigma}^{GW})^{\bold{k}}[\bold{G}] (i\omega_n)= (\bold{\Sigma}^{GW}_{\infty})^{\bold{k}}[\bold{G}] + (\tilde{\bold{\Sigma}}^{GW})^{\bold{k}}[\bold{G}](i\omega_n), 
\label{eqn:2cGW_selfenergy}
\end{align}
where $\bold{\Sigma}^{GW}_{\infty}$ is the static self-energy, and $\tilde{\bold{\Sigma}}^{GW}(i\omega_n)$ corresponds to the frequency-dependent contribution to the self-energy which is obtained via the summation of an infinite series of RPA-like `bubble' diagrams \cite{Hedin65}. 
The off-diagonal spin-orbit contributions enter the non-interacting Green's function and contribute to the interacting Green's functions and self-energies through Eq.~\ref{Eq:Dyson} and Eq.~\ref{eqn:2cGW_selfenergy}. 

The static part of the $GW$ self-energy ($\bold{\Sigma}^{GW}_{\infty}$) contains the Hartree- ($\bold{J}$) and exchange-like ($\bold{K}$) terms, 
\begin{align}
(\Sigma^{GW}_{\infty})^{\bold{k}}_{i\sigma,j\sigma'} = J^{\bold{k}}_{i\sigma,j\sigma'} + K^{\bold{k}}_{i\sigma,j\sigma'}, 
\end{align}
which are 
\begin{align}
&J^{\bold{k}}_{i\sigma,j\sigma'} = \frac{\delta_{\sigma\sigma'}}{N_{k}}\sum_{\bold{k}'}\sum_{\sigma_{1}}\sum_{ab}U^{\bold{k}\bold{k}\bold{k}'\bold{k}'}_{i \ j\ a \  b}\gamma^{\bold{k}'}_{b\sigma_{1},a\sigma_{1}}, \label{eqn:HF_J}\\
&K^{\bold{k}}_{i\sigma,j\sigma'} = \frac{-1}{N_{k}}\sum_{\bold{k}'}\sum_{ab} U^{\bold{k}\bold{k}'\bold{k}'\bold{k}}_{i \ b\ a \ j}\gamma^{\bold{k}'}_{b\sigma,a\sigma'},\label{eqn:HF_K}
\end{align}
and $\gamma=-G(\tau=\beta^{-})$ is a correlated one-body spin-density matrix
where $G(\tau)$ is the Fourier transform of the Matsubara Green's function to imaginary time. 
$\tau = \beta^{-}$ is referred to as $\lim_{\tau \rightarrow \beta^{-}}$ since $G(\tau)$ has a discontinuity at $\tau=\beta$ due to its anti-periodicity. 

The dynamic part of the $GW$ self-energy reads
\begin{align}
(\tilde{\Sigma}^{GW}&)^{\bold{k}}_{i\sigma,j\sigma'}(\tau) = \nonumber\\
&-\frac{1}{N_{k}}\sum_{\bold{q}}\sum_{ab} G^{\bold{k-q}}_{a\sigma,b\sigma'}(\tau)\tilde{W}^{\bold{k},\bold{k-q},\bold{k-q},\bold{k}}_{\ i\ \ a \ \ \ \ b \ \ \ j}(-\tau), 
\label{eqn:Sigma_GW}
\end{align}
where $\tilde{W}$ is the spin-free effective screened interaction which contains contributions beyond the static bare interaction $U$ and neglects the vertex corrections in the polarization function. In practice, we employ a density fitting decomposition for the two-body Coulomb integrals~\cite{RI_HF_GW_MP2_2012,RSDF_HongZhou2021}. The decomposition reads 
\begin{align}
U^{\bold{k}_{1}\bold{k}_{2}\bold{k}_{3}\bold{k}_{4}}_{\ i\ j\ \ k\ l} = \sum_{Q}V^{\bold{k}_{1}\bold{k}_{2}}_{\ i\ \ j}(Q)V^{\bold{k}_{3}\bold{k}_{4}}_{\ k\ \ l}(Q), 
\end{align}
where $Q$ is an auxiliary scalar Gaussian basis index and $V^{\bold{k}_{1}\bold{k}_{4}}_{\ i\ \ l}(Q)$ is the three-point integral defined in Eq. 14 from Ref.~\cite{SEET_Sergei20}. This decomposition allows us to express the effective screened interaction $\tilde{W}$ as 
\begin{align}
\tilde{W}^{\bold{k},\bold{k-q},\bold{k-q},\bold{k}}_{\ i\ \ a \ \ \ \ b \ \ \ j}&(\tau) =\nonumber \\ &\sum_{Q,Q'}V^{\bold{k},\bold{k-q}}_{\ i \ \ a}(Q)\tilde{P}^{\bold{q}}_{QQ'}(\tau)V^{\bold{k-q},\bold{k}}_{\ \ b \ \ \ j}(Q') ,
\end{align}
\begin{align}
&\tilde{P}^{\bold{q}}_{QQ}(\tau) = \frac{1}{\beta}\sum_{n}\tilde{P}^{\bold{q}}_{QQ'}(i\Omega_{n})e^{-i\Omega_{n}\tau},\label{Eq:Piw_to_Pt}
\end{align}
where $\tilde{P}^{\bold{q}}(i\Omega_{n})$ is a renormalized auxiliary function, an $N_{Q}$ by $N_{Q}$ matrix for each momentum $\bold{q}$ and bosonic Matsubara  frequency  $\Omega_{n} = 2n\pi/\beta$ ($n=0, \pm1, ...$), given by 
\begin{align}
\tilde{P}^{\bold{q}}(i\Omega_{n}) &= \sum_{m=1}^{\infty}[\tilde{P}^{\bold{q}}_{0}(i\Omega_{n})]^{m} \nonumber\\
&= [I_{Q} - \tilde{P}^{\bold{q}}_{0}(i\Omega_{n})]^{-1}\tilde{P}^{\bold{q}}_{0}(i\Omega_{n})
\end{align}
and 
\begin{align}
\tilde{P}^{\bold{q}}_{0,QQ'}(i\Omega_{n}&) = \int_{0}^{\beta}d\tau \tilde{P}^{\bold{q}}_{0,QQ'}(\tau)e^{i\Omega_{n}\tau}, \label{Eq:P0t_to_P0w}\\
\tilde{P}^{\bold{q}}_{0,QQ'}(\tau)& = \frac{-1}{N_{k}}\sum_{\bold{k}}\sum_{\sigma\sigma'}\sum_{abcd}V^{\bold{k},\bold{k+q}}_{\ d \ a}(Q)\nonumber\\
&\times G^{\bold{k}}_{c\sigma',d\sigma}(-\tau)G^{\bold{k+q}}_{a\sigma \ b\sigma'}(\tau)V^{\bold{k+q},\bold{k}}_{\ \ b\  \ \ c}(Q').\label{Eq:P0_tau}
\end{align}
Note that $\tilde{P}^{\bold{q}}_{0}$ is a non-interacting auxiliary function which differs from a conventional non-interacting polarization function with an additional square root of the Coulomb integrals multiplied from both sides. The same holds for the renormalized auxiliary function $\tilde{P}^{\bold{q}}$. 
Although $\tilde{P}^{\bold{q}}_{0}$ is spin-free, it is affected by SOC effect via the summation over spin indices $\sigma$ and $\sigma'$ in Eq.~\ref{Eq:P0_tau}. 

\subsection{Integrable divergence treatment in two-component sc$GW$ \label{subsec:integrable_div}}
Due to the singularity at the bare Coulomb potential at $\bold{q}=0$ in Eq.~\ref{Eq:V_ijkl}, both the HF exchange potential (Eq.~\ref{eqn:HF_K}) and the dynamic part of the $GW$ self-energy (Eq.~\ref{eqn:Sigma_GW}) have a integrable divergence when any finite $k$-point mesh is used. A simple workaround is to manually exclude the $\bold{G} = 0$ singularity at $\bold{q} = 0$ in the evaluation of two-electron Coulomb integrals, Eq.~\ref{Eq:V_ijkl}. 
However, this will result in a slow convergence to the thermodynamic limit with respect to the number of $k$-points ($N_{k}$). In practical calculations, an additional finite-size correction is crucial to facilitate the convergence to the thermodynamic limit. 

Rewriting the effective screened interaction in the plane-wave basis, we have 
\begin{align}
\tilde{W}&^{\bold{k},\bold{k-q},\bold{k-q},\bold{k}}_{\ i\ \ a \ \ \ \ b \ \ \ j}(\tau)=\frac{1}{\Omega}\sum_{\bold{G}\bold{G}'}\rho^{\bold{k-q}\bold{k}*}_{\ \ a \ \ i}(\bold{G})\frac{\sqrt{4\pi}}{|\bold{q+G}|}\nonumber\\
&\times [\epsilon^{\bold{q},-1}_{\bold{G}\bold{G}'}(\tau) - \delta_{\bold{G}\bold{G}'}]\frac{\sqrt{4\pi}}{|\bold{q+G}|}\rho^{\bold{k-q}\bold{k}}_{\ \ b \ \ j}(\bold{G}'), 
\label{eqn:W_GG'}
\end{align}
where $\rho^{\bold{k-q}\bold{k}}_{\ \ a \ \ i}(\bold{G})$ is the Fourier transform of the pair density $\rho^{\bold{k-q}\bold{k}}_{\ \ a \  \ i}(\bold{r}) = g^{\bold{k-q}*}_{a}(\bold{r})g^{\bold{k}}_{i}(\bold{r})$, and $\epsilon^{\bold{q}}_{\bold{G}\bold{G}'}(\tau)$ is the dielectric function in the plane-wave basis. 
Inserting Eq.~\ref{eqn:W_GG'} into Eq.~\ref{eqn:Sigma_GW}, we arrive at 
\begin{align}
(&\tilde{\Sigma}^{GW})^{\bold{k}}_{i\sigma,j\sigma'}(\tau)=\frac{-1}{N_{k}\Omega}\sum_{\bold{q}}\sum_{ab}\sum_{\bold{G}\bold{G}'}G^{\bold{k-q}}_{a\sigma,b\sigma'}(\tau)\rho^{\bold{k-q}\bold{k}*}_{\ \ a \ \ i}(\bold{G})\nonumber\\
&\times  \frac{\sqrt{4\pi}}{|\bold{q+G}|}[\epsilon^{\bold{q},-1}_{\bold{G}\bold{G}'}(\tau) - \delta_{\bold{G}\bold{G}'}]\frac{\sqrt{4\pi}}{|\bold{q+G'}|}\rho^{\bold{k-q}\bold{k}}_{\ \ b\  \ j}(\bold{G}'). 
\label{eqn:Sigma_GW_PW}
\end{align}
Eq.~\ref{eqn:Sigma_GW_PW} has singularities on the right-hand side at $\bold{q}=\bold{0}$ whenever $\bold{G}=\bold{0}$ or $\bold{G}' = \bold{0}$. However, these divergences are integrable in the limit of infinite $k$-points, $\sum_{\bold{q}}\rightarrow\frac{\Omega N_{\bold{k}}}{2\pi^{3}}\int d\bold{q}$. 
For any finite $k$-point mesh, the singularities can be circumvented by manually neglecting the singularity of Coulomb potential $\sim 1/|\bold{q+G}|^{2}$ at $\bold{q}=\bold{G}=\bold{0}$. The resulting leading-order error is typically referred to as the head correction that comes from $\bold{q}=\bold{G}=\bold{G}'=\bold{0}$. 
Following the procedure proposed by Ref.~\onlinecite{Exdiv_Gygi_1986}, the singularity at the $\bold{q}=\bold{G}=\bold{G}'=\bold{0}$ is eliminated by subtracting and adding an auxiliary function with the same $1/\bold{q}^{2}$ divergence on the right-hand side of Eq.~\ref{eqn:Sigma_GW_PW}. The subtracted term eliminates the divergence which makes the right-hand side of Eq.~\ref{eqn:Sigma_GW_PW} be evaluated accurately using a finite number of $k$-points. The singularity is effectively transferred to the added term which will be evaluated through analytical integration~\cite{Exdiv_Gygi_1986, Exdiv_Carrier_2007,ERI_correction_2009}. 
We use the same auxiliary function proposed in Ref.~\onlinecite{ERI_correction_2009}. 
The head correction of the dynamic part of the $GW$ self-energy can be be expressed as 
\begin{align}
(\Delta&^{GW})^{\bold{k}}_{i\sigma,j\sigma'}(\tau) = -\chi \sum_{ab} G^{\bold{k-q}}_{a\sigma,b\sigma'}(\tau)\rho^{\bold{k-q}\bold{k}*}_{\ \ a \ \ i}(\bold{G})\nonumber\\
&\times[\epsilon^{\bold{q},-1}_{\bold{G}\bold{G}'}(-\tau) - \delta_{\bold{G}\bold{G}'}]\rho^{\bold{k-q}\bold{k}}_{\ \ b \ \ j}(\bold{G}')\Big|_{\bold{q}=\bold{G}=\bold{G}'=\bold{0}}\\
&= -\chi [\epsilon^{\bold{0},-1}_{\bold{00}}(-\tau) - 1] \sum_{ab}S^{\bold{k}}_{ia}G^{\bold{k}}_{a\sigma,b\sigma'}(\tau)S^{\bold{k}}_{bj}, 
\label{eqn:GW_head_corr}
\end{align}
where $\chi$ is the supercell Madelung constant~\cite{ERI_correction_2009}. 
In the present work, $\epsilon^{\bold{q}=\bold{0},-1}_{\bold{G=0},\bold{G'=0}}(\tau)$ is obtained by extrapolating using a least-square fit from finite $q$-points around the $\Gamma$-point. 
Similarly, the static finite-size correction to the HF exchange potential reads~\cite{ERI_correction_2009}
\begin{align}
(\Delta^{\mathrm{HF}})^{\bold{k}}_{i\sigma,j\sigma'} = -\chi \sum_{ab}S^{\bold{k}}_{ib}\gamma^{\bold{k}}_{b\sigma,a\sigma'}S^{\bold{k}}_{aj}, 
\end{align}
where $\gamma$ is a correlated one-body spin-density matrix. 

\section{Computational details} \label{sec:compdet}
We apply both relativistic and non-relativistic~\cite{SEET_Sergei20} sc$GW$ to the electronic structure of silver halides $\mathrm{AgX}$ ($\mathrm{X} = \mathrm{Cl, Br, I}$).  
Silver halides are semiconductors with small indirect band gaps that crystallize in a rock salt structure~\cite{Peralta_JCP2005,Rundong_JCP2016,Kadek_PRB2019}. 
They exhibit a large scalar relativistic effect and an increasing SOC contribution as the halogen is changed from $\mathrm{Cl}$ to $\mathrm{I}$. 
Several calculations are available for comparison~\cite{Rundong_JCP2016,Kadek_PRB2019}.

The equilibrium lattice constant $a_{0}$ of the silver halides are either taken from experiment~\cite{AgCl_expt_a_1955,AgBr_expt_a_2009} or from PBE calculations~\cite{Rundong_JCP2016}. 
Note that the theoretical rock-salt structure optimized at the level of PBE~\cite{Rundong_JCP2016} is adopted for AgI in order to have a direct comparison with Ref.~\onlinecite{Rundong_JCP2016, Kadek_PRB2019}. 
All sc$GW$ calculations are done at the inverse temperature $\beta=700$ a.u.$^{-1}$ ($\sim$ 451 K) with a $6\times 6\times 6$ $k$-mesh in the first Brillouin zone. 

For relativistic calculations, we use the all-electron triple-$\zeta$ bases optimized with respect to X2C Hamiltonians (x2c-TZVPall)~\cite{x2c_cgto_Pollak}. 
The fully uncontracted basis is employed during the constructions of the Dirac Hamiltonian $\mathcal{H}^{\bold{k}}$, the coupling matrix $\bold{X}^{\bold{k}}$, and the X2C1e electronic Hamiltonian $\bold{H}^{\mathrm{X2C1e}}_{+}$. 
Once the X2C electronic Hamiltonian is computed, it is then transformed back to the contracted basis and combined with the two-electron Coulomb interactions. 
For non-relativistic calculations, the all-electron pob-TZVP bases of triple-$\zeta$ quality optimized for solid-state calculations~\cite{pob_crystal_basis_2019} are used for Cl atoms. Since pob-TZVP bases are not available for heavy elements, we employ the all-electron double-$\zeta$ basis sets of \textcite{dzvp_Godbout1992} for Ag, Br, and I. 
Even-tempered Gaussian bases are chosen to decompose the two-body Coulomb integrals using the periodic range-separated Gaussian density fitting recently developed by \textcite{RSDF_HongZhou2021}. 
The number of even-tempered Gaussian basis is found to be converged for orbital energies within 0.001 a.u.. 

Integrals for the periodic X2C1e Hamiltonians and the density-fitted two-body Coulomb integrals, as well as the generalized DFT calculations, are evaluated in a modified version of \texttt{PySCF}~\cite{PySCF_2020}. 
The finite-size effects of HF and $GW$ exchange diagrams are corrected by a supercell Madelung constant using the procedure described in Ref.~\cite{ERI_correction_2005,ERI_correction_2009}. 
A Gaussian nucleus model proposed by Visscher and Dyall~\cite{Gaussian_nucleus_model_Visscher1997} is employed in relativistic calculations. 

To compute the spectral function, the converged single-particle Green's function is analytically continued from the imaginary to the real frequency axis using the Nevanlinna analytical continuation~\cite{Nevanlinna_Jiani_2021} along a high-symmetry $k$-path. This continuation method guarantees causality of the continued function \cite{Nevanlinna_Jiani_2021,Fei21}. A broadening parameter of $\eta = 0.007$ a.u. is used for all calculations. We found that reducing $\eta$ to $\eta=0.005$ and $\eta=0.001$ sharpens the quasiparticle structure but does not result in quantitative differences. 
For a direct comparison to zero-temperature DFT results, a constant chemical potential shift of around 1.0 eV  is applied after the analytical continuation to all finite-temperature spectral function calculations in order to align the highest valence band with the Fermi energy. 

All dynamic quantities such as the Green's functions, self-energies, and polarization function are expanded into the compact intermediate representation (IR)~\cite{IR_Hiroshi_2017} with sparse sampling on both imaginary-time and Matsubara frequency axes~\cite{sparse_sampling_Jia_2020}. Sparse sampling greatly reduces the memory and computation requirements and accelerates the Fourier transforms between the imaginary-time and Matsubara frequency axis. 

\section{Results}\label{sec:res}

We choose to analyze the performance of the sc$GW$ based on the relativistic X2C1e Hamiltonian on a series of silver halides in a periodic code using Gaussian basis. These systems are difficult since both correlation and relativistic effects are assumed to play a significant role in reaching an agreement with experimental values. 

Several correlated $GW$ calculations for  silver halides were reported before using different versions of the $GW$ self-consistency. 
Quasiparticle $GW$ results for silver halides based on the one-shot $G_{0}W_{0}$ or $GW_{0}$ have been reported in Refs.~\cite{high_throughput_GW,GW_AgX_PRB_2018,AgX_LAPW_HLOs, AgCl_optical_2021}. 
In these works, the relativistic effects are treated either using the second variational approach~\cite{AgX_LAPW_HLOs,LAPW} or the pseudopotential~\cite{high_throughput_GW,GW_AgX_PRB_2018}. 
Both Ref.~\onlinecite{GW_AgX_PRB_2018,AgX_LAPW_HLOs} reached a good agreement with experimental values and the reported bandgaps were only slightly underestimated. This is possibly due to the error cancellation between the non-self-consistent approximation and the missing vertex corrections in $GW$~\cite{GW_vertex_Gruneis_2014,GW_vertex_Andrey_2017}. 
A more severe underestimation of band gaps is observed in Ref.~\onlinecite{high_throughput_GW} which could partially be due to choice of pseudopotential~\cite{GW100_benchmark_West_2018,GW_AgX_PRB_2018}. 
Since multiple theoretical differences such as the basis, the level of self-consistency, or the exact version of $GW$ (real axis vs imaginary axis), or even the level of inclusion of relativistic effects are present in these previous works, ultimately we refrain ourselves from comparing to their theoretical results and choose to compare our results to experimental values directly. 

\subsection{X2C1e approximation \label{subsec:x2c1e_AgI}}
We start the discussion of the two-component formalism by examining the quality of the X2C1e-Coulomb and sfX2C1e-Coulomb Hamiltonians (as defined in Sec.~\ref{subsec:x2c1e-Coulomb}) within DFT. This allows us to straightforwardly compare band structure effects to other DFT implementations.
We chose $\mathrm{AgI}$ as our test system because, within the family of silver halides,  both the scalar relativistic effect and the SOC are expected to be the largest for $\mathrm{AgI}$. 

We solve the Kohn-Sham (KS) equation with the PBE exchange-correlation functional for $\mathrm{AgI}$ based on the X2C1e-Coulomb and the sfX2C1e-Coulomb Hamiltonian, and compare these results against the DFT results with Dirac Kohn-Sham Coulomb (DKS-Coulomb), the sfX2C-Coulomb, and the X2C-Coulomb Hamiltonian listed in Refs.~\onlinecite{Rundong_JCP2016,Kadek_PRB2019}. 
The DKS-Coulomb Hamiltonian is solved as a one-electron DKS Hamiltonian whose external potential $V(\bold{r})$ is chosen to be the KS potential using the PBE exchange-correlation functional~\cite{Kadek_PRB2019}. 
The X2C-Coulomb and sfX2C-Coulomb are the corresponding X2C Hamiltonians with and without the SOC contribution from the small component potential~\cite{Rundong_JCP2016}. 
Note that all three Hamiltonians approximate the two-electron Coulomb interaction with the non-relativistic Coulomb operator, neglecting relativistic corrections to the electron-electron interaction. 

\begin{table}[bth]
\begin{ruledtabular}
\begin{tabular}{c|c|cccc}
Non-relativistic & $a_{0}$ & $L-L$ & $\Gamma-\Gamma$ & $X-X$ & $L-X$ \\
\hline
PBE & 6.280 & 4.16 & 3.10 & 3.55 & 1.64 \\
PBE~\cite{Kadek_PRB2019} & 6.280 & 3.99 & 3.11 & 3.54 & 1.59 \\
PBE~\cite{Rundong_JCP2016} & 6.280 & 3.91 & 3.14 & 3.56 & 1.60 \\ 
\hline
Scalar relativistic \\
\hline
sfX2C1e-Coulomb & 6.169 & 3.48 & 2.26 & 3.05 & 0.75 \\
sfX2C-Coulomb~\cite{Rundong_JCP2016} & 6.165 & 3.42 & 2.27 & 3.07 & 0.74 \\
pseudopotential & 6.169 & 3.47 & 2.24 & 3.06 & 0.74 \\
\hline
Fully relativistic \\
\hline
X2C1e-Coulomb & 6.169 & 3.22 & 1.88 & 2.75 & 0.51 \\
DKS-Coulomb~\cite{Kadek_PRB2019} & 6.169 & 3.25 & 1.88 & 2.74 & 0.49 \\
X2C-Coulomb~\cite{Rundong_JCP2016} & 6.169 & 3.17 & 1.90 & 2.76 & 0.49 \\ 
\end{tabular}
\end{ruledtabular}
\caption{Lattice constants $a_{0}$ and energy gaps (eV) of AgI calculated using the PBE functional with various Hamiltonians. 
\label{tab:AgI_tab}}
\end{table}

\begin{figure*}[tb!]
\includegraphics[width=0.32\textwidth]{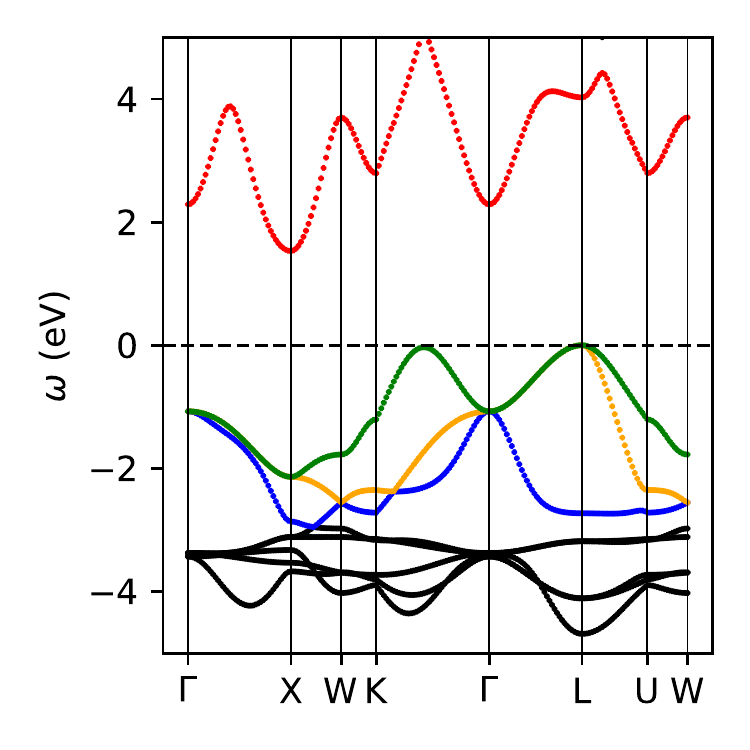}
\includegraphics[width=0.32\textwidth]{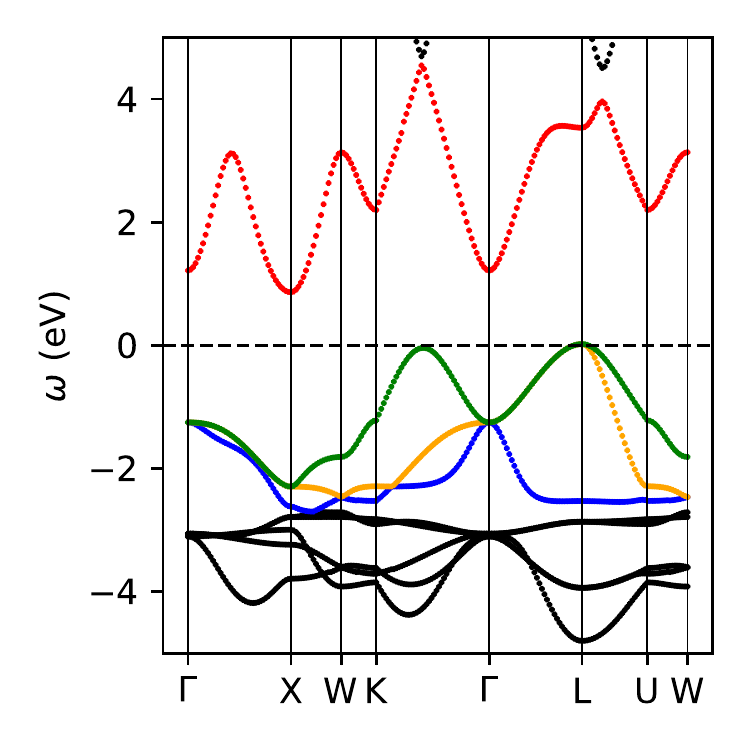}
\includegraphics[width=0.32\textwidth]{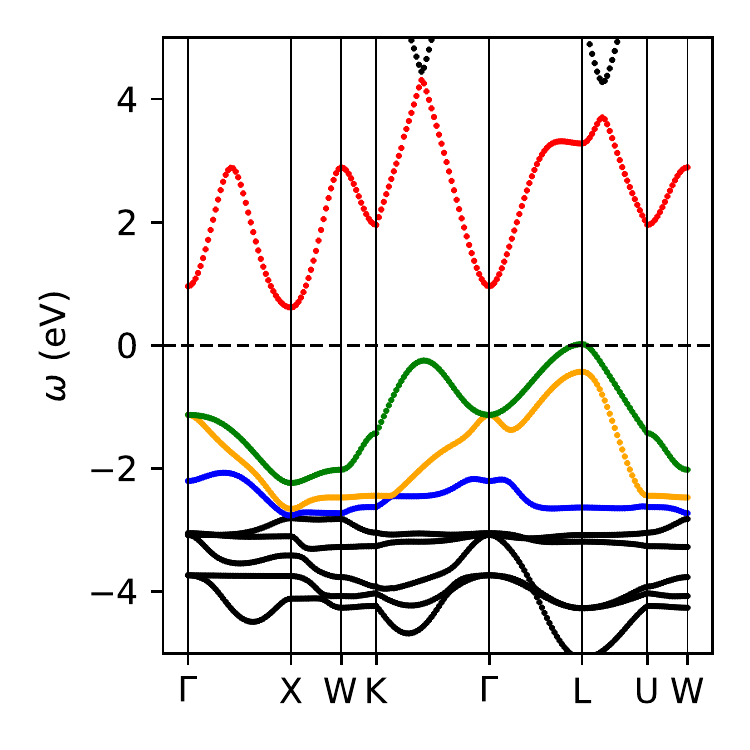}
\caption{The DFT band structure of $\mathrm{AgI}$ calculated using the PBE functional with various Hamiltonians. The left panel is obtained from the non-relativistic Hamiltonian ($a_{0} = 6.280$ \AA). The middle and the right panel show the result from the sfX2C1e-Coulomb and the X2C1e-Coulomb Hamiltonian ($a_{0} = 6.169$ \AA). The coloring of bands is employed to highlight orbitals close to the Fermi level. }\label{fig:AgI_bands}
\end{figure*}

Table~\ref{tab:AgI_tab} shows the band gaps of $\mathrm{AgI}$ at the selected special $k$-points calculated using the PBE functional with various Hamiltonians. Shown are non-relativistic, scalar relativistic, and fully relativistic DFT results. 

We first discuss the sfX2C1e-Coulomb results, where only scalar relativistic effects are included. 
Compared to the non-relativistic calculation, the scalar relativistic effects induce a band-gap narrowing on the order of 1 eV. 
Compared to the more precise sfX2C-Coulomb Hamiltonian, which  additionally includes small component potentials from the non-relativistic Coulomb operator and the PBE exchange-correlation operator (thereby eliminating the picture-change error), the agreement within DFT is excellent.
This quantitative agreement between sfX2C-Coulomb and sfX2C1e-Coulomb suggests that the effect of picture-change error is negligible in the valence bands of $\mathrm{AgI}$. 

\begin{figure}[tb]
\includegraphics[width=0.32\textwidth]{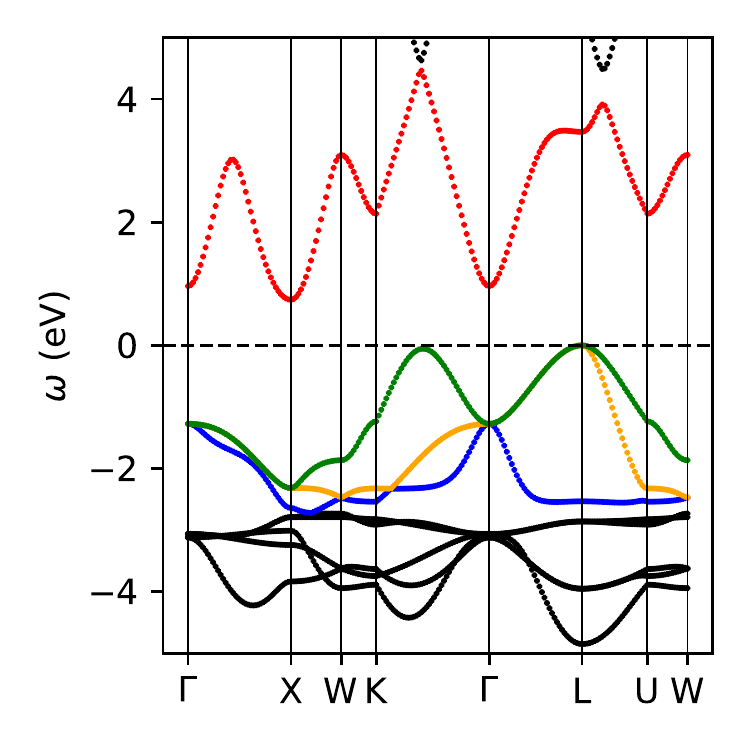}
\caption{The DFT band structure of $\mathrm{AgI}$ calculated using the PBE functional in the Gaussian \emph{gth-tzvp-molopt-sr} basis~\cite{gth_GTO_VandeVondele_2007} and the \emph{gth-pbe} pseudopotential~\cite{GTH_pseudo_1996}. $a_{0} = 6.169$ \AA. The coloring of bands is employed to highlight orbitals close to the Fermi level.}\label{fig:AgI_bands_pp}
\end{figure}

We have also conducted a non-relativistic calculation for $\mathrm{AgI}$ in the Gaussian \emph{gth-tzvp-molopt-sr} basis~\cite{gth_GTO_VandeVondele_2007} and the \emph{gth-pbe} pseudopotential~\cite{GTH_pseudo_1996}. 
The corresponding band gaps (see Table.~\ref{tab:AgI_tab}) and the band structure (see Fig.~\ref{fig:AgI_bands_pp}) are found to be very similar to the ones from the sfX2C1e-Coulomb Hamiltonian. 
Since the \emph{gth-pbe} pseudopotential is optimized in relativistic atomic calculations, it contains the atomic scalar relativistic effects from the core electrons. 
The agreement between the results from the sfX2C1e-Coulomb Hamiltonian and the non-relativistic pseudopotential calculation confirms that the scalar relativistic effects in this system are mostly atomic-like. This agreement justifies the wide usage of such pseudopotentials in real-material simulations. 

Next, we discuss results that include the SOC term in the X2C1e-Coulomb Hamiltonian. 
As shown in Table~\ref{tab:AgI_tab}, band gaps at the selected special $k$-points are again in excellent agreements with results from both the DKS-Coulomb~\cite{Kadek_PRB2019} and the X2C-Coulomb Hamiltonian~\cite{Rundong_JCP2016}. 
Besides the additional band-gap narrowing, the SOC induces non-negligible spin-orbit splittings around the Fermi energy along the high-symmetry $k$ path as shown in the right panel of Fig.~\ref{fig:AgI_bands}.  
Such splittings are found to be both qualitatively and quantitatively consistent with the ones from the more sophisticated four-component DKS-Coulomb Hamiltonian~\cite{Kadek_PRB2019}. 
Similar analysis for AgCl and AgBr can be found in Appendix~\ref{appendix:AgCl_AgBr_PBE_gaps}. 

Fig.~\ref{fig:AgI_bands} illustrates the changes of the band-structure of $\mathrm{AgI}$ as relativistic effects are considered. The left panel shows results from a non-relativistic calculation. The middle panel includes scalar relativistic effects, and the right panel additionally includes SOC. 
As alluded to by Tab.~\ref{tab:AgI_tab}, scalar relativistic effects lead to large changes in the band gap, and spin-orbit coupling to an additional adjustment of the gap and to a remarkable splitting of the orbital degeneracies at the $\Gamma,$ $X$ and $L$ points (see colored bands).

Fig.~\ref{fig:AgI_bands_pp}, which should be compared to the middle panel of Fig.~\ref{fig:AgI_bands}, further illustrates how results from the non-relativistic pseudopotential calculation recover the (scalar relativistic) effects absorbed in the pseudopotential.

\subsection{Relativistic sc$GW$}

\begin{figure*}[tbh]
\includegraphics[width=0.32\textwidth]{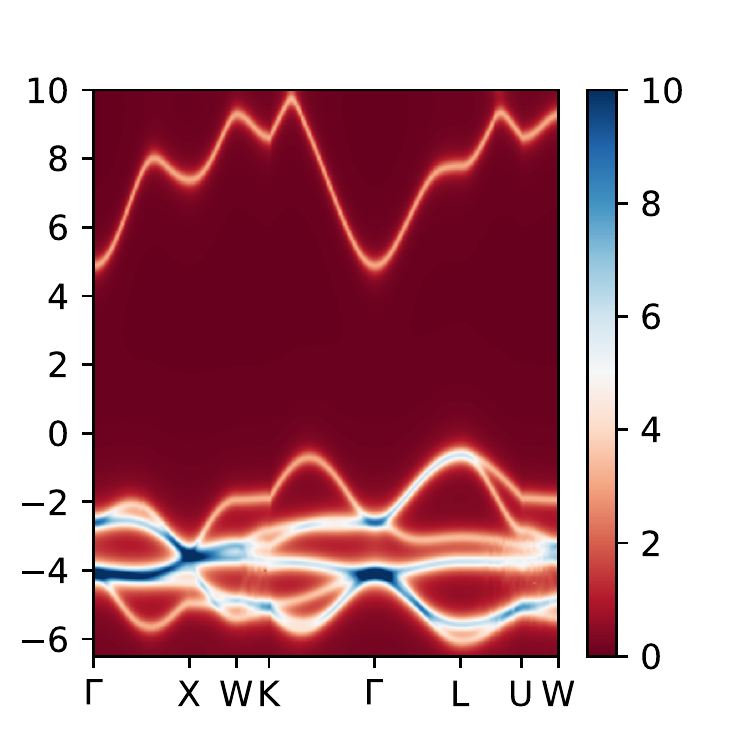}
\includegraphics[width=0.32\textwidth]{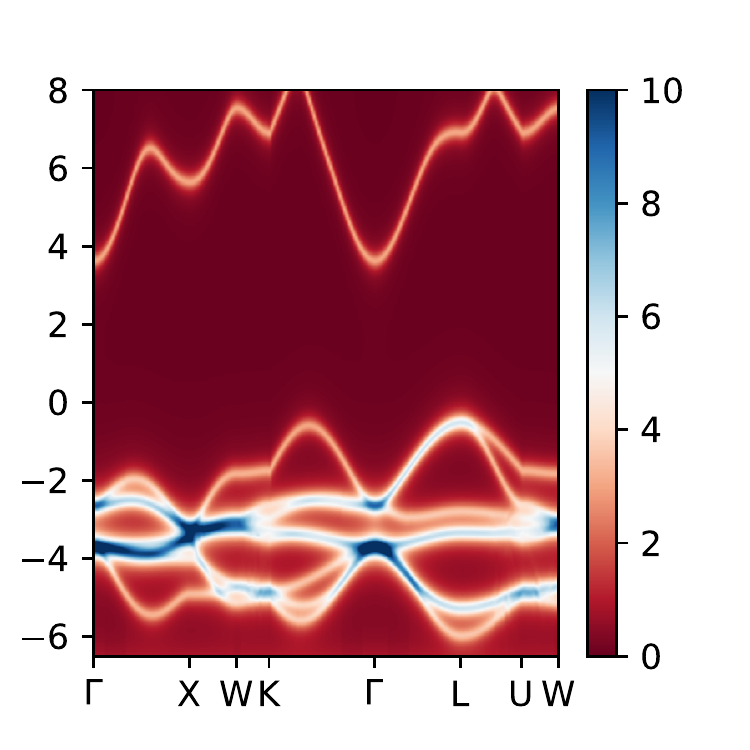}
\includegraphics[width=0.32\textwidth]{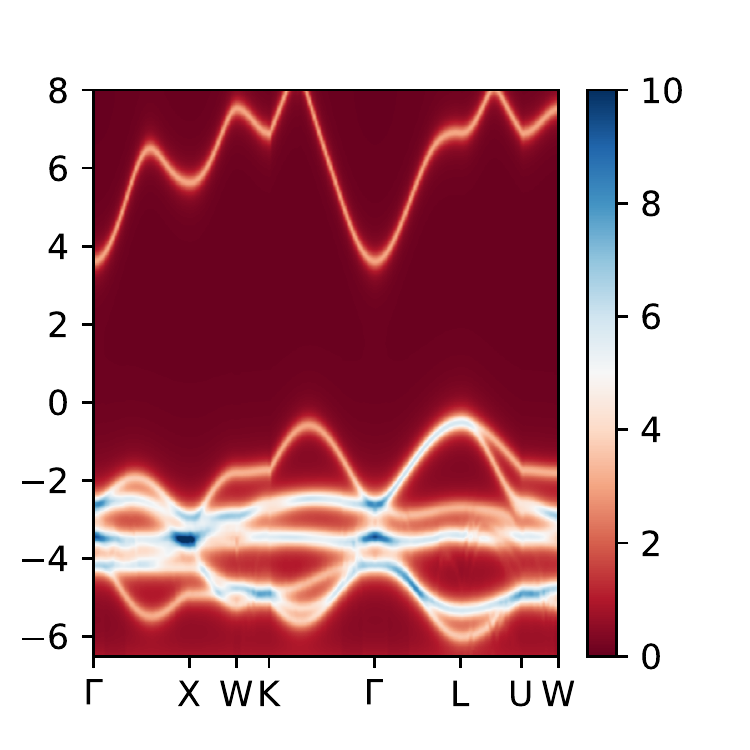} \\
\includegraphics[width=0.32\textwidth]{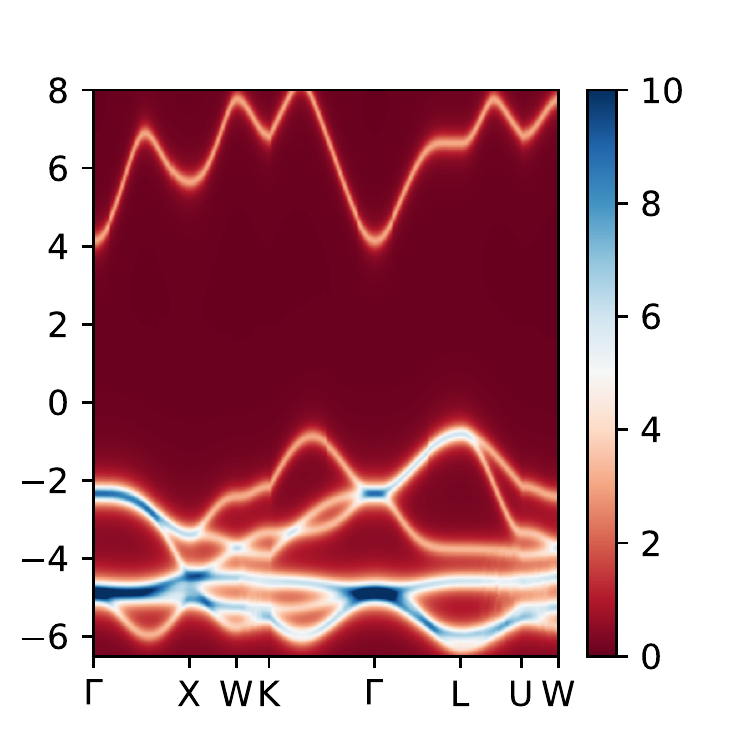}
\includegraphics[width=0.32\textwidth]{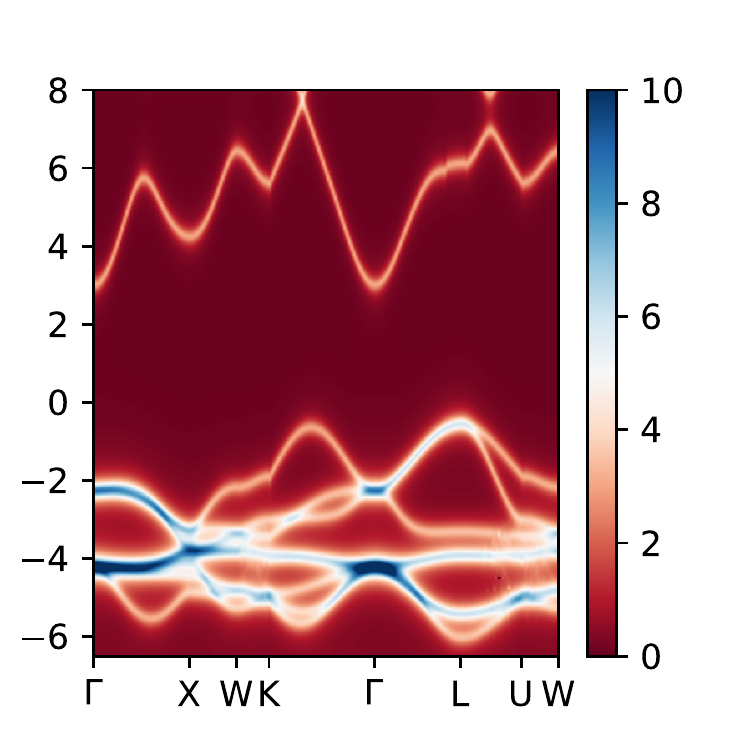}
\includegraphics[width=0.32\textwidth]{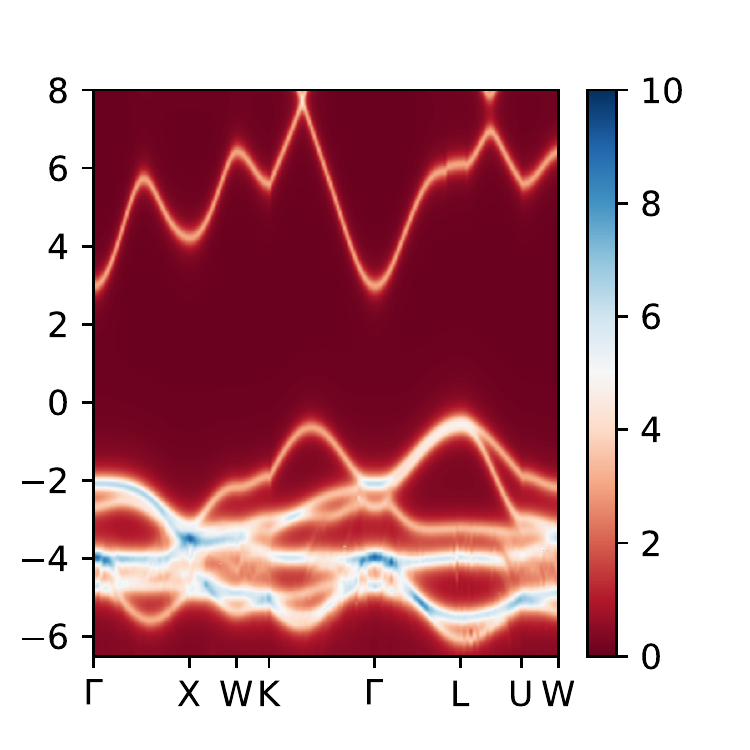}\\
\includegraphics[width=0.32\textwidth]{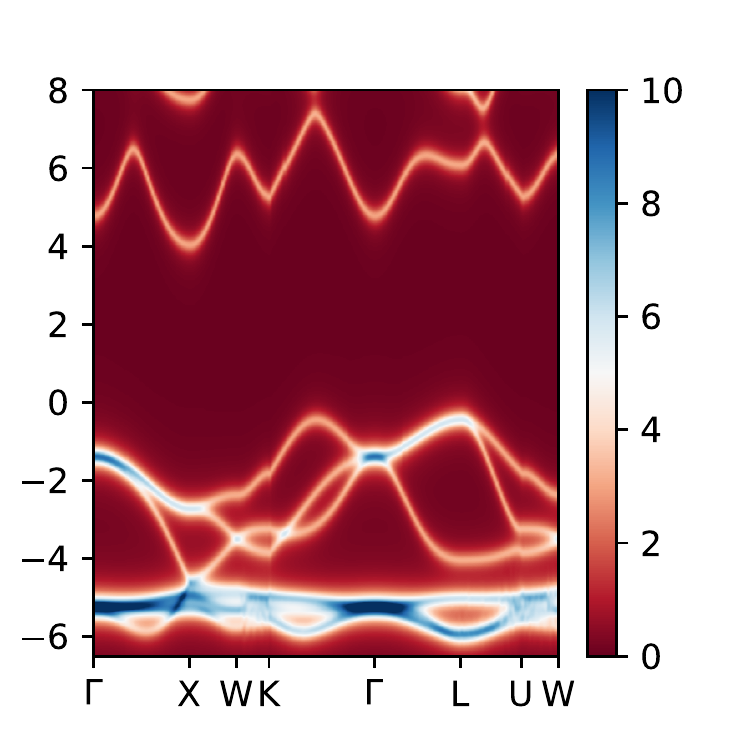}
\includegraphics[width=0.32\textwidth]{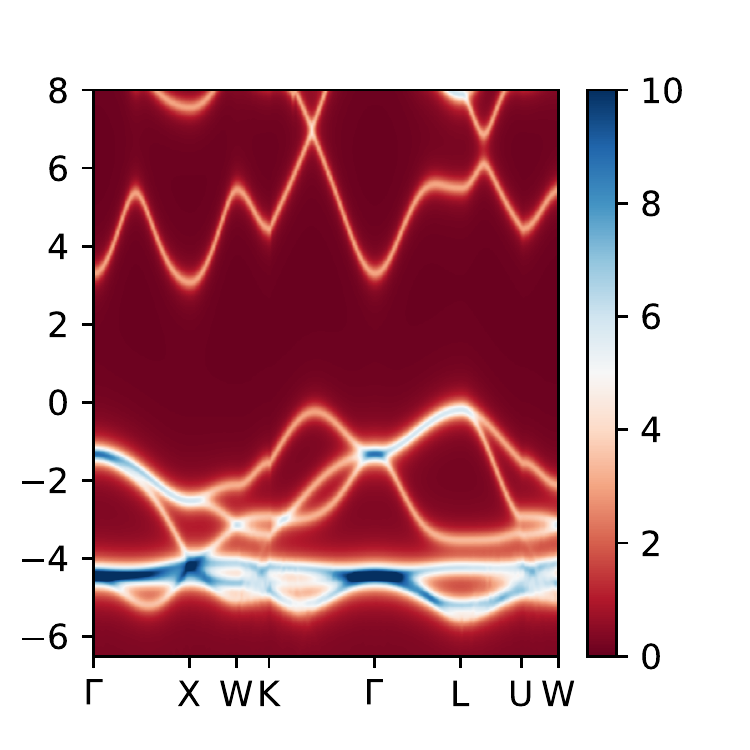}
\includegraphics[width=0.32\textwidth]{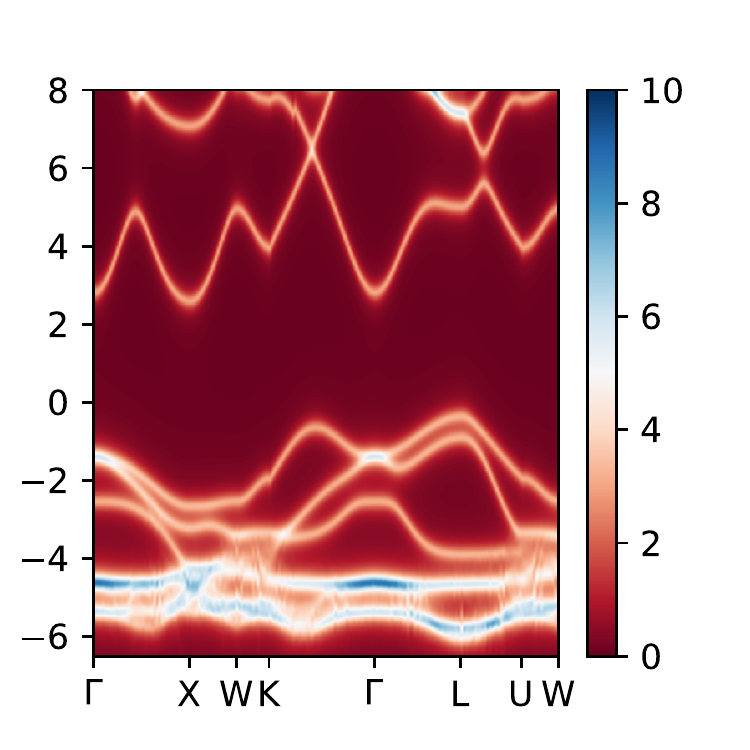}
\caption{The sc$GW$ $k$-resolved spectral functions of $\mathrm{AgCl}$ (top row), $\mathrm{AgBr}$ (middle row), and $\mathrm{AgI}$ (bottom row) calculating from the non-relativistic (left), the sfX2C1e-Coulomb (middle), and the X2C1e-Coulomb (right) Hamiltonian ($a_{0}$ = 5.550, 5.774, and 6.169 \AA \ respectively).}\label{fig:AgX_GW_DOS}
\end{figure*}

Having established the quality of the X2C1e approximation, we now show results from fully self-consistent finite-temperature $GW$ perturbation theory and compare them with available experimental data~\cite{Expt_silver_halides,AgCl_indirect_gap_1983,AgBr_AgCl_indirect_gap,AgBr_direct_gap_1965,AgBr_indirect_gap_1984}. We emphasize that our results are fully self-consistent and conserving solutions of Hedin's equations and contain no further approximations such as quasi-particle or $G_0W_0$ approximations. 
All $GW$ results reported here include finite-size corrections in both the HF and $GW$ exchange diagrams as discussed in Sec.~\ref{subsec:integrable_div}. 
A comparison with and without the finite-size corrections is discussed in Appendix~\ref{app:finite_size}. 
Table~\ref{tab:AgX_GW_gaps} illustrates band gaps at the selected $k$-points for the three compounds $\mathrm{AgCl}$ (top row), $\mathrm{AgBr}$ (middle row), and $\mathrm{AgI}$ (bottom row) calculated using sc$GW$. In Fig.~\ref{fig:AgX_GW_DOS}, left panels show non-relativistic $k$-resolved spectral function calculations; middle panels show spin-free calculations; and right panels show the effects of spin-orbit coupling. 

\begin{table}[bth]
\begin{ruledtabular}
\begin{tabular}{c|cccc}
$\mathrm{AgCl; a_0=5.550}$& $L-L$ & $\Gamma-\Gamma$ & $X-X$ & $L-\Gamma$ \\
\hline
Non-relativistic & 8.42 & 7.42 & 10.92 & 5.45  \\
sfX2C1e-Coulomb & 7.42 & 6.23 & 8.78 & 4.10 \\
X2C1e-Coulomb & 7.42 & 6.18 & 8.50 & 4.09 \\
Expt  & &5.2~\cite{Expt_silver_halides} & & 3.2~\cite{AgCl_indirect_gap_1983}, 3.0~\cite{AgBr_AgCl_indirect_gap}\\
\hline
$\mathrm{AgBr; a_{0}=5.774}$& $L-L$ & $\Gamma-\Gamma$ & $X-X$ & $L-\Gamma$  \\
\hline
Non-relativistic  & 7.48 & 6.47 & 8.94 & 4.98 \\
sfX2C1e-Coulomb  & 6.49 & 5.21 & 7.50 & 3.51 \\
X2C1e-Coulomb  & 6.41 & 5.01 & 7.33 & 3.44 \\
Expt & & 4.3~\cite{AgBr_direct_gap_1965} & & 2.7~\cite{AgBr_indirect_gap_1984}, 2.5~\cite{AgBr_AgCl_indirect_gap} \\
\hline
$\mathrm{AgI}; a_{0}=6.169$& $L-L$ & $\Gamma-\Gamma$ & $X-X$ & $L-X$\\
\hline
Non-relativistic & 6.53 & 6.07 & 6.69 & 4.42 \\
sfX2C1e-Coulomb & 5.60 & 4.48 & 5.54 & 3.22 \\
X2C1e-Coulomb & 5.27 & 4.05 & 5.18 & 2.90 \\
\end{tabular}
\end{ruledtabular}
\caption{Band gaps of AgCl, AgBr, and AgI at special $k$-points calculated using sc$GW$ for Hamiltonians indicated. \label{tab:AgX_GW_gaps}}
\end{table}

\begin{table*}[bth]
\begin{ruledtabular}
\begin{tabular}{cccccccc}
Systems & PBE & PBE & PBE & sc$GW$ & sc$GW$ & sc$GW$ & Expt \\
& NR & sfX2C1e-Coulomb & X2C1e-Coulomb & NR & sfX2c1e-Coulomb & X2C1e-Coulomb \\ 
\hline
AgCl & 1.71 & 0.93 & 0.88 & 5.45 & 4.10 & 4.09 & 3.2~\cite{AgCl_indirect_gap_1983}, 3.0~\cite{AgBr_AgCl_indirect_gap}\\
AgBr & 1.62 & 0.69 & 0.61 & 4.98 & 3.51 & 3.44 & 2.7~\cite{AgBr_indirect_gap_1984}, 2.5~\cite{AgBr_AgCl_indirect_gap}\\ 
AgI & 1.56 & 0.75 & 0.51 & 4.42 & 3.22 & 2.90

\end{tabular}
\end{ruledtabular}
\caption{Indirect band gaps of AgCl, AgBr, and AgI calculated using PBE and sc$GW$ for the non-relativistic (NR), sfX2C1e-Coulomb, and X2C1e-Coulomb Hamiltonians. The indirect band gap occurs between the $L$ and $\Gamma$ point for AgCl, AgBr, and between the $L$ and $X$ point for the rock-salt AgI. Experimental lattice constants are used for AgCl (5.550 \AA) and AgBr (5.774 \AA) while an PBE optimized one is taken for AgI (6.169 \AA). \label{tab:AgX_gaps}}
\end{table*}

We start our discussion with the non-relativistic sc$GW$ $k$-resolved spectral functions as shown in the left column of Fig.~\ref{fig:AgX_GW_DOS}. 
sc$GW$ correctly predicts the indirect band gaps between the $\Gamma$ and the $L$ point for $\mathrm{AgCl}$ and $\mathrm{AgBr}$, and between the $L$ and the $X$ point for $\mathrm{AgI}$. This is consistent with DFT results~\cite{AgX_DFT_Peralta_2005,Rundong_JCP2016,Kadek_PRB2019}. 
An orbitally-resolved analysis of the $GW$ spectral functions suggests that the features around $-2$ eV are mainly of halogen $p$ character, while features around $-4$ eV are dominated by $\mathrm{Ag}$ $d$ orbitals. 
As the weight of the halogen is increased, the hybridization between the halogen $p$ orbitals and the $\mathrm{Ag}$ $d$ orbitals gradually decreases. 
As shown in Table.~\ref{tab:AgX_gaps}, within DFT, the non-relativistic indirect band gaps of $\mathrm{AgX}$ are almost independent of the halogen. In contrast, $GW$ results show increasing indirect band gaps from $\mathrm{I}$ to $\mathrm{Br}$, and to $\mathrm{Cl}$. 
The maximum band-gap widening found in $\mathrm{AgCl}$ is about 3.5 eV compared to the non-relativistic DFT result. 
As compared to experiment, non-relativistic sc$GW$ consistently overestimates the band gaps by up to 2 eV~\cite{Expt_silver_halides,AgCl_indirect_gap_1983,AgBr_indirect_gap_1984,AgBr_AgCl_indirect_gap}.

Next, we discuss the inclusion of scalar relativistic effects through the sfX2C1e-Coulomb Hamiltonian. Similar to what is observed in PBE calculations for $\mathrm{AgI}$, the scalar relativistic effect induces strong band-gap narrowing in all silver halides, rendering the $GW$ predictions closer to the experimental data compared to their non-relativistic counterparts, as shown in Table~\ref{tab:AgX_GW_gaps} and~\ref{tab:AgX_gaps}. 
On the other hand, a similar band-gap narrowing effect pushes the PBE gap values even far away from the experimental values, see Table.~\ref{tab:AgX_gaps}. 
The band-gap narrowing is mainly caused by the orbital contraction of the $\mathrm{Ag}$ $5s$ orbital, which lowers the energy of the lowest conduction band. 
If one measures the scalar relativistic effect in terms of the band-gap narrowing compared to non-relativistic calculations, a similar magnitude of the scalar relativistic effect is observed in all silver halides even though the atomic number of halogens increases from $\mathrm{Cl}$ to $\mathrm{Br}$, and then to $\mathrm{I}$. 
We suspect that a similar band-gap narrowing reflects the strong scalar relativistic effect in the Ag atoms, especially in the conduction bands which corresponds to the strong Ag $5s$ orbital contractions. 

\begin{figure}[tbh]
\includegraphics[width=0.45\textwidth]{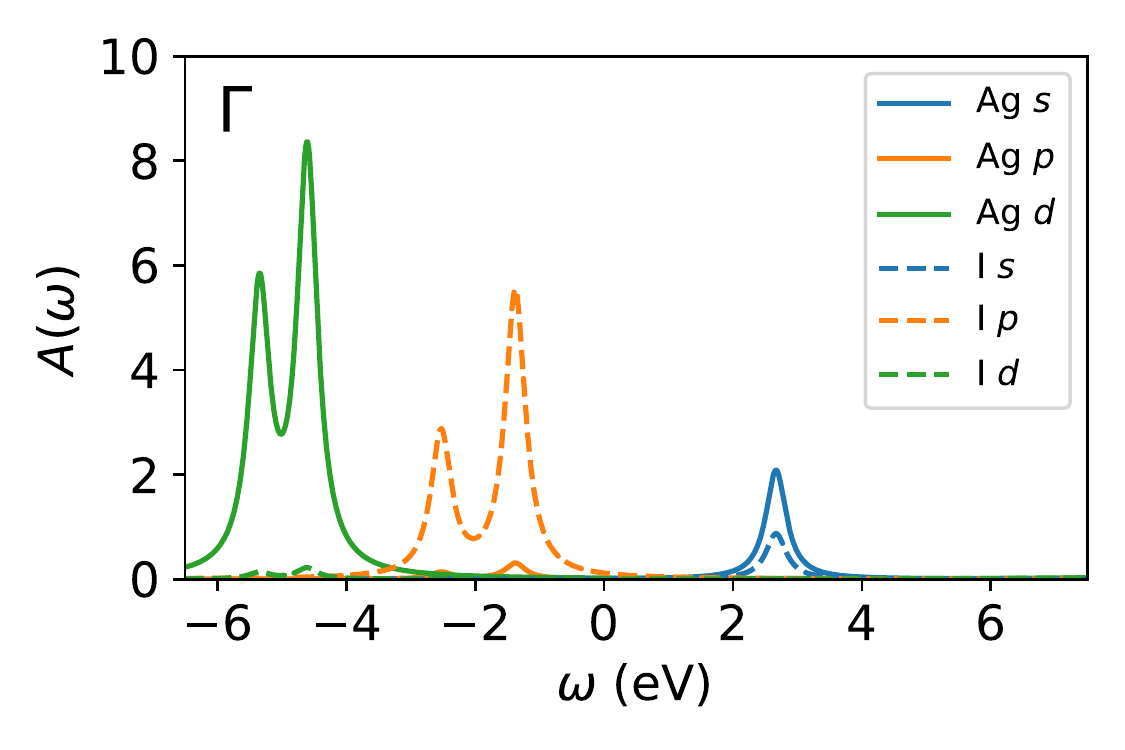}
\includegraphics[width=0.45\textwidth]{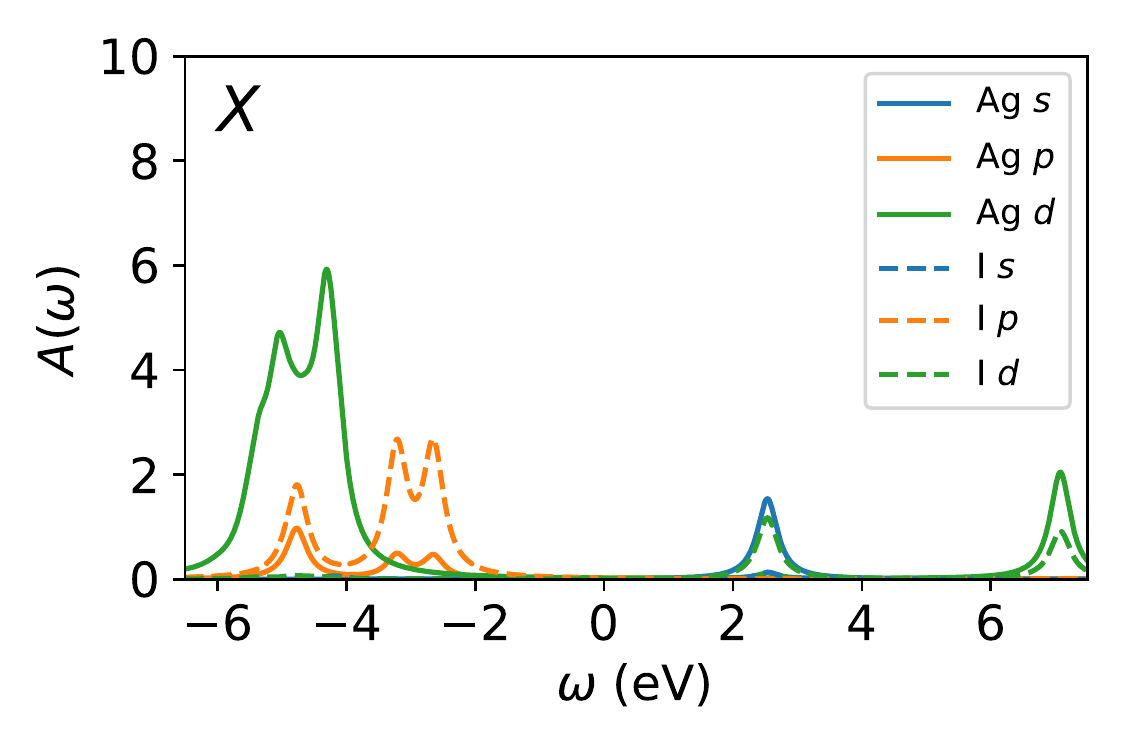}
\includegraphics[width=0.45\textwidth]{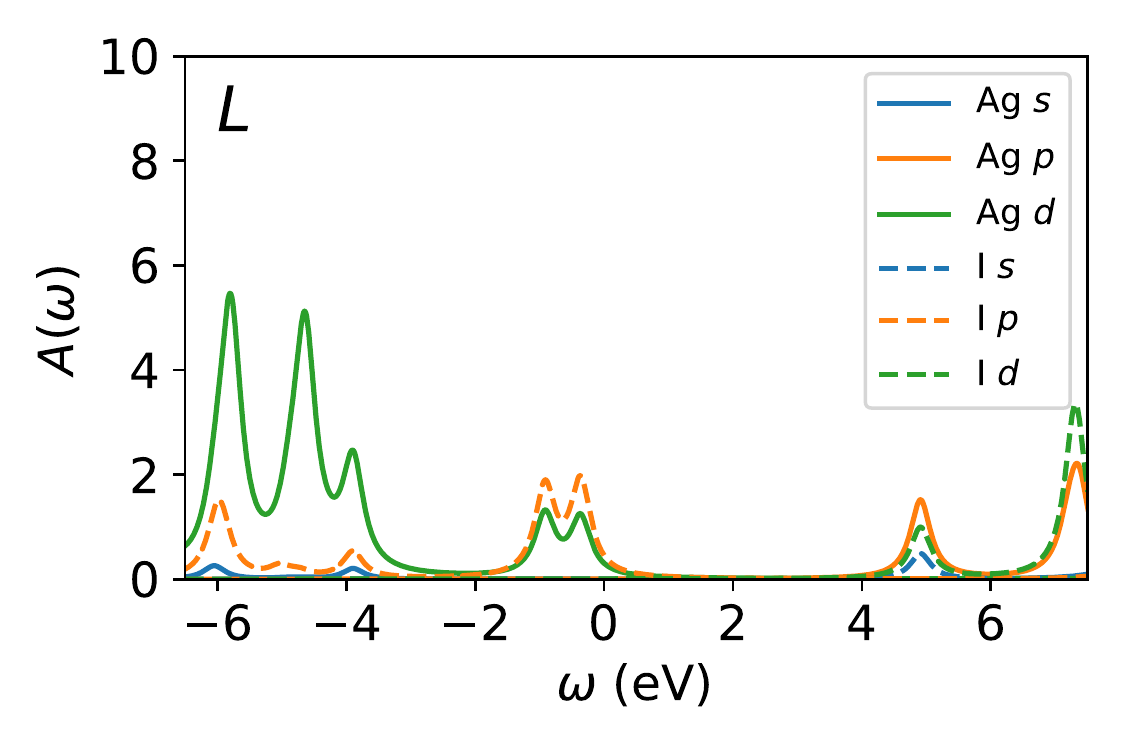}
\caption{The orbital-resolved $k$-dependent spectral functions of $\mathrm{AgI}$ at the $\Gamma$, $X$, and $L$ point, calculated using sc$GW$ based on the X2C1e-Coulomb Hamiltonian. $a_{0} = 6.169$ \AA.}\label{fig:AgI_GW_PDOS}
\end{figure}

Finally, using Fig.~\ref{fig:AgX_GW_DOS},~\ref{fig:AgI_GW_PDOS}, and Table~\ref{tab:SOC_splitting}, we discuss the SOC contribution. 
As expected, spin-orbit effects in $\mathrm{AgCl}$ are weakest. The largest difference of the band gaps occurs at the $X$ point, where the spin-orbit splitting is caused by the $\mathrm{Ag}$ $d$ orbitals.  $\mathrm{Cl}$ $p$ orbitals, which dominate the remainder of the states near the Fermi energy, exhibit much less SOC. 

In $\mathrm{AgBr}$ and $\mathrm{AgI}$, SOC within the X2C1e-Coulomb approximation further reduces the band gaps and causes a substantial spin-orbit splitting around the Fermi energy, rendering the sc$GW$ AgBr band gaps slightly closer to the experimental values. 
Table~\ref{tab:SOC_splitting} shows the spin-orbit splitting gap calculated from PBE, HF, and sc$GW$ at the $\Gamma$ point, which we define as the gap between the $p_{3/2}$  and the $p_{1/2}$ bands. Also shown are splittings at the $X$ and the $L$ points, which are  defined  as the splitting of the $p_{3/2}$ bands due to the cubic crystal field.  Due to the thermal broadening in the finite-temperature sc$GW$ and the broadening introduced by the analytical continuation, we are not able to resolve the small $GW$ orbital splitting of $\mathrm{AgBr}$ at the $X$ point.  Consistently, HF predicts the largest spin-orbit splittings while the ones from PBE are the smallest. 
The differences between HF and sc$GW$ are exclusively due to the additional electron correlation illustrated at the level of $GW$ and its interplay with relativistic effects. 

\begin{table}[tb]
\begin{ruledtabular}
\begin{tabular}{c|ccc|ccc}
&    &  $\mathrm{AgBr}$ & & & $\mathrm{AgI}$ \\
& PBE & HF  & sc$GW$ & PBE & HF & sc$GW$ \\
\hline
$\Gamma$ & 0.55 & 0.71 & 0.60 & 1.07 & 1.34 & 1.15 \\
$X$ & 0.11 & 0.39 & - & 0.42 & 0.66 &  0.56 \\ 
$L$ & 0.13& 0.28 & 0.19 & 0.45 & 0.71 & 0.55 \\
\end{tabular}
\end{ruledtabular}
\caption{Spin-orbit splittings of $\mathrm{AgBr}$ and $\mathrm{AgI}$ at the $\Gamma$, $X$, and $L$ points.}
\label{tab:SOC_splitting}
\end{table}

Fig.~\ref{fig:AgI_GW_PDOS} shows the orbital-resolved $k$-dependent spectral functions of $\mathrm{AgI}$ at the $\Gamma$, the $X$, and the $L$ point,  obtained via $scGW$. The atomic orbital character is defined in terms of symmetrized atomic orbitals (SAO)~\cite{Lowdin_sym_orth_1970} constructed from Gaussian Bloch orbitals. 
Orbitals with the same atomic symmetry are then added up. The height of such a partial orbital-summed spectral function will then reflect the corresponding degeneracy. 
The characters of the low-lying bands varies in the Brillouin zone and  involves $s$, $p$, and $d$-type orbitals from both $\mathrm{Ag}$ and $\mathrm{I}$. 
Different types of orbital admixtures are found at different $k$-points. 
However, for all the $k$-points analyzed here, features between $-4$ to $-6$ eV are dominated by orbitals with $\mathrm{Ag}$ $d$ character. 

At the $\Gamma$ point, the highest two valence bands are dominated by the $\mathrm{I}$ $p$ orbitals. Their six-fold degeneracy is broken in the presence of SOC, resulting in two two-fold degenerate $p_{3/2}$ bands and one two-fold degenerate $p_{1/2}$ band. 
The tallest orange dotted feature corresponds to the $p_{3/2}$ bands. 

At the $X$ point, besides the spin-orbit splitting, the cubic crystal field further splits the $p_{3/2}$ bands into two eigenstates ($m_{j} = \pm3/2$, $\pm1/2$), resulting in a three-peak structure  where the three peaks have a similar peak height. 
In contrast to the $\Gamma$ point, the $\mathrm{I}$ $p$ orbitals hybridize with the $\mathrm{Ag}$ $p$ orbitals with the same orbital splitting pattern.  
Additional orbital mixture is also found between the $\mathrm{Ag}$ $s$ and the $\mathrm{I}$ $d$ orbitals at the lowest conduction band. 

At the $L$ point, $\mathrm{Ag}$ $d$ orbitals start to contribute to the highest valence bands and the $\mathrm{Ag}$ $p$ orbitals hybridize with the lowest conduction band. 
Similar observations can be made for  $\mathrm{AgCl}$ and $\mathrm{AgBr}$. 

The strong $k$-dependence of the orbitals involved in the low-energy physics implies that special care needs to be taken when  low-energy effective model systems are constructed, such as those needed in DMFT and other embedding theories. 

Overall, we found that relativistic effects result in large quantitative differences in the electronic band structure as well as the band gap values. 
While non-relativistic all-electron sc$GW$ tends to significantly overestimate band gaps, relativistic all-electron sc$GW$ renders the theoretical band gaps closer to the experimental values. 
Note that the basis convergence of silver halides has been recently found to be particularly slow due to the Ag $d$ orbitals~\cite{GW_AgX_PRB_2018,AgX_LAPW_HLOs}. 
An analysis of the basis set convergence of our GTOs basis set, as shown in the Appendix~\ref{appendix:basis_set_conv}, suggests a further band-gap narrowing of about $0.1\sim0.2$ eV for our theoretical indirect band gaps from a triple-$\zeta$ basis set (x2c-TZVPall) to a quadruple-$\zeta$ basis set (x2c-QZVPall).  


\section{Conclusions}\label{sec:conc} 
In this paper, we present a formulation of relativistic all-electron sc$GW$ for periodic systems where relativistic effects are treated in the X2C1e approximation. 
The formulation is able to capture electron correlations, one-electron relativistic effects, as well as the interplay of correlations with relativistic effects. It is fully \emph{ab initio}, in the sense that no adjustable parameters are used. 
For systems with weak SOC, the spin separation in the X2C theory provides a promising spin-free approximation whose computational complexity is identical to non-relativistic calculations. 

We present results from the newly implemented methodology  for the silver halides $\mathrm{AgCl}$, $\mathrm{AgBr}$, and $\mathrm{AgI}$. These materials form a sequence of semiconductors with small indirect band gaps where relativistic effects are systematically increasing.
To validate the X2C1e-Coulomb approximation, we test DFT with the X2C1e-Coulomb and sfX2C1e-Coulomb Hamiltonians against reference  4-component DFT calculations and obtain excellent agreement with this more sophisticated approximation. 

By systematically adding relativistic effects in sc$GW$ calculations, we find that electron correlation, relativistics, and their interplay are essential to describe the near-Fermi-surface orbitals. For $\mathrm{AgCl}$ and $\mathrm{AgBr}$, the relativistic sc$GW$ treatment consistently improves agreement with experimental data (no such data is available for $\mathrm{AgI}$). 

The remaining deviations from the experimental values are likely due to a combination of correlation effects ({\it i.e.} beyond-$GW$ diagrammatics), basis-set effects, finite size effects, picture-change errors, and relativistic approximations on the two-particle level. We believe that, of those, the correlation effects form the dominant contribution. Embedding theories such as DMFT~\cite{Kotliar06,Antoine_DMFT_RevModPhys_1996} or SEET~\cite{Alexie_SEET_PRB2015,Dominika_SEET_2017,Lan_Generalized_SEET_2017} provide promising routes to include some of these correlations, at least where they are local. While the {\it ab initio} inclusion of these terms within four-component theories requires major changes to impurity solvers and self-consistencies, as well as additional approximations, we emphasize that one of the main advances
of the X2C1e-Coulomb Hamiltonian is that two-body terms remain unchanged from the non-relativistic version. Non-relativistic diagrammatic implementations of methods such as $GW$, DMFT, or SEET can therefore directly be applied to relativistic problems.

\begin{acknowledgments}
CNY and DZ acknowledged support from NSF grant CHE-1453894. AS and EG were supported by the Simons foundation via the Simons Collaboration on the many-electron problem. 
This research used resources of the National Energy Research Scientific Computing Center (NERSC), a U. S. Department of Energy Office of Science User Facility operated under Contract No. DE-AC02-05CH11231.
\end{acknowledgments}

\bibliography{refs}

\begin{thebibliography}{101}%
\makeatletter
\providecommand \@ifxundefined [1]{%
 \@ifx{#1\undefined}
}%
\providecommand \@ifnum [1]{%
 \ifnum #1\expandafter \@firstoftwo
 \else \expandafter \@secondoftwo
 \fi
}%
\providecommand \@ifx [1]{%
 \ifx #1\expandafter \@firstoftwo
 \else \expandafter \@secondoftwo
 \fi
}%
\providecommand \natexlab [1]{#1}%
\providecommand \enquote  [1]{``#1''}%
\providecommand \bibnamefont  [1]{#1}%
\providecommand \bibfnamefont [1]{#1}%
\providecommand \citenamefont [1]{#1}%
\providecommand \href@noop [0]{\@secondoftwo}%
\providecommand \href [0]{\begingroup \@sanitize@url \@href}%
\providecommand \@href[1]{\@@startlink{#1}\@@href}%
\providecommand \@@href[1]{\endgroup#1\@@endlink}%
\providecommand \@sanitize@url [0]{\catcode `\\12\catcode `\$12\catcode
  `\&12\catcode `\#12\catcode `\^12\catcode `\_12\catcode `\%12\relax}%
\providecommand \@@startlink[1]{}%
\providecommand \@@endlink[0]{}%
\providecommand \url  [0]{\begingroup\@sanitize@url \@url }%
\providecommand \@url [1]{\endgroup\@href {#1}{\urlprefix }}%
\providecommand \urlprefix  [0]{URL }%
\providecommand \Eprint [0]{\href }%
\providecommand \doibase [0]{https://doi.org/}%
\providecommand \selectlanguage [0]{\@gobble}%
\providecommand \bibinfo  [0]{\@secondoftwo}%
\providecommand \bibfield  [0]{\@secondoftwo}%
\providecommand \translation [1]{[#1]}%
\providecommand \BibitemOpen [0]{}%
\providecommand \bibitemStop [0]{}%
\providecommand \bibitemNoStop [0]{.\EOS\space}%
\providecommand \EOS [0]{\spacefactor3000\relax}%
\providecommand \BibitemShut  [1]{\csname bibitem#1\endcsname}%
\let\auto@bib@innerbib\@empty
\bibitem [{\citenamefont {Rachel}(2018)}]{Rachel_2018}%
  \BibitemOpen
  \bibfield  {author} {\bibinfo {author} {\bibfnamefont {S.}~\bibnamefont
  {Rachel}},\ }\bibfield  {title} {\bibinfo {title} {{Interacting Topological
  Insulators: A Review}},\ }\href {https://doi.org/10.1088/1361-6633/aad6a6}
  {\bibfield  {journal} {\bibinfo  {journal} {Reports on Progress in Physics}\
  }\textbf {\bibinfo {volume} {81}},\ \bibinfo {pages} {116501} (\bibinfo
  {year} {2018})}\BibitemShut {NoStop}%
\bibitem [{\citenamefont {Sato}\ and\ \citenamefont {Ando}(2017)}]{Sato_2017}%
  \BibitemOpen
  \bibfield  {author} {\bibinfo {author} {\bibfnamefont {M.}~\bibnamefont
  {Sato}}\ and\ \bibinfo {author} {\bibfnamefont {Y.}~\bibnamefont {Ando}},\
  }\bibfield  {title} {\bibinfo {title} {{Topological Superconductors: A
  Review}},\ }\href {https://doi.org/10.1088/1361-6633/aa6ac7} {\bibfield
  {journal} {\bibinfo  {journal} {Reports on Progress in Physics}\ }\textbf
  {\bibinfo {volume} {80}},\ \bibinfo {pages} {076501} (\bibinfo {year}
  {2017})}\BibitemShut {NoStop}%
\bibitem [{\citenamefont {Zhou}\ \emph {et~al.}(2017)\citenamefont {Zhou},
  \citenamefont {Kanoda},\ and\ \citenamefont {Ng}}]{Zhou_RMP_2017}%
  \BibitemOpen
  \bibfield  {author} {\bibinfo {author} {\bibfnamefont {Y.}~\bibnamefont
  {Zhou}}, \bibinfo {author} {\bibfnamefont {K.}~\bibnamefont {Kanoda}},\ and\
  \bibinfo {author} {\bibfnamefont {T.-K.}\ \bibnamefont {Ng}},\ }\bibfield
  {title} {\bibinfo {title} {{Quantum Spin Liquid States}},\ }\href
  {https://doi.org/10.1103/RevModPhys.89.025003} {\bibfield  {journal}
  {\bibinfo  {journal} {Rev. Mod. Phys.}\ }\textbf {\bibinfo {volume} {89}},\
  \bibinfo {pages} {025003} (\bibinfo {year} {2017})}\BibitemShut {NoStop}%
\bibitem [{\citenamefont {Lv}\ \emph {et~al.}(2021)\citenamefont {Lv},
  \citenamefont {Qian},\ and\ \citenamefont {Ding}}]{Lv_RMP_2021}%
  \BibitemOpen
  \bibfield  {author} {\bibinfo {author} {\bibfnamefont {B.~Q.}\ \bibnamefont
  {Lv}}, \bibinfo {author} {\bibfnamefont {T.}~\bibnamefont {Qian}},\ and\
  \bibinfo {author} {\bibfnamefont {H.}~\bibnamefont {Ding}},\ }\bibfield
  {title} {\bibinfo {title} {{Experimental Perspective on Three-Dimensional
  Topological Semimetals}},\ }\href
  {https://doi.org/10.1103/RevModPhys.93.025002} {\bibfield  {journal}
  {\bibinfo  {journal} {Rev. Mod. Phys.}\ }\textbf {\bibinfo {volume} {93}},\
  \bibinfo {pages} {025002} (\bibinfo {year} {2021})}\BibitemShut {NoStop}%
\bibitem [{\citenamefont {Kim}\ \emph {et~al.}(2008)\citenamefont {Kim},
  \citenamefont {Jin}, \citenamefont {Moon}, \citenamefont {Kim}, \citenamefont
  {Park}, \citenamefont {Leem}, \citenamefont {Yu}, \citenamefont {Noh},
  \citenamefont {Kim}, \citenamefont {Oh}, \citenamefont {Park}, \citenamefont
  {Durairaj}, \citenamefont {Cao},\ and\ \citenamefont
  {Rotenberg}}]{Kim_iridate_PRL_2008}%
  \BibitemOpen
  \bibfield  {author} {\bibinfo {author} {\bibfnamefont {B.~J.}\ \bibnamefont
  {Kim}}, \bibinfo {author} {\bibfnamefont {H.}~\bibnamefont {Jin}}, \bibinfo
  {author} {\bibfnamefont {S.~J.}\ \bibnamefont {Moon}}, \bibinfo {author}
  {\bibfnamefont {J.-Y.}\ \bibnamefont {Kim}}, \bibinfo {author} {\bibfnamefont
  {B.-G.}\ \bibnamefont {Park}}, \bibinfo {author} {\bibfnamefont {C.~S.}\
  \bibnamefont {Leem}}, \bibinfo {author} {\bibfnamefont {J.}~\bibnamefont
  {Yu}}, \bibinfo {author} {\bibfnamefont {T.~W.}\ \bibnamefont {Noh}},
  \bibinfo {author} {\bibfnamefont {C.}~\bibnamefont {Kim}}, \bibinfo {author}
  {\bibfnamefont {S.-J.}\ \bibnamefont {Oh}}, \bibinfo {author} {\bibfnamefont
  {J.-H.}\ \bibnamefont {Park}}, \bibinfo {author} {\bibfnamefont
  {V.}~\bibnamefont {Durairaj}}, \bibinfo {author} {\bibfnamefont
  {G.}~\bibnamefont {Cao}},\ and\ \bibinfo {author} {\bibfnamefont
  {E.}~\bibnamefont {Rotenberg}},\ }\bibfield  {title} {\bibinfo {title}
  {{Novel ${J}_{\mathrm{eff}}=1/2$ Mott State Induced by Relativistic
  Spin-Orbit Coupling in} $\mathrm{Sr_{2}IrO_{4}}$},\ }\href
  {https://doi.org/10.1103/PhysRevLett.101.076402} {\bibfield  {journal}
  {\bibinfo  {journal} {Phys. Rev. Lett.}\ }\textbf {\bibinfo {volume} {101}},\
  \bibinfo {pages} {076402} (\bibinfo {year} {2008})}\BibitemShut {NoStop}%
\bibitem [{\citenamefont {Bramberger}\ \emph {et~al.}(2021)\citenamefont
  {Bramberger}, \citenamefont {Mravlje}, \citenamefont {Grundner},
  \citenamefont {Schollw\"ock},\ and\ \citenamefont
  {Zingl}}]{DMFT_BaOaO3_2021}%
  \BibitemOpen
  \bibfield  {author} {\bibinfo {author} {\bibfnamefont {M.}~\bibnamefont
  {Bramberger}}, \bibinfo {author} {\bibfnamefont {J.}~\bibnamefont {Mravlje}},
  \bibinfo {author} {\bibfnamefont {M.}~\bibnamefont {Grundner}}, \bibinfo
  {author} {\bibfnamefont {U.}~\bibnamefont {Schollw\"ock}},\ and\ \bibinfo
  {author} {\bibfnamefont {M.}~\bibnamefont {Zingl}},\ }\bibfield  {title}
  {\bibinfo {title} {{$\mathrm{BaOsO_{3}}$: A Hund's Metal in the Presence of
  Strong Spin-Orbit Coupling}},\ }\href
  {https://doi.org/10.1103/PhysRevB.103.165133} {\bibfield  {journal} {\bibinfo
   {journal} {Phys. Rev. B}\ }\textbf {\bibinfo {volume} {103}},\ \bibinfo
  {pages} {165133} (\bibinfo {year} {2021})}\BibitemShut {NoStop}%
\bibitem [{\citenamefont {Fiebig}\ \emph {et~al.}(2016)\citenamefont {Fiebig},
  \citenamefont {Lottermoser}, \citenamefont {Meier},\ and\ \citenamefont
  {Trassin}}]{Fiebig2016}%
  \BibitemOpen
  \bibfield  {author} {\bibinfo {author} {\bibfnamefont {M.}~\bibnamefont
  {Fiebig}}, \bibinfo {author} {\bibfnamefont {T.}~\bibnamefont {Lottermoser}},
  \bibinfo {author} {\bibfnamefont {D.}~\bibnamefont {Meier}},\ and\ \bibinfo
  {author} {\bibfnamefont {M.}~\bibnamefont {Trassin}},\ }\bibfield  {title}
  {\bibinfo {title} {{The Evolution of Multiferroics}},\ }\bibfield  {journal}
  {\bibinfo  {journal} {Nature Reviews Materials}\ }\textbf {\bibinfo {volume}
  {1}},\ \href {https://doi.org/10.1038/natrevmats.2016.46}
  {10.1038/natrevmats.2016.46} (\bibinfo {year} {2016})\BibitemShut {NoStop}%
\bibitem [{\citenamefont {Amin}\ \emph {et~al.}(2015)\citenamefont {Amin},
  \citenamefont {Singh},\ and\ \citenamefont
  {Schwingenschl\"ogl}}]{PhysRevB.92.075439}%
  \BibitemOpen
  \bibfield  {author} {\bibinfo {author} {\bibfnamefont {B.}~\bibnamefont
  {Amin}}, \bibinfo {author} {\bibfnamefont {N.}~\bibnamefont {Singh}},\ and\
  \bibinfo {author} {\bibfnamefont {U.}~\bibnamefont {Schwingenschl\"ogl}},\
  }\bibfield  {title} {\bibinfo {title} {{Heterostructures of Transition Metal
  Dichalcogenides}},\ }\href {https://doi.org/10.1103/PhysRevB.92.075439}
  {\bibfield  {journal} {\bibinfo  {journal} {Phys. Rev. B}\ }\textbf {\bibinfo
  {volume} {92}},\ \bibinfo {pages} {075439} (\bibinfo {year}
  {2015})}\BibitemShut {NoStop}%
\bibitem [{\citenamefont {Dirac}\ and\ \citenamefont
  {Fowler}(1928)}]{Dirac_eqn_1928}%
  \BibitemOpen
  \bibfield  {author} {\bibinfo {author} {\bibfnamefont {P.~A.~M.}\
  \bibnamefont {Dirac}}\ and\ \bibinfo {author} {\bibfnamefont {R.~H.}\
  \bibnamefont {Fowler}},\ }\bibfield  {title} {\bibinfo {title} {{The Quantum
  Theory of The Electron}},\ }\href {https://doi.org/10.1098/rspa.1928.0023}
  {\bibfield  {journal} {\bibinfo  {journal} {Proceedings of the Royal Society
  of London. Series A, Containing Papers of a Mathematical and Physical
  Character}\ }\textbf {\bibinfo {volume} {117}},\ \bibinfo {pages} {610}
  (\bibinfo {year} {1928})}\BibitemShut {NoStop}%
\bibitem [{\citenamefont {Saue}\ \emph {et~al.}(2020)\citenamefont {Saue},
  \citenamefont {Bast}, \citenamefont {Gomes}, \citenamefont {Jensen},
  \citenamefont {Visscher}, \citenamefont {Aucar}, \citenamefont {Remigio},
  \citenamefont {Dyall}, \citenamefont {Eliav}, \citenamefont {Fasshauer},
  \citenamefont {Fleig}, \citenamefont {Halbert}, \citenamefont {Hedeg{\aa}rd},
  \citenamefont {Helmich-Paris}, \citenamefont {Ilia{\v{s}}}, \citenamefont
  {Jacob}, \citenamefont {Knecht}, \citenamefont {Laerdahl}, \citenamefont
  {Vidal}, \citenamefont {Nayak}, \citenamefont {Olejniczak}, \citenamefont
  {Olsen}, \citenamefont {Pernpointner}, \citenamefont {Senjean}, \citenamefont
  {Shee}, \citenamefont {Sunaga},\ and\ \citenamefont {van
  Stralen}}]{Dirac_2020}%
  \BibitemOpen
  \bibfield  {author} {\bibinfo {author} {\bibfnamefont {T.}~\bibnamefont
  {Saue}}, \bibinfo {author} {\bibfnamefont {R.}~\bibnamefont {Bast}}, \bibinfo
  {author} {\bibfnamefont {A.~S.~P.}\ \bibnamefont {Gomes}}, \bibinfo {author}
  {\bibfnamefont {H.~J.~A.}\ \bibnamefont {Jensen}}, \bibinfo {author}
  {\bibfnamefont {L.}~\bibnamefont {Visscher}}, \bibinfo {author}
  {\bibfnamefont {I.~A.}\ \bibnamefont {Aucar}}, \bibinfo {author}
  {\bibfnamefont {R.~D.}\ \bibnamefont {Remigio}}, \bibinfo {author}
  {\bibfnamefont {K.~G.}\ \bibnamefont {Dyall}}, \bibinfo {author}
  {\bibfnamefont {E.}~\bibnamefont {Eliav}}, \bibinfo {author} {\bibfnamefont
  {E.}~\bibnamefont {Fasshauer}}, \bibinfo {author} {\bibfnamefont
  {T.}~\bibnamefont {Fleig}}, \bibinfo {author} {\bibfnamefont
  {L.}~\bibnamefont {Halbert}}, \bibinfo {author} {\bibfnamefont {E.~D.}\
  \bibnamefont {Hedeg{\aa}rd}}, \bibinfo {author} {\bibfnamefont
  {B.}~\bibnamefont {Helmich-Paris}}, \bibinfo {author} {\bibfnamefont
  {M.}~\bibnamefont {Ilia{\v{s}}}}, \bibinfo {author} {\bibfnamefont {C.~R.}\
  \bibnamefont {Jacob}}, \bibinfo {author} {\bibfnamefont {S.}~\bibnamefont
  {Knecht}}, \bibinfo {author} {\bibfnamefont {J.~K.}\ \bibnamefont
  {Laerdahl}}, \bibinfo {author} {\bibfnamefont {M.~L.}\ \bibnamefont {Vidal}},
  \bibinfo {author} {\bibfnamefont {M.~K.}\ \bibnamefont {Nayak}}, \bibinfo
  {author} {\bibfnamefont {M.}~\bibnamefont {Olejniczak}}, \bibinfo {author}
  {\bibfnamefont {J.~M.~H.}\ \bibnamefont {Olsen}}, \bibinfo {author}
  {\bibfnamefont {M.}~\bibnamefont {Pernpointner}}, \bibinfo {author}
  {\bibfnamefont {B.}~\bibnamefont {Senjean}}, \bibinfo {author} {\bibfnamefont
  {A.}~\bibnamefont {Shee}}, \bibinfo {author} {\bibfnamefont {A.}~\bibnamefont
  {Sunaga}},\ and\ \bibinfo {author} {\bibfnamefont {J.~N.~P.}\ \bibnamefont
  {van Stralen}},\ }\bibfield  {title} {\bibinfo {title} {{The DIRAC code for
  Relativistic Molecular Calculations}},\ }\href
  {https://doi.org/10.1063/5.0004844} {\bibfield  {journal} {\bibinfo
  {journal} {The Journal of Chemical Physics}\ }\textbf {\bibinfo {volume}
  {152}},\ \bibinfo {pages} {204104} (\bibinfo {year} {2020})}\BibitemShut
  {NoStop}%
\bibitem [{\citenamefont {Zhang}\ \emph {et~al.}(2020)\citenamefont {Zhang},
  \citenamefont {Suo}, \citenamefont {Wang}, \citenamefont {Zhang},
  \citenamefont {Li}, \citenamefont {Lei}, \citenamefont {Zou}, \citenamefont
  {Gao}, \citenamefont {Peng}, \citenamefont {Pu}, \citenamefont {Xiao},
  \citenamefont {Sun}, \citenamefont {Wang}, \citenamefont {Ma}, \citenamefont
  {Wang}, \citenamefont {Guo},\ and\ \citenamefont {Liu}}]{BDF_2020}%
  \BibitemOpen
  \bibfield  {author} {\bibinfo {author} {\bibfnamefont {Y.}~\bibnamefont
  {Zhang}}, \bibinfo {author} {\bibfnamefont {B.}~\bibnamefont {Suo}}, \bibinfo
  {author} {\bibfnamefont {Z.}~\bibnamefont {Wang}}, \bibinfo {author}
  {\bibfnamefont {N.}~\bibnamefont {Zhang}}, \bibinfo {author} {\bibfnamefont
  {Z.}~\bibnamefont {Li}}, \bibinfo {author} {\bibfnamefont {Y.}~\bibnamefont
  {Lei}}, \bibinfo {author} {\bibfnamefont {W.}~\bibnamefont {Zou}}, \bibinfo
  {author} {\bibfnamefont {J.}~\bibnamefont {Gao}}, \bibinfo {author}
  {\bibfnamefont {D.}~\bibnamefont {Peng}}, \bibinfo {author} {\bibfnamefont
  {Z.}~\bibnamefont {Pu}}, \bibinfo {author} {\bibfnamefont {Y.}~\bibnamefont
  {Xiao}}, \bibinfo {author} {\bibfnamefont {Q.}~\bibnamefont {Sun}}, \bibinfo
  {author} {\bibfnamefont {F.}~\bibnamefont {Wang}}, \bibinfo {author}
  {\bibfnamefont {Y.}~\bibnamefont {Ma}}, \bibinfo {author} {\bibfnamefont
  {X.}~\bibnamefont {Wang}}, \bibinfo {author} {\bibfnamefont {Y.}~\bibnamefont
  {Guo}},\ and\ \bibinfo {author} {\bibfnamefont {W.}~\bibnamefont {Liu}},\
  }\bibfield  {title} {\bibinfo {title} {{BDF: A Relativistic Electronic
  Structure Program Package}},\ }\href {https://doi.org/10.1063/1.5143173}
  {\bibfield  {journal} {\bibinfo  {journal} {The Journal of Chemical Physics}\
  }\textbf {\bibinfo {volume} {152}},\ \bibinfo {pages} {064113} (\bibinfo
  {year} {2020})}\BibitemShut {NoStop}%
\bibitem [{\citenamefont {Sun}\ \emph {et~al.}(2020)\citenamefont {Sun},
  \citenamefont {Zhang}, \citenamefont {Banerjee}, \citenamefont {Bao},
  \citenamefont {Barbry}, \citenamefont {Blunt}, \citenamefont {Bogdanov},
  \citenamefont {Booth}, \citenamefont {Chen}, \citenamefont {Cui},
  \citenamefont {Eriksen}, \citenamefont {Gao}, \citenamefont {Guo},
  \citenamefont {Hermann}, \citenamefont {Hermes}, \citenamefont {Koh},
  \citenamefont {Koval}, \citenamefont {Lehtola}, \citenamefont {Li},
  \citenamefont {Liu}, \citenamefont {Mardirossian}, \citenamefont {McClain},
  \citenamefont {Motta}, \citenamefont {Mussard}, \citenamefont {Pham},
  \citenamefont {Pulkin}, \citenamefont {Purwanto}, \citenamefont {Robinson},
  \citenamefont {Ronca}, \citenamefont {Sayfutyarova}, \citenamefont
  {Scheurer}, \citenamefont {Schurkus}, \citenamefont {Smith}, \citenamefont
  {Sun}, \citenamefont {Sun}, \citenamefont {Upadhyay}, \citenamefont {Wagner},
  \citenamefont {Wang}, \citenamefont {White}, \citenamefont {Whitfield},
  \citenamefont {Williamson}, \citenamefont {Wouters}, \citenamefont {Yang},
  \citenamefont {Yu}, \citenamefont {Zhu}, \citenamefont {Berkelbach},
  \citenamefont {Sharma}, \citenamefont {Sokolov},\ and\ \citenamefont
  {Chan}}]{PySCF_2020}%
  \BibitemOpen
  \bibfield  {author} {\bibinfo {author} {\bibfnamefont {Q.}~\bibnamefont
  {Sun}}, \bibinfo {author} {\bibfnamefont {X.}~\bibnamefont {Zhang}}, \bibinfo
  {author} {\bibfnamefont {S.}~\bibnamefont {Banerjee}}, \bibinfo {author}
  {\bibfnamefont {P.}~\bibnamefont {Bao}}, \bibinfo {author} {\bibfnamefont
  {M.}~\bibnamefont {Barbry}}, \bibinfo {author} {\bibfnamefont {N.~S.}\
  \bibnamefont {Blunt}}, \bibinfo {author} {\bibfnamefont {N.~A.}\ \bibnamefont
  {Bogdanov}}, \bibinfo {author} {\bibfnamefont {G.~H.}\ \bibnamefont {Booth}},
  \bibinfo {author} {\bibfnamefont {J.}~\bibnamefont {Chen}}, \bibinfo {author}
  {\bibfnamefont {Z.-H.}\ \bibnamefont {Cui}}, \bibinfo {author} {\bibfnamefont
  {J.~J.}\ \bibnamefont {Eriksen}}, \bibinfo {author} {\bibfnamefont
  {Y.}~\bibnamefont {Gao}}, \bibinfo {author} {\bibfnamefont {S.}~\bibnamefont
  {Guo}}, \bibinfo {author} {\bibfnamefont {J.}~\bibnamefont {Hermann}},
  \bibinfo {author} {\bibfnamefont {M.~R.}\ \bibnamefont {Hermes}}, \bibinfo
  {author} {\bibfnamefont {K.}~\bibnamefont {Koh}}, \bibinfo {author}
  {\bibfnamefont {P.}~\bibnamefont {Koval}}, \bibinfo {author} {\bibfnamefont
  {S.}~\bibnamefont {Lehtola}}, \bibinfo {author} {\bibfnamefont
  {Z.}~\bibnamefont {Li}}, \bibinfo {author} {\bibfnamefont {J.}~\bibnamefont
  {Liu}}, \bibinfo {author} {\bibfnamefont {N.}~\bibnamefont {Mardirossian}},
  \bibinfo {author} {\bibfnamefont {J.~D.}\ \bibnamefont {McClain}}, \bibinfo
  {author} {\bibfnamefont {M.}~\bibnamefont {Motta}}, \bibinfo {author}
  {\bibfnamefont {B.}~\bibnamefont {Mussard}}, \bibinfo {author} {\bibfnamefont
  {H.~Q.}\ \bibnamefont {Pham}}, \bibinfo {author} {\bibfnamefont
  {A.}~\bibnamefont {Pulkin}}, \bibinfo {author} {\bibfnamefont
  {W.}~\bibnamefont {Purwanto}}, \bibinfo {author} {\bibfnamefont {P.~J.}\
  \bibnamefont {Robinson}}, \bibinfo {author} {\bibfnamefont {E.}~\bibnamefont
  {Ronca}}, \bibinfo {author} {\bibfnamefont {E.~R.}\ \bibnamefont
  {Sayfutyarova}}, \bibinfo {author} {\bibfnamefont {M.}~\bibnamefont
  {Scheurer}}, \bibinfo {author} {\bibfnamefont {H.~F.}\ \bibnamefont
  {Schurkus}}, \bibinfo {author} {\bibfnamefont {J.~E.~T.}\ \bibnamefont
  {Smith}}, \bibinfo {author} {\bibfnamefont {C.}~\bibnamefont {Sun}}, \bibinfo
  {author} {\bibfnamefont {S.-N.}\ \bibnamefont {Sun}}, \bibinfo {author}
  {\bibfnamefont {S.}~\bibnamefont {Upadhyay}}, \bibinfo {author}
  {\bibfnamefont {L.~K.}\ \bibnamefont {Wagner}}, \bibinfo {author}
  {\bibfnamefont {X.}~\bibnamefont {Wang}}, \bibinfo {author} {\bibfnamefont
  {A.}~\bibnamefont {White}}, \bibinfo {author} {\bibfnamefont {J.~D.}\
  \bibnamefont {Whitfield}}, \bibinfo {author} {\bibfnamefont {M.~J.}\
  \bibnamefont {Williamson}}, \bibinfo {author} {\bibfnamefont
  {S.}~\bibnamefont {Wouters}}, \bibinfo {author} {\bibfnamefont
  {J.}~\bibnamefont {Yang}}, \bibinfo {author} {\bibfnamefont {J.~M.}\
  \bibnamefont {Yu}}, \bibinfo {author} {\bibfnamefont {T.}~\bibnamefont
  {Zhu}}, \bibinfo {author} {\bibfnamefont {T.~C.}\ \bibnamefont {Berkelbach}},
  \bibinfo {author} {\bibfnamefont {S.}~\bibnamefont {Sharma}}, \bibinfo
  {author} {\bibfnamefont {A.~Y.}\ \bibnamefont {Sokolov}},\ and\ \bibinfo
  {author} {\bibfnamefont {G.~K.-L.}\ \bibnamefont {Chan}},\ }\bibfield
  {title} {\bibinfo {title} {{Recent Developments in the PySCF Program
  Package}},\ }\href {https://doi.org/10.1063/5.0006074} {\bibfield  {journal}
  {\bibinfo  {journal} {The Journal of Chemical Physics}\ }\textbf {\bibinfo
  {volume} {153}},\ \bibinfo {pages} {024109} (\bibinfo {year}
  {2020})}\BibitemShut {NoStop}%
\bibitem [{\citenamefont {Repisky}\ \emph {et~al.}(2020)\citenamefont
  {Repisky}, \citenamefont {Komorovsky}, \citenamefont {Kadek}, \citenamefont
  {Konecny}, \citenamefont {Ekstr\"{o}m}, \citenamefont {Malkin}, \citenamefont
  {Kaupp}, \citenamefont {Ruud}, \citenamefont {Malkina},\ and\ \citenamefont
  {Malkin}}]{ReSpect_Repisky2020}%
  \BibitemOpen
  \bibfield  {author} {\bibinfo {author} {\bibfnamefont {M.}~\bibnamefont
  {Repisky}}, \bibinfo {author} {\bibfnamefont {S.}~\bibnamefont {Komorovsky}},
  \bibinfo {author} {\bibfnamefont {M.}~\bibnamefont {Kadek}}, \bibinfo
  {author} {\bibfnamefont {L.}~\bibnamefont {Konecny}}, \bibinfo {author}
  {\bibfnamefont {U.}~\bibnamefont {Ekstr\"{o}m}}, \bibinfo {author}
  {\bibfnamefont {E.}~\bibnamefont {Malkin}}, \bibinfo {author} {\bibfnamefont
  {M.}~\bibnamefont {Kaupp}}, \bibinfo {author} {\bibfnamefont
  {K.}~\bibnamefont {Ruud}}, \bibinfo {author} {\bibfnamefont {O.~L.}\
  \bibnamefont {Malkina}},\ and\ \bibinfo {author} {\bibfnamefont {V.~G.}\
  \bibnamefont {Malkin}},\ }\bibfield  {title} {\bibinfo {title} {{ReSpect:
  Relativistic Spectroscopy DFT Program Package}},\ }\href
  {https://doi.org/10.1063/5.0005094} {\bibfield  {journal} {\bibinfo
  {journal} {The Journal of Chemical Physics}\ }\textbf {\bibinfo {volume}
  {152}},\ \bibinfo {pages} {184101} (\bibinfo {year} {2020})}\BibitemShut
  {NoStop}%
\bibitem [{\citenamefont {Visscher}(2002)}]{review_Visscher2002}%
  \BibitemOpen
  \bibfield  {author} {\bibinfo {author} {\bibfnamefont {L.}~\bibnamefont
  {Visscher}},\ }\bibfield  {title} {\bibinfo {title} {{Chapter 6 Post
  Dirac-Hartree-Fock Methods-Electron Correlation}},\ }in\ \href
  {https://doi.org/10.1016/s1380-7323(02)80032-2} {\emph {\bibinfo {booktitle}
  {Theoretical and Computational Chemistry}}}\ (\bibinfo  {publisher}
  {Elsevier},\ \bibinfo {year} {2002})\ pp.\ \bibinfo {pages}
  {291--331}\BibitemShut {NoStop}%
\bibitem [{\citenamefont
  {Fleig}(2012)}]{review_relativistic_correlated_method_Fleig2012}%
  \BibitemOpen
  \bibfield  {author} {\bibinfo {author} {\bibfnamefont {T.}~\bibnamefont
  {Fleig}},\ }\bibfield  {title} {\bibinfo {title} {{Invited review:
  Relativistic Wave-Function based Electron Correlation Methods}},\ }\href
  {https://doi.org/10.1016/j.chemphys.2011.06.032} {\bibfield  {journal}
  {\bibinfo  {journal} {Chemical Physics}\ }\textbf {\bibinfo {volume} {395}},\
  \bibinfo {pages} {2} (\bibinfo {year} {2012})}\BibitemShut {NoStop}%
\bibitem [{\citenamefont {Liu}(2010)}]{Liu_relativistic_review_2010}%
  \BibitemOpen
  \bibfield  {author} {\bibinfo {author} {\bibfnamefont {W.}~\bibnamefont
  {Liu}},\ }\bibfield  {title} {\bibinfo {title} {{Ideas of Relativistic
  Quantum Chemistry}},\ }\href {https://doi.org/10.1080/00268971003781571}
  {\bibfield  {journal} {\bibinfo  {journal} {Molecular Physics}\ }\textbf
  {\bibinfo {volume} {108}},\ \bibinfo {pages} {1679} (\bibinfo {year}
  {2010})}\BibitemShut {NoStop}%
\bibitem [{\citenamefont {Saue}(2011)}]{Saue_Review}%
  \BibitemOpen
  \bibfield  {author} {\bibinfo {author} {\bibfnamefont {T.}~\bibnamefont
  {Saue}},\ }\bibfield  {title} {\bibinfo {title} {{Relativistic Hamiltonians
  for Chemistry: A Primer}},\ }\href
  {https://doi.org/https://doi.org/10.1002/cphc.201100682} {\bibfield
  {journal} {\bibinfo  {journal} {ChemPhysChem}\ }\textbf {\bibinfo {volume}
  {12}},\ \bibinfo {pages} {3077} (\bibinfo {year} {2011})}\BibitemShut
  {NoStop}%
\bibitem [{\citenamefont {Reiher}\ and\ \citenamefont {Wolf}(2004)}]{DKH_2004}%
  \BibitemOpen
  \bibfield  {author} {\bibinfo {author} {\bibfnamefont {M.}~\bibnamefont
  {Reiher}}\ and\ \bibinfo {author} {\bibfnamefont {A.}~\bibnamefont {Wolf}},\
  }\bibfield  {title} {\bibinfo {title} {{Exact Decoupling of the Dirac
  Hamiltonian. II. The Generalized Douglas-Kroll-Hess Transformation up to
  Arbitrary Order}},\ }\href {https://doi.org/10.1063/1.1818681} {\bibfield
  {journal} {\bibinfo  {journal} {The Journal of Chemical Physics}\ }\textbf
  {\bibinfo {volume} {121}},\ \bibinfo {pages} {10945} (\bibinfo {year}
  {2004})}\BibitemShut {NoStop}%
\bibitem [{\citenamefont {Chang}\ \emph {et~al.}(1986)\citenamefont {Chang},
  \citenamefont {Pelissier},\ and\ \citenamefont {Durand}}]{ZORA_1986}%
  \BibitemOpen
  \bibfield  {author} {\bibinfo {author} {\bibfnamefont {C.}~\bibnamefont
  {Chang}}, \bibinfo {author} {\bibfnamefont {M.}~\bibnamefont {Pelissier}},\
  and\ \bibinfo {author} {\bibfnamefont {P.}~\bibnamefont {Durand}},\
  }\bibfield  {title} {\bibinfo {title} {{Regular Two-Component Pauli-Like
  Effective Hamiltonians in Dirac Theory}},\ }\href
  {https://doi.org/10.1088/0031-8949/34/5/007} {\bibfield  {journal} {\bibinfo
  {journal} {Physica Scripta}\ }\textbf {\bibinfo {volume} {34}},\ \bibinfo
  {pages} {394} (\bibinfo {year} {1986})}\BibitemShut {NoStop}%
\bibitem [{\citenamefont {van Lenthe}\ \emph {et~al.}(1993)\citenamefont {van
  Lenthe}, \citenamefont {Baerends},\ and\ \citenamefont
  {Snijders}}]{ZORA_1993}%
  \BibitemOpen
  \bibfield  {author} {\bibinfo {author} {\bibfnamefont {E.}~\bibnamefont {van
  Lenthe}}, \bibinfo {author} {\bibfnamefont {E.~J.}\ \bibnamefont
  {Baerends}},\ and\ \bibinfo {author} {\bibfnamefont {J.~G.}\ \bibnamefont
  {Snijders}},\ }\bibfield  {title} {\bibinfo {title} {{Relativistic Regular
  Two-Component Hamiltonians}},\ }\href {https://doi.org/10.1063/1.466059}
  {\bibfield  {journal} {\bibinfo  {journal} {The Journal of Chemical Physics}\
  }\textbf {\bibinfo {volume} {99}},\ \bibinfo {pages} {4597} (\bibinfo {year}
  {1993})}\BibitemShut {NoStop}%
\bibitem [{\citenamefont {Cremer}\ \emph {et~al.}(2014)\citenamefont {Cremer},
  \citenamefont {Zou},\ and\ \citenamefont {Filatov}}]{NESC_Review_Cremer2014}%
  \BibitemOpen
  \bibfield  {author} {\bibinfo {author} {\bibfnamefont {D.}~\bibnamefont
  {Cremer}}, \bibinfo {author} {\bibfnamefont {W.}~\bibnamefont {Zou}},\ and\
  \bibinfo {author} {\bibfnamefont {M.}~\bibnamefont {Filatov}},\ }\bibfield
  {title} {\bibinfo {title} {{Dirac-Exact Relativistic Methods: the Normalized
  Elimination of the Small Component Method}},\ }\href
  {https://doi.org/10.1002/wcms.1181} {\bibfield  {journal} {\bibinfo
  {journal} {{WIREs} Computational Molecular Science}\ }\textbf {\bibinfo
  {volume} {4}},\ \bibinfo {pages} {436} (\bibinfo {year} {2014})}\BibitemShut
  {NoStop}%
\bibitem [{\citenamefont {Scherpelz}\ \emph {et~al.}(2016)\citenamefont
  {Scherpelz}, \citenamefont {Govoni}, \citenamefont {Hamada},\ and\
  \citenamefont {Galli}}]{2cG0W0_Scherpelz_2016}%
  \BibitemOpen
  \bibfield  {author} {\bibinfo {author} {\bibfnamefont {P.}~\bibnamefont
  {Scherpelz}}, \bibinfo {author} {\bibfnamefont {M.}~\bibnamefont {Govoni}},
  \bibinfo {author} {\bibfnamefont {I.}~\bibnamefont {Hamada}},\ and\ \bibinfo
  {author} {\bibfnamefont {G.}~\bibnamefont {Galli}},\ }\bibfield  {title}
  {\bibinfo {title} {{Implementation and Validation of Fully Relativistic $GW$
  Calculations: Spin-Orbit Coupling in Molecules, Nanocrystals, and Solids}},\
  }\href {https://doi.org/10.1021/acs.jctc.6b00114} {\bibfield  {journal}
  {\bibinfo  {journal} {Journal of Chemical Theory and Computation}\ }\textbf
  {\bibinfo {volume} {12}},\ \bibinfo {pages} {3523} (\bibinfo {year}
  {2016})}\BibitemShut {NoStop}%
\bibitem [{\citenamefont {Loucks}(1965)}]{relativistic_APW_1965}%
  \BibitemOpen
  \bibfield  {author} {\bibinfo {author} {\bibfnamefont {T.~L.}\ \bibnamefont
  {Loucks}},\ }\bibfield  {title} {\bibinfo {title} {{Relativistic Electronic
  Structure in Crystals. I. Theory}},\ }\href
  {https://doi.org/10.1103/PhysRev.139.A1333} {\bibfield  {journal} {\bibinfo
  {journal} {Phys. Rev.}\ }\textbf {\bibinfo {volume} {139}},\ \bibinfo {pages}
  {A1333} (\bibinfo {year} {1965})}\BibitemShut {NoStop}%
\bibitem [{\citenamefont {MacDonald}\ \emph {et~al.}(1980)\citenamefont
  {MacDonald}, \citenamefont {Picket},\ and\ \citenamefont
  {Koelling}}]{LAPW_MacDonald_1980}%
  \BibitemOpen
  \bibfield  {author} {\bibinfo {author} {\bibfnamefont {A.~H.}\ \bibnamefont
  {MacDonald}}, \bibinfo {author} {\bibfnamefont {W.~E.}\ \bibnamefont
  {Picket}},\ and\ \bibinfo {author} {\bibfnamefont {D.~D.}\ \bibnamefont
  {Koelling}},\ }\bibfield  {title} {\bibinfo {title} {{A Linearised
  Relativistic Augmented-Plane-Wave Method Utilising Approximate Pure Spin
  Basis Functions}},\ }\href {https://doi.org/10.1088/0022-3719/13/14/009}
  {\bibfield  {journal} {\bibinfo  {journal} {Journal of Physics C: Solid State
  Physics}\ }\textbf {\bibinfo {volume} {13}},\ \bibinfo {pages} {2675}
  (\bibinfo {year} {1980})}\BibitemShut {NoStop}%
\bibitem [{\citenamefont {Wimmer}\ \emph {et~al.}(1981)\citenamefont {Wimmer},
  \citenamefont {Krakauer}, \citenamefont {Weinert},\ and\ \citenamefont
  {Freeman}}]{LAPW_Wimmer_1981}%
  \BibitemOpen
  \bibfield  {author} {\bibinfo {author} {\bibfnamefont {E.}~\bibnamefont
  {Wimmer}}, \bibinfo {author} {\bibfnamefont {H.}~\bibnamefont {Krakauer}},
  \bibinfo {author} {\bibfnamefont {M.}~\bibnamefont {Weinert}},\ and\ \bibinfo
  {author} {\bibfnamefont {A.~J.}\ \bibnamefont {Freeman}},\ }\bibfield
  {title} {\bibinfo {title} {{Full-Potential Self-Consistent
  Linearized-Augmented-Plane-Wave Method for Calculating the Electronic
  Structure of Molecules and Surfaces:} ${\mathrm{o}}_{2}$ {Molecule}},\ }\href
  {https://doi.org/10.1103/PhysRevB.24.864} {\bibfield  {journal} {\bibinfo
  {journal} {Phys. Rev. B}\ }\textbf {\bibinfo {volume} {24}},\ \bibinfo
  {pages} {864} (\bibinfo {year} {1981})}\BibitemShut {NoStop}%
\bibitem [{\citenamefont {Ebert}(1988)}]{relativistic_LMTO_1988}%
  \BibitemOpen
  \bibfield  {author} {\bibinfo {author} {\bibfnamefont {H.}~\bibnamefont
  {Ebert}},\ }\bibfield  {title} {\bibinfo {title} {{Two Ways to Perform
  Spin-Polarized Relativistic Linear Muffin-Tin-Orbital Calculations}},\ }\href
  {https://doi.org/10.1103/PhysRevB.38.9390} {\bibfield  {journal} {\bibinfo
  {journal} {Phys. Rev. B}\ }\textbf {\bibinfo {volume} {38}},\ \bibinfo
  {pages} {9390} (\bibinfo {year} {1988})}\BibitemShut {NoStop}%
\bibitem [{\citenamefont {Ebert}\ \emph {et~al.}(1988)\citenamefont {Ebert},
  \citenamefont {Strange},\ and\ \citenamefont
  {Gyorffy}}]{relativistic_LMTO_second_1988}%
  \BibitemOpen
  \bibfield  {author} {\bibinfo {author} {\bibfnamefont {H.}~\bibnamefont
  {Ebert}}, \bibinfo {author} {\bibfnamefont {P.}~\bibnamefont {Strange}},\
  and\ \bibinfo {author} {\bibfnamefont {B.~L.}\ \bibnamefont {Gyorffy}},\
  }\bibfield  {title} {\bibinfo {title} {{Spin-Polarized Relativistic {LMTO}
  Method}},\ }\href {https://doi.org/10.1063/1.340893} {\bibfield  {journal}
  {\bibinfo  {journal} {Journal of Applied Physics}\ }\textbf {\bibinfo
  {volume} {63}},\ \bibinfo {pages} {3052} (\bibinfo {year}
  {1988})}\BibitemShut {NoStop}%
\bibitem [{\citenamefont {Dal~Corso}(2010)}]{relativistic_PAW_2010}%
  \BibitemOpen
  \bibfield  {author} {\bibinfo {author} {\bibfnamefont {A.}~\bibnamefont
  {Dal~Corso}},\ }\bibfield  {title} {\bibinfo {title} {{Projector
  Augmented-Wave Method: Application to Relativistic Spin-Density Functional
  Theory}},\ }\href {https://doi.org/10.1103/PhysRevB.82.075116} {\bibfield
  {journal} {\bibinfo  {journal} {Phys. Rev. B}\ }\textbf {\bibinfo {volume}
  {82}},\ \bibinfo {pages} {075116} (\bibinfo {year} {2010})}\BibitemShut
  {NoStop}%
\bibitem [{\citenamefont {Philipsen}\ \emph {et~al.}(1997)\citenamefont
  {Philipsen}, \citenamefont {van Lenthe}, \citenamefont {Snijders},\ and\
  \citenamefont {Baerends}}]{ZORA_BAND_1997}%
  \BibitemOpen
  \bibfield  {author} {\bibinfo {author} {\bibfnamefont {P.~H.~T.}\
  \bibnamefont {Philipsen}}, \bibinfo {author} {\bibfnamefont {E.}~\bibnamefont
  {van Lenthe}}, \bibinfo {author} {\bibfnamefont {J.~G.}\ \bibnamefont
  {Snijders}},\ and\ \bibinfo {author} {\bibfnamefont {E.~J.}\ \bibnamefont
  {Baerends}},\ }\bibfield  {title} {\bibinfo {title} {{Relativistic
  Calculations on the Adsorption of} $\mathrm{CO}$ {on the (111) Surfaces of}
  $\mathrm{Ni}$, $\mathrm{Pd}$, and $\mathrm{Pt}$ {within the Zeroth-Order
  Regular Approximation}},\ }\href {https://doi.org/10.1103/PhysRevB.56.13556}
  {\bibfield  {journal} {\bibinfo  {journal} {Phys. Rev. B}\ }\textbf {\bibinfo
  {volume} {56}},\ \bibinfo {pages} {13556} (\bibinfo {year}
  {1997})}\BibitemShut {NoStop}%
\bibitem [{\citenamefont {Zhao}\ \emph {et~al.}(2016)\citenamefont {Zhao},
  \citenamefont {Zhang}, \citenamefont {Xiao},\ and\ \citenamefont
  {Liu}}]{Rundong_JCP2016}%
  \BibitemOpen
  \bibfield  {author} {\bibinfo {author} {\bibfnamefont {R.}~\bibnamefont
  {Zhao}}, \bibinfo {author} {\bibfnamefont {Y.}~\bibnamefont {Zhang}},
  \bibinfo {author} {\bibfnamefont {Y.}~\bibnamefont {Xiao}},\ and\ \bibinfo
  {author} {\bibfnamefont {W.}~\bibnamefont {Liu}},\ }\bibfield  {title}
  {\bibinfo {title} {{Exact Two-Component Relativistic Energy Band Theory and
  Application}},\ }\href {https://doi.org/10.1063/1.4940140} {\bibfield
  {journal} {\bibinfo  {journal} {The Journal of Chemical Physics}\ }\textbf
  {\bibinfo {volume} {144}},\ \bibinfo {pages} {044105} (\bibinfo {year}
  {2016})}\BibitemShut {NoStop}%
\bibitem [{\citenamefont {Kadek}\ \emph {et~al.}(2019)\citenamefont {Kadek},
  \citenamefont {Repisky},\ and\ \citenamefont {Ruud}}]{Kadek_PRB2019}%
  \BibitemOpen
  \bibfield  {author} {\bibinfo {author} {\bibfnamefont {M.}~\bibnamefont
  {Kadek}}, \bibinfo {author} {\bibfnamefont {M.}~\bibnamefont {Repisky}},\
  and\ \bibinfo {author} {\bibfnamefont {K.}~\bibnamefont {Ruud}},\ }\bibfield
  {title} {\bibinfo {title} {{All-Electron Fully Relativistic Kohn-Sham Theory
  for Solids based on the Dirac-Coulomb Hamiltonian and Gaussian-Type
  Functions}},\ }\href {https://doi.org/10.1103/PhysRevB.99.205103} {\bibfield
  {journal} {\bibinfo  {journal} {Phys. Rev. B}\ }\textbf {\bibinfo {volume}
  {99}},\ \bibinfo {pages} {205103} (\bibinfo {year} {2019})}\BibitemShut
  {NoStop}%
\bibitem [{\citenamefont {Dyall}\ and\ \citenamefont
  {Faegri}(2007)}]{intro_relativistic_QC_Dyall_2007}%
  \BibitemOpen
  \bibfield  {author} {\bibinfo {author} {\bibfnamefont {K.~G.}\ \bibnamefont
  {Dyall}}\ and\ \bibinfo {author} {\bibfnamefont {K.}~\bibnamefont {Faegri}},\
  }\href {https://doi.org/10.1093/oso/9780195140866.001.0001} {\emph {\bibinfo
  {title} {{Introduction to Relativistic Quantum Chemistry}}}}\ (\bibinfo
  {publisher} {Oxford University Press},\ \bibinfo {year} {2007})\BibitemShut
  {NoStop}%
\bibitem [{\citenamefont {Kananenka}\ \emph {et~al.}(2015)\citenamefont
  {Kananenka}, \citenamefont {Gull},\ and\ \citenamefont
  {Zgid}}]{Alexie_SEET_PRB2015}%
  \BibitemOpen
  \bibfield  {author} {\bibinfo {author} {\bibfnamefont {A.~A.}\ \bibnamefont
  {Kananenka}}, \bibinfo {author} {\bibfnamefont {E.}~\bibnamefont {Gull}},\
  and\ \bibinfo {author} {\bibfnamefont {D.}~\bibnamefont {Zgid}},\ }\bibfield
  {title} {\bibinfo {title} {{Systematically Improvable Multiscale Solver for
  Correlated Electron Systems}},\ }\href
  {https://doi.org/10.1103/PhysRevB.91.121111} {\bibfield  {journal} {\bibinfo
  {journal} {Phys. Rev. B}\ }\textbf {\bibinfo {volume} {91}},\ \bibinfo
  {pages} {121111} (\bibinfo {year} {2015})}\BibitemShut {NoStop}%
\bibitem [{\citenamefont {Lan}\ and\ \citenamefont
  {Zgid}(2017)}]{Lan_Generalized_SEET_2017}%
  \BibitemOpen
  \bibfield  {author} {\bibinfo {author} {\bibfnamefont {T.~N.}\ \bibnamefont
  {Lan}}\ and\ \bibinfo {author} {\bibfnamefont {D.}~\bibnamefont {Zgid}},\
  }\bibfield  {title} {\bibinfo {title} {Generalized self-energy embedding
  theory},\ }\href {https://doi.org/10.1021/acs.jpclett.7b00689} {\bibfield
  {journal} {\bibinfo  {journal} {The Journal of Physical Chemistry Letters}\
  }\textbf {\bibinfo {volume} {8}},\ \bibinfo {pages} {2200} (\bibinfo {year}
  {2017})}\BibitemShut {NoStop}%
\bibitem [{\citenamefont {Zgid}\ and\ \citenamefont
  {Gull}(2017)}]{Dominika_SEET_2017}%
  \BibitemOpen
  \bibfield  {author} {\bibinfo {author} {\bibfnamefont {D.}~\bibnamefont
  {Zgid}}\ and\ \bibinfo {author} {\bibfnamefont {E.}~\bibnamefont {Gull}},\
  }\bibfield  {title} {\bibinfo {title} {{Finite Temperature Quantum Embedding
  Theories for Correlated Systems}},\ }\href
  {https://doi.org/10.1088/1367-2630/aa5d34} {\bibfield  {journal} {\bibinfo
  {journal} {New Journal of Physics}\ }\textbf {\bibinfo {volume} {19}},\
  \bibinfo {pages} {023047} (\bibinfo {year} {2017})}\BibitemShut {NoStop}%
\bibitem [{\citenamefont {Rusakov}\ \emph {et~al.}(2019)\citenamefont
  {Rusakov}, \citenamefont {Iskakov}, \citenamefont {Tran},\ and\ \citenamefont
  {Zgid}}]{Zgid_periodic_seet2019}%
  \BibitemOpen
  \bibfield  {author} {\bibinfo {author} {\bibfnamefont {A.~A.}\ \bibnamefont
  {Rusakov}}, \bibinfo {author} {\bibfnamefont {S.}~\bibnamefont {Iskakov}},
  \bibinfo {author} {\bibfnamefont {L.~N.}\ \bibnamefont {Tran}},\ and\
  \bibinfo {author} {\bibfnamefont {D.}~\bibnamefont {Zgid}},\ }\bibfield
  {title} {\bibinfo {title} {{Self-Energy Embedding Theory ($\mathrm{SEET}$)
  for Periodic Systems}},\ }\href {https://doi.org/10.1021/acs.jctc.8b00927}
  {\bibfield  {journal} {\bibinfo  {journal} {Journal of Chemical Theory and
  Computation}\ }\textbf {\bibinfo {volume} {15}},\ \bibinfo {pages} {229}
  (\bibinfo {year} {2019})},\ \bibinfo {note} {pMID: 30540474}\BibitemShut
  {NoStop}%
\bibitem [{\citenamefont {Iskakov}\ \emph {et~al.}(2020)\citenamefont
  {Iskakov}, \citenamefont {Yeh}, \citenamefont {Gull},\ and\ \citenamefont
  {Zgid}}]{SEET_Sergei20}%
  \BibitemOpen
  \bibfield  {author} {\bibinfo {author} {\bibfnamefont {S.}~\bibnamefont
  {Iskakov}}, \bibinfo {author} {\bibfnamefont {C.-N.}\ \bibnamefont {Yeh}},
  \bibinfo {author} {\bibfnamefont {E.}~\bibnamefont {Gull}},\ and\ \bibinfo
  {author} {\bibfnamefont {D.}~\bibnamefont {Zgid}},\ }\bibfield  {title}
  {\bibinfo {title} {{\emph{Ab Initio} Self-Energy Embedding for the
  Photoemission Spectra of NiO and MnO}},\ }\href
  {https://doi.org/10.1103/PhysRevB.102.085105} {\bibfield  {journal} {\bibinfo
   {journal} {Phys. Rev. B}\ }\textbf {\bibinfo {volume} {102}},\ \bibinfo
  {pages} {085105} (\bibinfo {year} {2020})}\BibitemShut {NoStop}%
\bibitem [{\citenamefont {Kotliar}\ \emph {et~al.}(2006)\citenamefont
  {Kotliar}, \citenamefont {Savrasov}, \citenamefont {Haule}, \citenamefont
  {Oudovenko}, \citenamefont {Parcollet},\ and\ \citenamefont
  {Marianetti}}]{Kotliar06}%
  \BibitemOpen
  \bibfield  {author} {\bibinfo {author} {\bibfnamefont {G.}~\bibnamefont
  {Kotliar}}, \bibinfo {author} {\bibfnamefont {S.~Y.}\ \bibnamefont
  {Savrasov}}, \bibinfo {author} {\bibfnamefont {K.}~\bibnamefont {Haule}},
  \bibinfo {author} {\bibfnamefont {V.~S.}\ \bibnamefont {Oudovenko}}, \bibinfo
  {author} {\bibfnamefont {O.}~\bibnamefont {Parcollet}},\ and\ \bibinfo
  {author} {\bibfnamefont {C.~A.}\ \bibnamefont {Marianetti}},\ }\bibfield
  {title} {\bibinfo {title} {{Electronic Structure Calculations with Dynamical
  Mean-Field Theory}},\ }\href {https://doi.org/10.1103/RevModPhys.78.865}
  {\bibfield  {journal} {\bibinfo  {journal} {Rev. Mod. Phys.}\ }\textbf
  {\bibinfo {volume} {78}},\ \bibinfo {pages} {865} (\bibinfo {year}
  {2006})}\BibitemShut {NoStop}%
\bibitem [{\citenamefont {Martins}\ \emph {et~al.}(2011)\citenamefont
  {Martins}, \citenamefont {Aichhorn}, \citenamefont {Vaugier},\ and\
  \citenamefont {Biermann}}]{LDA_DMFT_Sr2IrO4_2011}%
  \BibitemOpen
  \bibfield  {author} {\bibinfo {author} {\bibfnamefont {C.}~\bibnamefont
  {Martins}}, \bibinfo {author} {\bibfnamefont {M.}~\bibnamefont {Aichhorn}},
  \bibinfo {author} {\bibfnamefont {L.}~\bibnamefont {Vaugier}},\ and\ \bibinfo
  {author} {\bibfnamefont {S.}~\bibnamefont {Biermann}},\ }\bibfield  {title}
  {\bibinfo {title} {{Reduced Effective Spin-Orbital Degeneracy and
  Spin-Orbital Ordering in Paramagnetic Transition-Metal Oxides:}
  $\mathrm{Sr_{2}IrO_{4}}$ {versus} $\mathrm{Sr_{2}RhO_{4}}$},\ }\href
  {https://doi.org/10.1103/PhysRevLett.107.266404} {\bibfield  {journal}
  {\bibinfo  {journal} {Phys. Rev. Lett.}\ }\textbf {\bibinfo {volume} {107}},\
  \bibinfo {pages} {266404} (\bibinfo {year} {2011})}\BibitemShut {NoStop}%
\bibitem [{\citenamefont {Aichhorn}\ \emph {et~al.}(2016)\citenamefont
  {Aichhorn}, \citenamefont {Pourovskii}, \citenamefont {Seth}, \citenamefont
  {Vildosola}, \citenamefont {Zingl}, \citenamefont {Peil}, \citenamefont
  {Deng}, \citenamefont {Mravlje}, \citenamefont {Kraberger}, \citenamefont
  {Martins}, \citenamefont {Ferrero},\ and\ \citenamefont
  {Parcollet}}]{TRIQS_Aichhorn2016}%
  \BibitemOpen
  \bibfield  {author} {\bibinfo {author} {\bibfnamefont {M.}~\bibnamefont
  {Aichhorn}}, \bibinfo {author} {\bibfnamefont {L.}~\bibnamefont
  {Pourovskii}}, \bibinfo {author} {\bibfnamefont {P.}~\bibnamefont {Seth}},
  \bibinfo {author} {\bibfnamefont {V.}~\bibnamefont {Vildosola}}, \bibinfo
  {author} {\bibfnamefont {M.}~\bibnamefont {Zingl}}, \bibinfo {author}
  {\bibfnamefont {O.~E.}\ \bibnamefont {Peil}}, \bibinfo {author}
  {\bibfnamefont {X.}~\bibnamefont {Deng}}, \bibinfo {author} {\bibfnamefont
  {J.}~\bibnamefont {Mravlje}}, \bibinfo {author} {\bibfnamefont {G.~J.}\
  \bibnamefont {Kraberger}}, \bibinfo {author} {\bibfnamefont {C.}~\bibnamefont
  {Martins}}, \bibinfo {author} {\bibfnamefont {M.}~\bibnamefont {Ferrero}},\
  and\ \bibinfo {author} {\bibfnamefont {O.}~\bibnamefont {Parcollet}},\
  }\bibfield  {title} {\bibinfo {title} {{TRIQS/DFTTools: A TRIQS Application
  for \emph{Ab Initio} Calculations of Correlated Materials}},\ }\href
  {https://doi.org/10.1016/j.cpc.2016.03.014} {\bibfield  {journal} {\bibinfo
  {journal} {Computer Physics Communications}\ }\textbf {\bibinfo {volume}
  {204}},\ \bibinfo {pages} {200} (\bibinfo {year} {2016})}\BibitemShut
  {NoStop}%
\bibitem [{\citenamefont {Kim}\ \emph {et~al.}(2018)\citenamefont {Kim},
  \citenamefont {Mravlje}, \citenamefont {Ferrero}, \citenamefont {Parcollet},\
  and\ \citenamefont {Georges}}]{DMFT_Sr2RuO4_2018}%
  \BibitemOpen
  \bibfield  {author} {\bibinfo {author} {\bibfnamefont {M.}~\bibnamefont
  {Kim}}, \bibinfo {author} {\bibfnamefont {J.}~\bibnamefont {Mravlje}},
  \bibinfo {author} {\bibfnamefont {M.}~\bibnamefont {Ferrero}}, \bibinfo
  {author} {\bibfnamefont {O.}~\bibnamefont {Parcollet}},\ and\ \bibinfo
  {author} {\bibfnamefont {A.}~\bibnamefont {Georges}},\ }\bibfield  {title}
  {\bibinfo {title} {{Spin-Orbit Coupling and Electronic Correlations in}
  $\mathrm{Sr_{2}RuO_{4}}$},\ }\href
  {https://doi.org/10.1103/PhysRevLett.120.126401} {\bibfield  {journal}
  {\bibinfo  {journal} {Phys. Rev. Lett.}\ }\textbf {\bibinfo {volume} {120}},\
  \bibinfo {pages} {126401} (\bibinfo {year} {2018})}\BibitemShut {NoStop}%
\bibitem [{\citenamefont {Tamai}\ \emph {et~al.}(2019)\citenamefont {Tamai},
  \citenamefont {Zingl}, \citenamefont {Rozbicki}, \citenamefont {Cappelli},
  \citenamefont {Ricc\`o}, \citenamefont {de~la Torre}, \citenamefont
  {McKeown~Walker}, \citenamefont {Bruno}, \citenamefont {King}, \citenamefont
  {Meevasana}, \citenamefont {Shi}, \citenamefont
  {Radovi\ifmmode~\acute{c}\else \'{c}\fi{}}, \citenamefont {Plumb},
  \citenamefont {Gibbs}, \citenamefont {Mackenzie}, \citenamefont {Berthod},
  \citenamefont {Strand}, \citenamefont {Kim}, \citenamefont {Georges},\ and\
  \citenamefont {Baumberger}}]{Expt_DMFT_Sr2RuO4_2019}%
  \BibitemOpen
  \bibfield  {author} {\bibinfo {author} {\bibfnamefont {A.}~\bibnamefont
  {Tamai}}, \bibinfo {author} {\bibfnamefont {M.}~\bibnamefont {Zingl}},
  \bibinfo {author} {\bibfnamefont {E.}~\bibnamefont {Rozbicki}}, \bibinfo
  {author} {\bibfnamefont {E.}~\bibnamefont {Cappelli}}, \bibinfo {author}
  {\bibfnamefont {S.}~\bibnamefont {Ricc\`o}}, \bibinfo {author} {\bibfnamefont
  {A.}~\bibnamefont {de~la Torre}}, \bibinfo {author} {\bibfnamefont
  {S.}~\bibnamefont {McKeown~Walker}}, \bibinfo {author} {\bibfnamefont
  {F.~Y.}\ \bibnamefont {Bruno}}, \bibinfo {author} {\bibfnamefont {P.~D.~C.}\
  \bibnamefont {King}}, \bibinfo {author} {\bibfnamefont {W.}~\bibnamefont
  {Meevasana}}, \bibinfo {author} {\bibfnamefont {M.}~\bibnamefont {Shi}},
  \bibinfo {author} {\bibfnamefont {M.}~\bibnamefont
  {Radovi\ifmmode~\acute{c}\else \'{c}\fi{}}}, \bibinfo {author} {\bibfnamefont
  {N.~C.}\ \bibnamefont {Plumb}}, \bibinfo {author} {\bibfnamefont {A.~S.}\
  \bibnamefont {Gibbs}}, \bibinfo {author} {\bibfnamefont {A.~P.}\ \bibnamefont
  {Mackenzie}}, \bibinfo {author} {\bibfnamefont {C.}~\bibnamefont {Berthod}},
  \bibinfo {author} {\bibfnamefont {H.~U.~R.}\ \bibnamefont {Strand}}, \bibinfo
  {author} {\bibfnamefont {M.}~\bibnamefont {Kim}}, \bibinfo {author}
  {\bibfnamefont {A.}~\bibnamefont {Georges}},\ and\ \bibinfo {author}
  {\bibfnamefont {F.}~\bibnamefont {Baumberger}},\ }\bibfield  {title}
  {\bibinfo {title} {{High-Resolution Photoemission on $\mathrm{Sr_{2}RuO_{4}}$
  Reveals Correlation-Enhanced Effective Spin-Orbit Coupling and Dominantly
  Local Self-Energies}},\ }\href {https://doi.org/10.1103/PhysRevX.9.021048}
  {\bibfield  {journal} {\bibinfo  {journal} {Phys. Rev. X}\ }\textbf {\bibinfo
  {volume} {9}},\ \bibinfo {pages} {021048} (\bibinfo {year}
  {2019})}\BibitemShut {NoStop}%
\bibitem [{\citenamefont {Sakuma}\ \emph {et~al.}(2011)\citenamefont {Sakuma},
  \citenamefont {Friedrich}, \citenamefont {Miyake}, \citenamefont {Bl\"ugel},\
  and\ \citenamefont {Aryasetiawan}}]{4cG0W0_Sakuma_2011}%
  \BibitemOpen
  \bibfield  {author} {\bibinfo {author} {\bibfnamefont {R.}~\bibnamefont
  {Sakuma}}, \bibinfo {author} {\bibfnamefont {C.}~\bibnamefont {Friedrich}},
  \bibinfo {author} {\bibfnamefont {T.}~\bibnamefont {Miyake}}, \bibinfo
  {author} {\bibfnamefont {S.}~\bibnamefont {Bl\"ugel}},\ and\ \bibinfo
  {author} {\bibfnamefont {F.}~\bibnamefont {Aryasetiawan}},\ }\bibfield
  {title} {\bibinfo {title} {{${GW}$ Calculations Including Spin-Orbit
  Coupling: Application to Hg Chalcogenides}},\ }\href
  {https://doi.org/10.1103/PhysRevB.84.085144} {\bibfield  {journal} {\bibinfo
  {journal} {Phys. Rev. B}\ }\textbf {\bibinfo {volume} {84}},\ \bibinfo
  {pages} {085144} (\bibinfo {year} {2011})}\BibitemShut {NoStop}%
\bibitem [{\citenamefont {Kutepov}\ \emph {et~al.}(2012)\citenamefont
  {Kutepov}, \citenamefont {Haule}, \citenamefont {Savrasov},\ and\
  \citenamefont {Kotliar}}]{4c_scGW_Kutepov_2012}%
  \BibitemOpen
  \bibfield  {author} {\bibinfo {author} {\bibfnamefont {A.}~\bibnamefont
  {Kutepov}}, \bibinfo {author} {\bibfnamefont {K.}~\bibnamefont {Haule}},
  \bibinfo {author} {\bibfnamefont {S.~Y.}\ \bibnamefont {Savrasov}},\ and\
  \bibinfo {author} {\bibfnamefont {G.}~\bibnamefont {Kotliar}},\ }\bibfield
  {title} {\bibinfo {title} {{Electronic Structure of $\mathrm{Pu}$ and
  $\mathrm{Am}$ Metals by Self-Consistent Relativistic ${GW}$ Method}},\ }\href
  {https://doi.org/10.1103/PhysRevB.85.155129} {\bibfield  {journal} {\bibinfo
  {journal} {Phys. Rev. B}\ }\textbf {\bibinfo {volume} {85}},\ \bibinfo
  {pages} {155129} (\bibinfo {year} {2012})}\BibitemShut {NoStop}%
\bibitem [{\citenamefont {Gell-Mann}(1956)}]{minial_coupling_Gell-Mann1956}%
  \BibitemOpen
  \bibfield  {author} {\bibinfo {author} {\bibfnamefont {M.}~\bibnamefont
  {Gell-Mann}},\ }\bibfield  {title} {\bibinfo {title} {{The Interpretation of
  the New Particles as Displaced Charge Multiplets}},\ }\href
  {https://doi.org/10.1007/BF02748000} {\bibfield  {journal} {\bibinfo
  {journal} {Il Nuovo Cimento (1955-1965)}\ }\textbf {\bibinfo {volume} {4}},\
  \bibinfo {pages} {848} (\bibinfo {year} {1956})}\BibitemShut {NoStop}%
\bibitem [{\citenamefont {Stanton}\ and\ \citenamefont
  {Havriliak}(1984)}]{RKB_JCP}%
  \BibitemOpen
  \bibfield  {author} {\bibinfo {author} {\bibfnamefont {R.~E.}\ \bibnamefont
  {Stanton}}\ and\ \bibinfo {author} {\bibfnamefont {S.}~\bibnamefont
  {Havriliak}},\ }\bibfield  {title} {\bibinfo {title} {{Kinetic Balance: A
  Partial Solution to the Problem of Variational Safety in Dirac
  Calculations}},\ }\href {https://doi.org/10.1063/1.447865} {\bibfield
  {journal} {\bibinfo  {journal} {The Journal of Chemical Physics}\ }\textbf
  {\bibinfo {volume} {81}},\ \bibinfo {pages} {1910} (\bibinfo {year}
  {1984})}\BibitemShut {NoStop}%
\bibitem [{\citenamefont {Ishikawa}\ \emph {et~al.}(1985)\citenamefont
  {Ishikawa}, \citenamefont {Baretty},\ and\ \citenamefont
  {Binning}}]{RKB_ISHIKAWA1985}%
  \BibitemOpen
  \bibfield  {author} {\bibinfo {author} {\bibfnamefont {Y.}~\bibnamefont
  {Ishikawa}}, \bibinfo {author} {\bibfnamefont {R.}~\bibnamefont {Baretty}},\
  and\ \bibinfo {author} {\bibfnamefont {R.}~\bibnamefont {Binning}},\
  }\bibfield  {title} {\bibinfo {title} {{Relativistic Gaussian Basis Set
  Calculations on One-Electron Ions with a Nucleus of Finite Extent}},\ }\href
  {https://doi.org/https://doi.org/10.1016/0009-2614(85)87169-4} {\bibfield
  {journal} {\bibinfo  {journal} {Chemical Physics Letters}\ }\textbf {\bibinfo
  {volume} {121}},\ \bibinfo {pages} {130} (\bibinfo {year}
  {1985})}\BibitemShut {NoStop}%
\bibitem [{\citenamefont {Dyall}\ and\ \citenamefont
  {Faegri}(1990)}]{RKB_DYALL1990}%
  \BibitemOpen
  \bibfield  {author} {\bibinfo {author} {\bibfnamefont {K.~G.}\ \bibnamefont
  {Dyall}}\ and\ \bibinfo {author} {\bibfnamefont {K.}~\bibnamefont {Faegri}},\
  }\bibfield  {title} {\bibinfo {title} {{Kinetic Balance and Variational
  Bounds Failure in the Solution of the Dirac Equation in a Finite Gaussian
  Basis Set}},\ }\href
  {https://doi.org/https://doi.org/10.1016/0009-2614(90)85321-3} {\bibfield
  {journal} {\bibinfo  {journal} {Chemical Physics Letters}\ }\textbf {\bibinfo
  {volume} {174}},\ \bibinfo {pages} {25} (\bibinfo {year} {1990})}\BibitemShut
  {NoStop}%
\bibitem [{\citenamefont {Kutzelnigg}(1984)}]{modified_Dirac_Kutzelnigg1984}%
  \BibitemOpen
  \bibfield  {author} {\bibinfo {author} {\bibfnamefont {W.}~\bibnamefont
  {Kutzelnigg}},\ }\bibfield  {title} {\bibinfo {title} {{Basis Set Expansion
  of the Dirac Operator without Variational Collapse}},\ }\href
  {https://doi.org/10.1002/qua.560250112} {\bibfield  {journal} {\bibinfo
  {journal} {International Journal of Quantum Chemistry}\ }\textbf {\bibinfo
  {volume} {25}},\ \bibinfo {pages} {107} (\bibinfo {year} {1984})}\BibitemShut
  {NoStop}%
\bibitem [{\citenamefont {Dyall}(1994)}]{modified_Dirac_Dyall1994}%
  \BibitemOpen
  \bibfield  {author} {\bibinfo {author} {\bibfnamefont {K.~G.}\ \bibnamefont
  {Dyall}},\ }\bibfield  {title} {\bibinfo {title} {{An Exact Separation of the
  Spin-Free and Spin-Dependent Terms of the Dirac-Coulomb-Breit Hamiltonian}},\
  }\href {https://doi.org/10.1063/1.466508} {\bibfield  {journal} {\bibinfo
  {journal} {The Journal of Chemical Physics}\ }\textbf {\bibinfo {volume}
  {100}},\ \bibinfo {pages} {2118} (\bibinfo {year} {1994})}\BibitemShut
  {NoStop}%
\bibitem [{\citenamefont {Dyall}(2001)}]{x2c1e_Dyall_2001}%
  \BibitemOpen
  \bibfield  {author} {\bibinfo {author} {\bibfnamefont {K.~G.}\ \bibnamefont
  {Dyall}},\ }\bibfield  {title} {\bibinfo {title} {{Interfacing Relativistic
  and Nonrelativistic Methods. IV. One- and Two-Electron Scalar
  Approximations}},\ }\href {https://doi.org/10.1063/1.1413512} {\bibfield
  {journal} {\bibinfo  {journal} {The Journal of Chemical Physics}\ }\textbf
  {\bibinfo {volume} {115}},\ \bibinfo {pages} {9136} (\bibinfo {year}
  {2001})}\BibitemShut {NoStop}%
\bibitem [{\citenamefont {Dyall}(2002)}]{x2c1e_Dyall_2002}%
  \BibitemOpen
  \bibfield  {author} {\bibinfo {author} {\bibfnamefont {K.~G.}\ \bibnamefont
  {Dyall}},\ }\bibfield  {title} {\bibinfo {title} {{A Systematic Sequence of
  Relativistic Approximations}},\ }\href {https://doi.org/10.1002/jcc.10048}
  {\bibfield  {journal} {\bibinfo  {journal} {Journal of Computational
  Chemistry}\ }\textbf {\bibinfo {volume} {23}},\ \bibinfo {pages} {786}
  (\bibinfo {year} {2002})}\BibitemShut {NoStop}%
\bibitem [{\citenamefont {Foldy}\ and\ \citenamefont
  {Wouthuysen}(1950)}]{FW_transformation_1950}%
  \BibitemOpen
  \bibfield  {author} {\bibinfo {author} {\bibfnamefont {L.~L.}\ \bibnamefont
  {Foldy}}\ and\ \bibinfo {author} {\bibfnamefont {S.~A.}\ \bibnamefont
  {Wouthuysen}},\ }\bibfield  {title} {\bibinfo {title} {{On the Dirac Theory
  of Spin 1/2 Particles and Its Non-Relativistic Limit}},\ }\href
  {https://doi.org/10.1103/PhysRev.78.29} {\bibfield  {journal} {\bibinfo
  {journal} {Phys. Rev.}\ }\textbf {\bibinfo {volume} {78}},\ \bibinfo {pages}
  {29} (\bibinfo {year} {1950})}\BibitemShut {NoStop}%
\bibitem [{\citenamefont {Dyall}(1997)}]{NESC_Dyall_1997}%
  \BibitemOpen
  \bibfield  {author} {\bibinfo {author} {\bibfnamefont {K.~G.}\ \bibnamefont
  {Dyall}},\ }\bibfield  {title} {\bibinfo {title} {{Interfacing Relativistic
  and Nonrelativistic Methods. I. Normalized Elimination of the Small Component
  in the Modified Dirac Equation}},\ }\href {https://doi.org/10.1063/1.473860}
  {\bibfield  {journal} {\bibinfo  {journal} {The Journal of Chemical Physics}\
  }\textbf {\bibinfo {volume} {106}},\ \bibinfo {pages} {9618} (\bibinfo {year}
  {1997})}\BibitemShut {NoStop}%
\bibitem [{\citenamefont {Kutzelnigg}(1999)}]{FW_transformation_2_1999}%
  \BibitemOpen
  \bibfield  {author} {\bibinfo {author} {\bibfnamefont {W.}~\bibnamefont
  {Kutzelnigg}},\ }\bibfield  {title} {\bibinfo {title} {{Effective
  Hamiltonians for Degenerate and Quasidegenerate Direct Perturbation Theory of
  Relativistic Effects}},\ }\href {https://doi.org/10.1063/1.478739} {\bibfield
   {journal} {\bibinfo  {journal} {The Journal of Chemical Physics}\ }\textbf
  {\bibinfo {volume} {110}},\ \bibinfo {pages} {8283} (\bibinfo {year}
  {1999})}\BibitemShut {NoStop}%
\bibitem [{\citenamefont {Boys}\ and\ \citenamefont
  {Egerton}(1950)}]{Boys_Gaussian_basis_1950}%
  \BibitemOpen
  \bibfield  {author} {\bibinfo {author} {\bibfnamefont {S.~F.}\ \bibnamefont
  {Boys}}\ and\ \bibinfo {author} {\bibfnamefont {A.~C.}\ \bibnamefont
  {Egerton}},\ }\bibfield  {title} {\bibinfo {title} {{Electronic Wave
  Functions - I. A General Method of Calculation for the Stationary States of
  any Molecular System}},\ }\href {https://doi.org/10.1098/rspa.1950.0036}
  {\bibfield  {journal} {\bibinfo  {journal} {Proceedings of the Royal Society
  of London. Series A. Mathematical and Physical Sciences}\ }\textbf {\bibinfo
  {volume} {200}},\ \bibinfo {pages} {542} (\bibinfo {year}
  {1950})}\BibitemShut {NoStop}%
\bibitem [{\citenamefont {Dyall}\ \emph {et~al.}(1984)\citenamefont {Dyall},
  \citenamefont {Grant},\ and\ \citenamefont
  {Wilson}}]{variational_collapse_1984}%
  \BibitemOpen
  \bibfield  {author} {\bibinfo {author} {\bibfnamefont {K.~G.}\ \bibnamefont
  {Dyall}}, \bibinfo {author} {\bibfnamefont {I.~P.}\ \bibnamefont {Grant}},\
  and\ \bibinfo {author} {\bibfnamefont {S.}~\bibnamefont {Wilson}},\
  }\bibfield  {title} {\bibinfo {title} {{Matrix Representation of Operator
  Products}},\ }\href {https://doi.org/10.1088/0022-3700/17/4/006} {\bibfield
  {journal} {\bibinfo  {journal} {Journal of Physics B: Atomic and Molecular
  Physics}\ }\textbf {\bibinfo {volume} {17}},\ \bibinfo {pages} {493}
  (\bibinfo {year} {1984})}\BibitemShut {NoStop}%
\bibitem [{\citenamefont {Liu}\ and\ \citenamefont
  {Peng}(2009)}]{X2C_revisit_Liu_2009}%
  \BibitemOpen
  \bibfield  {author} {\bibinfo {author} {\bibfnamefont {W.}~\bibnamefont
  {Liu}}\ and\ \bibinfo {author} {\bibfnamefont {D.}~\bibnamefont {Peng}},\
  }\bibfield  {title} {\bibinfo {title} {{Exact Two-Component Hamiltonians
  Revisited}},\ }\href {https://doi.org/10.1063/1.3159445} {\bibfield
  {journal} {\bibinfo  {journal} {The Journal of Chemical Physics}\ }\textbf
  {\bibinfo {volume} {131}},\ \bibinfo {pages} {031104} (\bibinfo {year}
  {2009})}\BibitemShut {NoStop}%
\bibitem [{\citenamefont {Breit}(1929)}]{Coulomb_Breit_1929}%
  \BibitemOpen
  \bibfield  {author} {\bibinfo {author} {\bibfnamefont {G.}~\bibnamefont
  {Breit}},\ }\bibfield  {title} {\bibinfo {title} {{The Effect of Retardation
  on the Interaction of Two Electrons}},\ }\href
  {https://doi.org/10.1103/PhysRev.34.553} {\bibfield  {journal} {\bibinfo
  {journal} {Phys. Rev.}\ }\textbf {\bibinfo {volume} {34}},\ \bibinfo {pages}
  {553} (\bibinfo {year} {1929})}\BibitemShut {NoStop}%
\bibitem [{\citenamefont {Hedin}(1965)}]{Hedin65}%
  \BibitemOpen
  \bibfield  {author} {\bibinfo {author} {\bibfnamefont {L.}~\bibnamefont
  {Hedin}},\ }\bibfield  {title} {\bibinfo {title} {{New Method for Calculating
  the One-Particle Green's Function with Application to the Electron-Gas
  Problem}},\ }\href {https://doi.org/10.1103/PhysRev.139.A796} {\bibfield
  {journal} {\bibinfo  {journal} {Phys. Rev.}\ }\textbf {\bibinfo {volume}
  {139}},\ \bibinfo {pages} {A796} (\bibinfo {year} {1965})}\BibitemShut
  {NoStop}%
\bibitem [{\citenamefont {Aryasetiawan}\ and\ \citenamefont
  {Biermann}(2008)}]{Aryasetiawan_Generalized_Hedin_2008}%
  \BibitemOpen
  \bibfield  {author} {\bibinfo {author} {\bibfnamefont {F.}~\bibnamefont
  {Aryasetiawan}}\ and\ \bibinfo {author} {\bibfnamefont {S.}~\bibnamefont
  {Biermann}},\ }\bibfield  {title} {\bibinfo {title} {{Generalized Hedin's
  Equations for Quantum Many-Body Systems with Spin-Dependent Interactions}},\
  }\href {https://doi.org/10.1103/PhysRevLett.100.116402} {\bibfield  {journal}
  {\bibinfo  {journal} {Phys. Rev. Lett.}\ }\textbf {\bibinfo {volume} {100}},\
  \bibinfo {pages} {116402} (\bibinfo {year} {2008})}\BibitemShut {NoStop}%
\bibitem [{\citenamefont {Aryasetiawan}\ and\ \citenamefont
  {Biermann}(2009)}]{Aryasetiawan_Generalized_GW_2009}%
  \BibitemOpen
  \bibfield  {author} {\bibinfo {author} {\bibfnamefont {F.}~\bibnamefont
  {Aryasetiawan}}\ and\ \bibinfo {author} {\bibfnamefont {S.}~\bibnamefont
  {Biermann}},\ }\bibfield  {title} {\bibinfo {title} {{Generalized Hedin
  Equations and {$\sigma G\sigma W$} Approximation for Quantum Many-Body
  Systems with Spin-Dependent Interactions}},\ }\href
  {https://doi.org/10.1088/0953-8984/21/6/064232} {\bibfield  {journal}
  {\bibinfo  {journal} {Journal of Physics: Condensed Matter}\ }\textbf
  {\bibinfo {volume} {21}},\ \bibinfo {pages} {064232} (\bibinfo {year}
  {2009})}\BibitemShut {NoStop}%
\bibitem [{\citenamefont {Rusakov}\ and\ \citenamefont
  {Zgid}(2016)}]{Rusakov16}%
  \BibitemOpen
  \bibfield  {author} {\bibinfo {author} {\bibfnamefont {A.~A.}\ \bibnamefont
  {Rusakov}}\ and\ \bibinfo {author} {\bibfnamefont {D.}~\bibnamefont {Zgid}},\
  }\bibfield  {title} {\bibinfo {title} {{Self-Consistent Second-Order Green's
  Function Perturbation Theory for Periodic Systems}},\ }\href
  {https://doi.org/10.1063/1.4940900} {\bibfield  {journal} {\bibinfo
  {journal} {The Journal of Chemical Physics}\ }\textbf {\bibinfo {volume}
  {144}},\ \bibinfo {pages} {054106} (\bibinfo {year} {2016})}\BibitemShut
  {NoStop}%
\bibitem [{\citenamefont {Phillips}\ and\ \citenamefont
  {Zgid}(2014)}]{Phillips14}%
  \BibitemOpen
  \bibfield  {author} {\bibinfo {author} {\bibfnamefont {J.~J.}\ \bibnamefont
  {Phillips}}\ and\ \bibinfo {author} {\bibfnamefont {D.}~\bibnamefont
  {Zgid}},\ }\bibfield  {title} {\bibinfo {title} {{Communication: The
  Description of Strong Correlation within Self-Consistent Green's Function
  Second-Order Perturbation Theory}},\ }\href
  {https://doi.org/10.1063/1.4884951} {\bibfield  {journal} {\bibinfo
  {journal} {The Journal of Chemical Physics}\ }\textbf {\bibinfo {volume}
  {140}},\ \bibinfo {pages} {241101} (\bibinfo {year} {2014})}\BibitemShut
  {NoStop}%
\bibitem [{\citenamefont {Iskakov}\ \emph {et~al.}(2019)\citenamefont
  {Iskakov}, \citenamefont {Rusakov}, \citenamefont {Zgid},\ and\ \citenamefont
  {Gull}}]{Iskakov19}%
  \BibitemOpen
  \bibfield  {author} {\bibinfo {author} {\bibfnamefont {S.}~\bibnamefont
  {Iskakov}}, \bibinfo {author} {\bibfnamefont {A.~A.}\ \bibnamefont
  {Rusakov}}, \bibinfo {author} {\bibfnamefont {D.}~\bibnamefont {Zgid}},\ and\
  \bibinfo {author} {\bibfnamefont {E.}~\bibnamefont {Gull}},\ }\bibfield
  {title} {\bibinfo {title} {{Effect of Propagator Renormalization on the Band
  Gap of Insulating Solids}},\ }\href
  {https://doi.org/10.1103/PhysRevB.100.085112} {\bibfield  {journal} {\bibinfo
   {journal} {Phys. Rev. B}\ }\textbf {\bibinfo {volume} {100}},\ \bibinfo
  {pages} {085112} (\bibinfo {year} {2019})}\BibitemShut {NoStop}%
\bibitem [{\citenamefont {Georges}\ \emph {et~al.}(1996)\citenamefont
  {Georges}, \citenamefont {Kotliar}, \citenamefont {Krauth},\ and\
  \citenamefont {Rozenberg}}]{Antoine_DMFT_RevModPhys_1996}%
  \BibitemOpen
  \bibfield  {author} {\bibinfo {author} {\bibfnamefont {A.}~\bibnamefont
  {Georges}}, \bibinfo {author} {\bibfnamefont {G.}~\bibnamefont {Kotliar}},
  \bibinfo {author} {\bibfnamefont {W.}~\bibnamefont {Krauth}},\ and\ \bibinfo
  {author} {\bibfnamefont {M.~J.}\ \bibnamefont {Rozenberg}},\ }\bibfield
  {title} {\bibinfo {title} {{Dynamical Mean-Field Theory of Strongly
  Correlated Fermion Systems and the Limit of Infinite Dimensions}},\ }\href
  {https://doi.org/10.1103/RevModPhys.68.13} {\bibfield  {journal} {\bibinfo
  {journal} {Rev. Mod. Phys.}\ }\textbf {\bibinfo {volume} {68}},\ \bibinfo
  {pages} {13} (\bibinfo {year} {1996})}\BibitemShut {NoStop}%
\bibitem [{\citenamefont {Ren}\ \emph {et~al.}(2012)\citenamefont {Ren},
  \citenamefont {Rinke}, \citenamefont {Blum}, \citenamefont {Wieferink},
  \citenamefont {Tkatchenko}, \citenamefont {Sanfilippo}, \citenamefont
  {Reuter},\ and\ \citenamefont {Scheffler}}]{RI_HF_GW_MP2_2012}%
  \BibitemOpen
  \bibfield  {author} {\bibinfo {author} {\bibfnamefont {X.}~\bibnamefont
  {Ren}}, \bibinfo {author} {\bibfnamefont {P.}~\bibnamefont {Rinke}}, \bibinfo
  {author} {\bibfnamefont {V.}~\bibnamefont {Blum}}, \bibinfo {author}
  {\bibfnamefont {J.}~\bibnamefont {Wieferink}}, \bibinfo {author}
  {\bibfnamefont {A.}~\bibnamefont {Tkatchenko}}, \bibinfo {author}
  {\bibfnamefont {A.}~\bibnamefont {Sanfilippo}}, \bibinfo {author}
  {\bibfnamefont {K.}~\bibnamefont {Reuter}},\ and\ \bibinfo {author}
  {\bibfnamefont {M.}~\bibnamefont {Scheffler}},\ }\bibfield  {title} {\bibinfo
  {title} {{Resolution-of-Identity Approach to Hartree-Fock, Hybrid Density
  Functionals, RPA, MP2 and ${GW}$ with Numeric Atom-Centered Orbital Basis
  Functions}},\ }\href {https://doi.org/10.1088/1367-2630/14/5/053020}
  {\bibfield  {journal} {\bibinfo  {journal} {New Journal of Physics}\ }\textbf
  {\bibinfo {volume} {14}},\ \bibinfo {pages} {053020} (\bibinfo {year}
  {2012})}\BibitemShut {NoStop}%
\bibitem [{\citenamefont {Ye}\ and\ \citenamefont
  {Berkelbach}(2021)}]{RSDF_HongZhou2021}%
  \BibitemOpen
  \bibfield  {author} {\bibinfo {author} {\bibfnamefont {H.-Z.}\ \bibnamefont
  {Ye}}\ and\ \bibinfo {author} {\bibfnamefont {T.~C.}\ \bibnamefont
  {Berkelbach}},\ }\bibfield  {title} {\bibinfo {title} {{Fast Periodic
  Gaussian Density Fitting by Range Separation}},\ }\href
  {https://doi.org/10.1063/5.0046617} {\bibfield  {journal} {\bibinfo
  {journal} {The Journal of Chemical Physics}\ }\textbf {\bibinfo {volume}
  {154}},\ \bibinfo {pages} {131104} (\bibinfo {year} {2021})}\BibitemShut
  {NoStop}%
\bibitem [{\citenamefont {Gygi}\ and\ \citenamefont
  {Baldereschi}(1986)}]{Exdiv_Gygi_1986}%
  \BibitemOpen
  \bibfield  {author} {\bibinfo {author} {\bibfnamefont {F.}~\bibnamefont
  {Gygi}}\ and\ \bibinfo {author} {\bibfnamefont {A.}~\bibnamefont
  {Baldereschi}},\ }\bibfield  {title} {\bibinfo {title} {{Self-Consistent
  Hartree-Fock and Screened-Exchange Calculations in Solids: Application to
  Silicon}},\ }\href {https://doi.org/10.1103/PhysRevB.34.4405} {\bibfield
  {journal} {\bibinfo  {journal} {Phys. Rev. B}\ }\textbf {\bibinfo {volume}
  {34}},\ \bibinfo {pages} {4405} (\bibinfo {year} {1986})}\BibitemShut
  {NoStop}%
\bibitem [{\citenamefont {Carrier}\ \emph {et~al.}(2007)\citenamefont
  {Carrier}, \citenamefont {Rohra},\ and\ \citenamefont
  {G\"orling}}]{Exdiv_Carrier_2007}%
  \BibitemOpen
  \bibfield  {author} {\bibinfo {author} {\bibfnamefont {P.}~\bibnamefont
  {Carrier}}, \bibinfo {author} {\bibfnamefont {S.}~\bibnamefont {Rohra}},\
  and\ \bibinfo {author} {\bibfnamefont {A.}~\bibnamefont {G\"orling}},\
  }\bibfield  {title} {\bibinfo {title} {{General Treatment of the
  Singularities in Hartree-Fock and Exact-Exchange Kohn-Sham Methods for
  Solids}},\ }\href {https://doi.org/10.1103/PhysRevB.75.205126} {\bibfield
  {journal} {\bibinfo  {journal} {Phys. Rev. B}\ }\textbf {\bibinfo {volume}
  {75}},\ \bibinfo {pages} {205126} (\bibinfo {year} {2007})}\BibitemShut
  {NoStop}%
\bibitem [{\citenamefont {Broqvist}\ \emph {et~al.}(2009)\citenamefont
  {Broqvist}, \citenamefont {Alkauskas},\ and\ \citenamefont
  {Pasquarello}}]{ERI_correction_2009}%
  \BibitemOpen
  \bibfield  {author} {\bibinfo {author} {\bibfnamefont {P.}~\bibnamefont
  {Broqvist}}, \bibinfo {author} {\bibfnamefont {A.}~\bibnamefont
  {Alkauskas}},\ and\ \bibinfo {author} {\bibfnamefont {A.}~\bibnamefont
  {Pasquarello}},\ }\bibfield  {title} {\bibinfo {title} {{Hybrid-Functional
  Calculations with Plane-Wave Basis Sets: Effect of Singularity Correction on
  Total Energies, Energy Eigenvalues, and Defect Energy Levels}},\ }\href
  {https://doi.org/10.1103/PhysRevB.80.085114} {\bibfield  {journal} {\bibinfo
  {journal} {Phys. Rev. B}\ }\textbf {\bibinfo {volume} {80}},\ \bibinfo
  {pages} {085114} (\bibinfo {year} {2009})}\BibitemShut {NoStop}%
\bibitem [{\citenamefont {Peralta}\ \emph
  {et~al.}(2005{\natexlab{a}})\citenamefont {Peralta}, \citenamefont {Uddin},\
  and\ \citenamefont {Scuseria}}]{Peralta_JCP2005}%
  \BibitemOpen
  \bibfield  {author} {\bibinfo {author} {\bibfnamefont {J.~E.}\ \bibnamefont
  {Peralta}}, \bibinfo {author} {\bibfnamefont {J.}~\bibnamefont {Uddin}},\
  and\ \bibinfo {author} {\bibfnamefont {G.~E.}\ \bibnamefont {Scuseria}},\
  }\bibfield  {title} {\bibinfo {title} {{Scalar Relativistic All-Electron
  Density Functional Calculations on Periodic Systems}},\ }\href
  {https://doi.org/10.1063/1.1851973} {\bibfield  {journal} {\bibinfo
  {journal} {The Journal of Chemical Physics}\ }\textbf {\bibinfo {volume}
  {122}},\ \bibinfo {pages} {084108} (\bibinfo {year}
  {2005}{\natexlab{a}})}\BibitemShut {NoStop}%
\bibitem [{\citenamefont {Berry}(1955)}]{AgCl_expt_a_1955}%
  \BibitemOpen
  \bibfield  {author} {\bibinfo {author} {\bibfnamefont {C.~R.}\ \bibnamefont
  {Berry}},\ }\bibfield  {title} {\bibinfo {title} {Physical defects in silver
  halides},\ }\href {https://doi.org/10.1103/PhysRev.97.676} {\bibfield
  {journal} {\bibinfo  {journal} {Phys. Rev.}\ }\textbf {\bibinfo {volume}
  {97}},\ \bibinfo {pages} {676} (\bibinfo {year} {1955})}\BibitemShut
  {NoStop}%
\bibitem [{\citenamefont {Wang}\ \emph {et~al.}(2009)\citenamefont {Wang},
  \citenamefont {Huang}, \citenamefont {Zhang}, \citenamefont {Qin},
  \citenamefont {Jin}, \citenamefont {Dai}, \citenamefont {Wang}, \citenamefont
  {Wei}, \citenamefont {Zhan}, \citenamefont {Wang}, \citenamefont {Wang},\
  and\ \citenamefont {Whangbo}}]{AgBr_expt_a_2009}%
  \BibitemOpen
  \bibfield  {author} {\bibinfo {author} {\bibfnamefont {P.}~\bibnamefont
  {Wang}}, \bibinfo {author} {\bibfnamefont {B.}~\bibnamefont {Huang}},
  \bibinfo {author} {\bibfnamefont {X.}~\bibnamefont {Zhang}}, \bibinfo
  {author} {\bibfnamefont {X.}~\bibnamefont {Qin}}, \bibinfo {author}
  {\bibfnamefont {H.}~\bibnamefont {Jin}}, \bibinfo {author} {\bibfnamefont
  {Y.}~\bibnamefont {Dai}}, \bibinfo {author} {\bibfnamefont {Z.}~\bibnamefont
  {Wang}}, \bibinfo {author} {\bibfnamefont {J.}~\bibnamefont {Wei}}, \bibinfo
  {author} {\bibfnamefont {J.}~\bibnamefont {Zhan}}, \bibinfo {author}
  {\bibfnamefont {S.}~\bibnamefont {Wang}}, \bibinfo {author} {\bibfnamefont
  {J.}~\bibnamefont {Wang}},\ and\ \bibinfo {author} {\bibfnamefont {M.-H.}\
  \bibnamefont {Whangbo}},\ }\bibfield  {title} {\bibinfo {title} {{Highly
  Efficient Visible-Light Plasmonic Photocatalyst Ag@AgBr}},\ }\href
  {https://doi.org/10.1002/chem.200802327} {\bibfield  {journal} {\bibinfo
  {journal} {Chemistry - A European Journal}\ }\textbf {\bibinfo {volume}
  {15}},\ \bibinfo {pages} {1821} (\bibinfo {year} {2009})}\BibitemShut
  {NoStop}%
\bibitem [{\citenamefont {Pollak}\ and\ \citenamefont
  {Weigend}(2017)}]{x2c_cgto_Pollak}%
  \BibitemOpen
  \bibfield  {author} {\bibinfo {author} {\bibfnamefont {P.}~\bibnamefont
  {Pollak}}\ and\ \bibinfo {author} {\bibfnamefont {F.}~\bibnamefont
  {Weigend}},\ }\bibfield  {title} {\bibinfo {title} {{Segmented Contracted
  Error-Consistent Basis Sets of Double- and Triple-$\zeta$ Valence Quality for
  One- and Two-Component Relativistic All-Electron Calculations}},\ }\href
  {https://doi.org/10.1021/acs.jctc.7b00593} {\bibfield  {journal} {\bibinfo
  {journal} {Journal of Chemical Theory and Computation}\ }\textbf {\bibinfo
  {volume} {13}},\ \bibinfo {pages} {3696} (\bibinfo {year}
  {2017})}\BibitemShut {NoStop}%
\bibitem [{\citenamefont {Vilela~Oliveira}\ \emph {et~al.}(2019)\citenamefont
  {Vilela~Oliveira}, \citenamefont {Laun}, \citenamefont {Peintinger},\ and\
  \citenamefont {Bredow}}]{pob_crystal_basis_2019}%
  \BibitemOpen
  \bibfield  {author} {\bibinfo {author} {\bibfnamefont {D.}~\bibnamefont
  {Vilela~Oliveira}}, \bibinfo {author} {\bibfnamefont {J.}~\bibnamefont
  {Laun}}, \bibinfo {author} {\bibfnamefont {M.~F.}\ \bibnamefont
  {Peintinger}},\ and\ \bibinfo {author} {\bibfnamefont {T.}~\bibnamefont
  {Bredow}},\ }\bibfield  {title} {\bibinfo {title} {{BSSE-Correction Scheme
  for Consistent Gaussian Basis Sets of Double- and Triple-Zeta Valence with
  Polarization Quality for Solid-State Calculations}},\ }\href
  {https://doi.org/https://doi.org/10.1002/jcc.26013} {\bibfield  {journal}
  {\bibinfo  {journal} {Journal of Computational Chemistry}\ }\textbf {\bibinfo
  {volume} {40}},\ \bibinfo {pages} {2364} (\bibinfo {year}
  {2019})}\BibitemShut {NoStop}%
\bibitem [{\citenamefont {Godbout}\ \emph {et~al.}(1992)\citenamefont
  {Godbout}, \citenamefont {Salahub}, \citenamefont {Andzelm},\ and\
  \citenamefont {Wimmer}}]{dzvp_Godbout1992}%
  \BibitemOpen
  \bibfield  {author} {\bibinfo {author} {\bibfnamefont {N.}~\bibnamefont
  {Godbout}}, \bibinfo {author} {\bibfnamefont {D.~R.}\ \bibnamefont
  {Salahub}}, \bibinfo {author} {\bibfnamefont {J.}~\bibnamefont {Andzelm}},\
  and\ \bibinfo {author} {\bibfnamefont {E.}~\bibnamefont {Wimmer}},\
  }\bibfield  {title} {\bibinfo {title} {{Optimization of Gaussian-Type Basis
  Sets for Local Spin Density Functional Calculations. Part I. Boron Through
  Neon, Optimization Technique and Validation}},\ }\href
  {https://doi.org/10.1139/v92-079} {\bibfield  {journal} {\bibinfo  {journal}
  {Canadian Journal of Chemistry}\ }\textbf {\bibinfo {volume} {70}},\ \bibinfo
  {pages} {560} (\bibinfo {year} {1992})}\BibitemShut {NoStop}%
\bibitem [{\citenamefont {Paier}\ \emph {et~al.}(2005)\citenamefont {Paier},
  \citenamefont {Hirschl}, \citenamefont {Marsman},\ and\ \citenamefont
  {Kresse}}]{ERI_correction_2005}%
  \BibitemOpen
  \bibfield  {author} {\bibinfo {author} {\bibfnamefont {J.}~\bibnamefont
  {Paier}}, \bibinfo {author} {\bibfnamefont {R.}~\bibnamefont {Hirschl}},
  \bibinfo {author} {\bibfnamefont {M.}~\bibnamefont {Marsman}},\ and\ \bibinfo
  {author} {\bibfnamefont {G.}~\bibnamefont {Kresse}},\ }\bibfield  {title}
  {\bibinfo {title} {{The Perdew-Burke-Ernzerhof Exchange-Correlation
  Functional Applied to the G2-1 Test Set Using a Plane-Wave Basis Set}},\
  }\href {https://doi.org/10.1063/1.1926272} {\bibfield  {journal} {\bibinfo
  {journal} {The Journal of Chemical Physics}\ }\textbf {\bibinfo {volume}
  {122}},\ \bibinfo {pages} {234102} (\bibinfo {year} {2005})}\BibitemShut
  {NoStop}%
\bibitem [{\citenamefont {Visscher}\ and\ \citenamefont
  {Dyall}(1997)}]{Gaussian_nucleus_model_Visscher1997}%
  \BibitemOpen
  \bibfield  {author} {\bibinfo {author} {\bibfnamefont {L.}~\bibnamefont
  {Visscher}}\ and\ \bibinfo {author} {\bibfnamefont {K.}~\bibnamefont
  {Dyall}},\ }\bibfield  {title} {\bibinfo {title} {{Dirac-Fock Atomic
  Electronic Structure Calculations using Different Nuclear Charge
  Distributions}},\ }\href
  {https://doi.org/https://doi.org/10.1006/adnd.1997.0751} {\bibfield
  {journal} {\bibinfo  {journal} {Atomic Data and Nuclear Data Tables}\
  }\textbf {\bibinfo {volume} {67}},\ \bibinfo {pages} {207} (\bibinfo {year}
  {1997})}\BibitemShut {NoStop}%
\bibitem [{\citenamefont {Fei}\ \emph {et~al.}(2021{\natexlab{a}})\citenamefont
  {Fei}, \citenamefont {Yeh},\ and\ \citenamefont
  {Gull}}]{Nevanlinna_Jiani_2021}%
  \BibitemOpen
  \bibfield  {author} {\bibinfo {author} {\bibfnamefont {J.}~\bibnamefont
  {Fei}}, \bibinfo {author} {\bibfnamefont {C.-N.}\ \bibnamefont {Yeh}},\ and\
  \bibinfo {author} {\bibfnamefont {E.}~\bibnamefont {Gull}},\ }\bibfield
  {title} {\bibinfo {title} {{Nevanlinna Analytical Continuation}},\ }\href
  {https://doi.org/10.1103/PhysRevLett.126.056402} {\bibfield  {journal}
  {\bibinfo  {journal} {Phys. Rev. Lett.}\ }\textbf {\bibinfo {volume} {126}},\
  \bibinfo {pages} {056402} (\bibinfo {year} {2021}{\natexlab{a}})}\BibitemShut
  {NoStop}%
\bibitem [{\citenamefont {Fei}\ \emph {et~al.}(2021{\natexlab{b}})\citenamefont
  {Fei}, \citenamefont {Yeh}, \citenamefont {Zgid},\ and\ \citenamefont
  {Gull}}]{Fei21}%
  \BibitemOpen
  \bibfield  {author} {\bibinfo {author} {\bibfnamefont {J.}~\bibnamefont
  {Fei}}, \bibinfo {author} {\bibfnamefont {C.-N.}\ \bibnamefont {Yeh}},
  \bibinfo {author} {\bibfnamefont {D.}~\bibnamefont {Zgid}},\ and\ \bibinfo
  {author} {\bibfnamefont {E.}~\bibnamefont {Gull}},\ }\bibfield  {title}
  {\bibinfo {title} {{Analytical Continuation of Matrix-Valued Functions:
  Carath\'eodory Formalism}},\ }\href
  {https://doi.org/10.1103/PhysRevB.104.165111} {\bibfield  {journal} {\bibinfo
   {journal} {Phys. Rev. B}\ }\textbf {\bibinfo {volume} {104}},\ \bibinfo
  {pages} {165111} (\bibinfo {year} {2021}{\natexlab{b}})}\BibitemShut
  {NoStop}%
\bibitem [{\citenamefont {Shinaoka}\ \emph {et~al.}(2017)\citenamefont
  {Shinaoka}, \citenamefont {Otsuki}, \citenamefont {Ohzeki},\ and\
  \citenamefont {Yoshimi}}]{IR_Hiroshi_2017}%
  \BibitemOpen
  \bibfield  {author} {\bibinfo {author} {\bibfnamefont {H.}~\bibnamefont
  {Shinaoka}}, \bibinfo {author} {\bibfnamefont {J.}~\bibnamefont {Otsuki}},
  \bibinfo {author} {\bibfnamefont {M.}~\bibnamefont {Ohzeki}},\ and\ \bibinfo
  {author} {\bibfnamefont {K.}~\bibnamefont {Yoshimi}},\ }\bibfield  {title}
  {\bibinfo {title} {{Compressing Green's Function using Intermediate
  Representation between Imaginary-Time and Real-Frequency Domains}},\ }\href
  {https://doi.org/10.1103/PhysRevB.96.035147} {\bibfield  {journal} {\bibinfo
  {journal} {Phys. Rev. B}\ }\textbf {\bibinfo {volume} {96}},\ \bibinfo
  {pages} {035147} (\bibinfo {year} {2017})}\BibitemShut {NoStop}%
\bibitem [{\citenamefont {Li}\ \emph {et~al.}(2020)\citenamefont {Li},
  \citenamefont {Wallerberger}, \citenamefont {Chikano}, \citenamefont {Yeh},
  \citenamefont {Gull},\ and\ \citenamefont
  {Shinaoka}}]{sparse_sampling_Jia_2020}%
  \BibitemOpen
  \bibfield  {author} {\bibinfo {author} {\bibfnamefont {J.}~\bibnamefont
  {Li}}, \bibinfo {author} {\bibfnamefont {M.}~\bibnamefont {Wallerberger}},
  \bibinfo {author} {\bibfnamefont {N.}~\bibnamefont {Chikano}}, \bibinfo
  {author} {\bibfnamefont {C.-N.}\ \bibnamefont {Yeh}}, \bibinfo {author}
  {\bibfnamefont {E.}~\bibnamefont {Gull}},\ and\ \bibinfo {author}
  {\bibfnamefont {H.}~\bibnamefont {Shinaoka}},\ }\bibfield  {title} {\bibinfo
  {title} {{Sparse Sampling Approach to Efficient \emph{Ab Initio} Calculations
  at Finite Temperature}},\ }\href
  {https://doi.org/10.1103/PhysRevB.101.035144} {\bibfield  {journal} {\bibinfo
   {journal} {Phys. Rev. B}\ }\textbf {\bibinfo {volume} {101}},\ \bibinfo
  {pages} {035144} (\bibinfo {year} {2020})}\BibitemShut {NoStop}%
\bibitem [{\citenamefont {van Setten}\ \emph {et~al.}(2017)\citenamefont {van
  Setten}, \citenamefont {Giantomassi}, \citenamefont {Gonze}, \citenamefont
  {Rignanese},\ and\ \citenamefont {Hautier}}]{high_throughput_GW}%
  \BibitemOpen
  \bibfield  {author} {\bibinfo {author} {\bibfnamefont {M.~J.}\ \bibnamefont
  {van Setten}}, \bibinfo {author} {\bibfnamefont {M.}~\bibnamefont
  {Giantomassi}}, \bibinfo {author} {\bibfnamefont {X.}~\bibnamefont {Gonze}},
  \bibinfo {author} {\bibfnamefont {G.-M.}\ \bibnamefont {Rignanese}},\ and\
  \bibinfo {author} {\bibfnamefont {G.}~\bibnamefont {Hautier}},\ }\bibfield
  {title} {\bibinfo {title} {{Automation Methodologies and Large-Scale
  Validation for $GW$: Towards High-Throughput $GW$ Calculations}},\ }\href
  {https://doi.org/10.1103/PhysRevB.96.155207} {\bibfield  {journal} {\bibinfo
  {journal} {Phys. Rev. B}\ }\textbf {\bibinfo {volume} {96}},\ \bibinfo
  {pages} {155207} (\bibinfo {year} {2017})}\BibitemShut {NoStop}%
\bibitem [{\citenamefont {Gao}\ \emph {et~al.}(2018)\citenamefont {Gao},
  \citenamefont {Xia}, \citenamefont {Wu}, \citenamefont {Ren}, \citenamefont
  {Gao},\ and\ \citenamefont {Zhang}}]{GW_AgX_PRB_2018}%
  \BibitemOpen
  \bibfield  {author} {\bibinfo {author} {\bibfnamefont {W.}~\bibnamefont
  {Gao}}, \bibinfo {author} {\bibfnamefont {W.}~\bibnamefont {Xia}}, \bibinfo
  {author} {\bibfnamefont {Y.}~\bibnamefont {Wu}}, \bibinfo {author}
  {\bibfnamefont {W.}~\bibnamefont {Ren}}, \bibinfo {author} {\bibfnamefont
  {X.}~\bibnamefont {Gao}},\ and\ \bibinfo {author} {\bibfnamefont
  {P.}~\bibnamefont {Zhang}},\ }\bibfield  {title} {\bibinfo {title}
  {{Quasiparticle Band Structures of CuCl, CuBr, AgCl, and AgBr: The extreme
  case}},\ }\href {https://doi.org/10.1103/PhysRevB.98.045108} {\bibfield
  {journal} {\bibinfo  {journal} {Phys. Rev. B}\ }\textbf {\bibinfo {volume}
  {98}},\ \bibinfo {pages} {045108} (\bibinfo {year} {2018})}\BibitemShut
  {NoStop}%
\bibitem [{\citenamefont {Zhang}\ and\ \citenamefont
  {Jiang}(2019)}]{AgX_LAPW_HLOs}%
  \BibitemOpen
  \bibfield  {author} {\bibinfo {author} {\bibfnamefont {M.-Y.}\ \bibnamefont
  {Zhang}}\ and\ \bibinfo {author} {\bibfnamefont {H.}~\bibnamefont {Jiang}},\
  }\bibfield  {title} {\bibinfo {title} {{Electronic Band Structure of Cuprous
  and Silver Halides: An All-Electron $GW$ Study}},\ }\href
  {https://doi.org/10.1103/PhysRevB.100.205123} {\bibfield  {journal} {\bibinfo
   {journal} {Phys. Rev. B}\ }\textbf {\bibinfo {volume} {100}},\ \bibinfo
  {pages} {205123} (\bibinfo {year} {2019})}\BibitemShut {NoStop}%
\bibitem [{\citenamefont {Lorin}\ \emph {et~al.}(2021)\citenamefont {Lorin},
  \citenamefont {Gatti}, \citenamefont {Reining},\ and\ \citenamefont
  {Sottile}}]{AgCl_optical_2021}%
  \BibitemOpen
  \bibfield  {author} {\bibinfo {author} {\bibfnamefont {A.}~\bibnamefont
  {Lorin}}, \bibinfo {author} {\bibfnamefont {M.}~\bibnamefont {Gatti}},
  \bibinfo {author} {\bibfnamefont {L.}~\bibnamefont {Reining}},\ and\ \bibinfo
  {author} {\bibfnamefont {F.}~\bibnamefont {Sottile}},\ }\bibfield  {title}
  {\bibinfo {title} {{First-Principles Study of Excitons in Optical Spectra of
  Silver Chloride}},\ }\href {https://doi.org/10.1103/PhysRevB.104.235149}
  {\bibfield  {journal} {\bibinfo  {journal} {Phys. Rev. B}\ }\textbf {\bibinfo
  {volume} {104}},\ \bibinfo {pages} {235149} (\bibinfo {year}
  {2021})}\BibitemShut {NoStop}%
\bibitem [{\citenamefont {Singh}\ and\ \citenamefont
  {Nordstr\"om}(2006)}]{LAPW}%
  \BibitemOpen
  \bibfield  {author} {\bibinfo {author} {\bibfnamefont {D.~J.}\ \bibnamefont
  {Singh}}\ and\ \bibinfo {author} {\bibfnamefont {L.}~\bibnamefont
  {Nordstr\"om}},\ }\href {https://doi.org/10.1007/978-0-387-29684-5} {\emph
  {\bibinfo {title} {{Planewaves, Pseudopotentials and the LAPW Method}}}}\
  (\bibinfo  {publisher} {Springer {US}},\ \bibinfo {year} {2006})\BibitemShut
  {NoStop}%
\bibitem [{\citenamefont {Gr\"uneis}\ \emph {et~al.}(2014)\citenamefont
  {Gr\"uneis}, \citenamefont {Kresse}, \citenamefont {Hinuma},\ and\
  \citenamefont {Oba}}]{GW_vertex_Gruneis_2014}%
  \BibitemOpen
  \bibfield  {author} {\bibinfo {author} {\bibfnamefont {A.}~\bibnamefont
  {Gr\"uneis}}, \bibinfo {author} {\bibfnamefont {G.}~\bibnamefont {Kresse}},
  \bibinfo {author} {\bibfnamefont {Y.}~\bibnamefont {Hinuma}},\ and\ \bibinfo
  {author} {\bibfnamefont {F.}~\bibnamefont {Oba}},\ }\bibfield  {title}
  {\bibinfo {title} {{Ionization Potentials of Solids: The Importance of Vertex
  Corrections}},\ }\href {https://doi.org/10.1103/PhysRevLett.112.096401}
  {\bibfield  {journal} {\bibinfo  {journal} {Phys. Rev. Lett.}\ }\textbf
  {\bibinfo {volume} {112}},\ \bibinfo {pages} {096401} (\bibinfo {year}
  {2014})}\BibitemShut {NoStop}%
\bibitem [{\citenamefont {Kutepov}(2017)}]{GW_vertex_Andrey_2017}%
  \BibitemOpen
  \bibfield  {author} {\bibinfo {author} {\bibfnamefont {A.~L.}\ \bibnamefont
  {Kutepov}},\ }\bibfield  {title} {\bibinfo {title} {{Self-Consistent Solution
  of Hedin's Equations: Semiconductors and Insulators}},\ }\href
  {https://doi.org/10.1103/PhysRevB.95.195120} {\bibfield  {journal} {\bibinfo
  {journal} {Phys. Rev. B}\ }\textbf {\bibinfo {volume} {95}},\ \bibinfo
  {pages} {195120} (\bibinfo {year} {2017})}\BibitemShut {NoStop}%
\bibitem [{\citenamefont {Govoni}\ and\ \citenamefont
  {Galli}(2018)}]{GW100_benchmark_West_2018}%
  \BibitemOpen
  \bibfield  {author} {\bibinfo {author} {\bibfnamefont {M.}~\bibnamefont
  {Govoni}}\ and\ \bibinfo {author} {\bibfnamefont {G.}~\bibnamefont {Galli}},\
  }\bibfield  {title} {\bibinfo {title} {{GW100: Comparison of Methods and
  Accuracy of Results Obtained with the WEST Code}},\ }\href
  {https://doi.org/10.1021/acs.jctc.7b00952} {\bibfield  {journal} {\bibinfo
  {journal} {Journal of Chemical Theory and Computation}\ }\textbf {\bibinfo
  {volume} {14}},\ \bibinfo {pages} {1895} (\bibinfo {year} {2018})},\ \bibinfo
  {note} {pMID: 29397712}\BibitemShut {NoStop}%
\bibitem [{\citenamefont {VandeVondele}\ and\ \citenamefont
  {Hutter}(2007)}]{gth_GTO_VandeVondele_2007}%
  \BibitemOpen
  \bibfield  {author} {\bibinfo {author} {\bibfnamefont {J.}~\bibnamefont
  {VandeVondele}}\ and\ \bibinfo {author} {\bibfnamefont {J.}~\bibnamefont
  {Hutter}},\ }\bibfield  {title} {\bibinfo {title} {{Gaussian Basis Sets for
  Accurate Calculations on Molecular Systems in Gas and Condensed Phases}},\
  }\href {https://doi.org/10.1063/1.2770708} {\bibfield  {journal} {\bibinfo
  {journal} {The Journal of Chemical Physics}\ }\textbf {\bibinfo {volume}
  {127}},\ \bibinfo {pages} {114105} (\bibinfo {year} {2007})}\BibitemShut
  {NoStop}%
\bibitem [{\citenamefont {Goedecker}\ \emph {et~al.}(1996)\citenamefont
  {Goedecker}, \citenamefont {Teter},\ and\ \citenamefont
  {Hutter}}]{GTH_pseudo_1996}%
  \BibitemOpen
  \bibfield  {author} {\bibinfo {author} {\bibfnamefont {S.}~\bibnamefont
  {Goedecker}}, \bibinfo {author} {\bibfnamefont {M.}~\bibnamefont {Teter}},\
  and\ \bibinfo {author} {\bibfnamefont {J.}~\bibnamefont {Hutter}},\
  }\bibfield  {title} {\bibinfo {title} {{Separable Dual-Space Gaussian
  Pseudopotentials}},\ }\href {https://doi.org/10.1103/PhysRevB.54.1703}
  {\bibfield  {journal} {\bibinfo  {journal} {Phys. Rev. B}\ }\textbf {\bibinfo
  {volume} {54}},\ \bibinfo {pages} {1703} (\bibinfo {year}
  {1996})}\BibitemShut {NoStop}%
\bibitem [{\citenamefont {Madelung}\ \emph {et~al.}(1999)\citenamefont
  {Madelung}, \citenamefont {R\"{o}ssler},\ and\ \citenamefont
  {Schulz}}]{Expt_silver_halides}%
  \BibitemOpen
  \bibinfo {editor} {\bibfnamefont {O.}~\bibnamefont {Madelung}}, \bibinfo
  {editor} {\bibfnamefont {U.}~\bibnamefont {R\"{o}ssler}},\ and\ \bibinfo
  {editor} {\bibfnamefont {M.}~\bibnamefont {Schulz}},\ eds.,\ \href
  {https://doi.org/10.1007/b71137} {\emph {\bibinfo {title} {{II-VI and I-VII
  Compounds: Semimagnetic Compounds}}}}\ (\bibinfo  {publisher}
  {Springer-Verlag},\ \bibinfo {year} {1999})\BibitemShut {NoStop}%
\bibitem [{\citenamefont {Nakamura}\ and\ \citenamefont {von~der
  Osten}(1983)}]{AgCl_indirect_gap_1983}%
  \BibitemOpen
  \bibfield  {author} {\bibinfo {author} {\bibfnamefont {K.}~\bibnamefont
  {Nakamura}}\ and\ \bibinfo {author} {\bibfnamefont {W.}~\bibnamefont {von~der
  Osten}},\ }\bibfield  {title} {\bibinfo {title} {{Exciton Relaxation in AgCl
  Studied by Resonant Raman Scattering}},\ }\href
  {https://doi.org/10.1088/0022-3719/16/34/016} {\bibfield  {journal} {\bibinfo
   {journal} {Journal of Physics C: Solid State Physics}\ }\textbf {\bibinfo
  {volume} {16}},\ \bibinfo {pages} {6669} (\bibinfo {year}
  {1983})}\BibitemShut {NoStop}%
\bibitem [{\citenamefont {Sommer}(1968)}]{AgBr_AgCl_indirect_gap}%
  \BibitemOpen
  \bibfield  {author} {\bibinfo {author} {\bibfnamefont {A.~H.}\ \bibnamefont
  {Sommer}},\ }\href@noop {} {\emph {\bibinfo {title} {{Photoemissive
  Materials}}}}\ (\bibinfo  {publisher} {Wiley, New York},\ \bibinfo {year}
  {1968})\BibitemShut {NoStop}%
\bibitem [{\citenamefont {Bassani}\ \emph {et~al.}(1965)\citenamefont
  {Bassani}, \citenamefont {Knox},\ and\ \citenamefont
  {Fowler}}]{AgBr_direct_gap_1965}%
  \BibitemOpen
  \bibfield  {author} {\bibinfo {author} {\bibfnamefont {F.}~\bibnamefont
  {Bassani}}, \bibinfo {author} {\bibfnamefont {R.~S.}\ \bibnamefont {Knox}},\
  and\ \bibinfo {author} {\bibfnamefont {W.~B.}\ \bibnamefont {Fowler}},\
  }\bibfield  {title} {\bibinfo {title} {{Band Structure and Electronic
  Properties of AgCl and AgBr}},\ }\href
  {https://doi.org/10.1103/PhysRev.137.A1217} {\bibfield  {journal} {\bibinfo
  {journal} {Phys. Rev.}\ }\textbf {\bibinfo {volume} {137}},\ \bibinfo {pages}
  {A1217} (\bibinfo {year} {1965})}\BibitemShut {NoStop}%
\bibitem [{\citenamefont {Sliwczuk}\ \emph {et~al.}(1984)\citenamefont
  {Sliwczuk}, \citenamefont {Stolz},\ and\ \citenamefont {von
  Dee~Osten}}]{AgBr_indirect_gap_1984}%
  \BibitemOpen
  \bibfield  {author} {\bibinfo {author} {\bibfnamefont {U.}~\bibnamefont
  {Sliwczuk}}, \bibinfo {author} {\bibfnamefont {H.}~\bibnamefont {Stolz}},\
  and\ \bibinfo {author} {\bibfnamefont {W.}~\bibnamefont {von Dee~Osten}},\
  }\bibfield  {title} {\bibinfo {title} {{Indirect-Forbidden Exciton
  Transitions in AgBr}},\ }\href {https://doi.org/10.1002/pssb.2221220124}
  {\bibfield  {journal} {\bibinfo  {journal} {physica status solidi (b)}\
  }\textbf {\bibinfo {volume} {122}},\ \bibinfo {pages} {203} (\bibinfo {year}
  {1984})}\BibitemShut {NoStop}%
\bibitem [{\citenamefont {Peralta}\ \emph
  {et~al.}(2005{\natexlab{b}})\citenamefont {Peralta}, \citenamefont {Uddin},\
  and\ \citenamefont {Scuseria}}]{AgX_DFT_Peralta_2005}%
  \BibitemOpen
  \bibfield  {author} {\bibinfo {author} {\bibfnamefont {J.~E.}\ \bibnamefont
  {Peralta}}, \bibinfo {author} {\bibfnamefont {J.}~\bibnamefont {Uddin}},\
  and\ \bibinfo {author} {\bibfnamefont {G.~E.}\ \bibnamefont {Scuseria}},\
  }\bibfield  {title} {\bibinfo {title} {{Scalar Relativistic All-Electron
  Density Functional Calculations on Periodic Systems}},\ }\href
  {https://doi.org/10.1063/1.1851973} {\bibfield  {journal} {\bibinfo
  {journal} {The Journal of Chemical Physics}\ }\textbf {\bibinfo {volume}
  {122}},\ \bibinfo {pages} {084108} (\bibinfo {year}
  {2005}{\natexlab{b}})}\BibitemShut {NoStop}%
\bibitem [{\citenamefont {L\"{o}wdin}(1970)}]{Lowdin_sym_orth_1970}%
  \BibitemOpen
  \bibfield  {author} {\bibinfo {author} {\bibfnamefont {P.-O.}\ \bibnamefont
  {L\"{o}wdin}},\ }\bibfield  {title} {\bibinfo {title} {{On the
  Nonorthogonality Problem}},\ }in\ \href
  {https://doi.org/10.1016/s0065-3276(08)60339-1} {\emph {\bibinfo {booktitle}
  {Advances in Quantum Chemistry Volume 5}}}\ (\bibinfo  {publisher}
  {Elsevier},\ \bibinfo {year} {1970})\ pp.\ \bibinfo {pages}
  {185--199}\BibitemShut {NoStop}%
\bibitem [{\citenamefont {Franzke}\ \emph {et~al.}(2020)\citenamefont
  {Franzke}, \citenamefont {Spiske}, \citenamefont {Pollak},\ and\
  \citenamefont {Weigend}}]{x2c_qzvo_Franzke}%
  \BibitemOpen
  \bibfield  {author} {\bibinfo {author} {\bibfnamefont {Y.~J.}\ \bibnamefont
  {Franzke}}, \bibinfo {author} {\bibfnamefont {L.}~\bibnamefont {Spiske}},
  \bibinfo {author} {\bibfnamefont {P.}~\bibnamefont {Pollak}},\ and\ \bibinfo
  {author} {\bibfnamefont {F.}~\bibnamefont {Weigend}},\ }\bibfield  {title}
  {\bibinfo {title} {{Segmented Contracted Error-Consistent Basis Sets of
  Quadruple-$\zeta$ Valence Quality for One- and Two-Component Relativistic
  All-Electron Calculations}},\ }\href
  {https://doi.org/10.1021/acs.jctc.0c00546} {\bibfield  {journal} {\bibinfo
  {journal} {Journal of Chemical Theory and Computation}\ }\textbf {\bibinfo
  {volume} {16}},\ \bibinfo {pages} {5658} (\bibinfo {year} {2020})},\ \bibinfo
  {note} {pMID: 32786897}\BibitemShut {NoStop}%
\end{thebibliography}%

\appendix
\section{Basis set convergence\label{appendix:basis_set_conv}}
\begin{table}[bth]
\begin{ruledtabular}
\begin{tabular}{c|c|cccc}
basis & orbitals & $L-L$ & $\Gamma-\Gamma$ & $X-X$ & $L-\Gamma$ \\
\hline
x2c-SV(P)all &78 & 6.86 & 5.64 & 8.03 & 3.94 \\
x2c-SVPall & 101 & 6.66 & 5.37 & 7.66 & 3.69 \\
x2c-TZVPall & 111 & 6.54 & 5.28 & 7.54 & 3.58 \\
x2c-TZVPPall & 127 & 6.55 & 5.15 & 7.35 & 3.54 \\
x2c-QZVPall & 185 & 6.47 & 5.14 & 7.29 & 3.50 \\
\end{tabular}
\end{ruledtabular}
\caption{sc$GW$ band gaps of AgBr calculated employing the sfX2C1e-Coulomb Hamiltonian in different basis sets~\cite{x2c_cgto_Pollak, x2c_qzvo_Franzke} at an $4\times4\times4$ $\Gamma$-centred $k$-mesh and inverse temperature $\beta=300$ Ha$^{-1}$. The total numbers of GTOs per cell for each basis set are listed in the second column. 
\label{tab:basis_convergence}}
\end{table}
Here, we investigate the basis convergence of the sc$GW$ band gaps. 
Note that, in the present work, no implicit orbital truncation is employed which means all the Bloch GTOs (both occupied and virtual orbitals) are included in every $GW$ evaluated expression. 
We adopt a family of all-electron Gaussian basis optimized with respect to X2C Hamiltonian~\cite{x2c_cgto_Pollak,x2c_qzvo_Franzke}. 
The basis set is systematically enlarged starting from the double-$\zeta$ (called here x2c-SV(P)all) to triple-$\zeta$ (x2c-TZVPall), and to quadruple-$\zeta$ (x2c-QZVPall) level. For the double-$\zeta$ and the triple-$\zeta$ basis, more polarized variants are also used (x2c-SVPall and x2c-TZVPPall, respectively) in which additional high-lying orbitals are added. 
A higher temperature is used to circumvent large IR grids on the imaginary axes for large basis set such as x2c-QZVPall basis. It is found, as expected for insulators, that temperature dependency of band gaps from $\beta=700$ to $300$ a.u.$^{-1}$ is consistently well below 0.01 eV.  

As suggested in Ref.~\onlinecite{AgX_LAPW_HLOs}, for silver halides similar to the well-known system ZnO,  a slow basis set convergence,  is expected due to the silver $d$ orbitals.
In Table~\ref{tab:basis_convergence}, we examine such basis set effects and observe a systematic basis set convergence of the sc$GW$ band gaps.
From x2c-SV(P)all to x2c-TZVPall basis, a maximum difference of $\sim 0.5$ eV for $X-X$ band gap is observed. All other bandgaps ($L-L$, $\Gamma-\Gamma$, $L-\Gamma$) result in smaller differences. Further adding more high-lying localized orbitals, a maximum difference of $\sim 0.25$ eV is observed for the $X$-to-$X$ gap when going from x2c-TZVPall to x2c-QZVPall basis. Even smaller differences are observed for the other gaps.
Note that from x2c-SV(P)all to x2c-TZVPall, and to x2c-QZVPall basis set, the number of GTO orbitals in the unit cell increases from 78 to 111, and finally to 185. 
Consequently, while we cannot attest that our results are converged completely with the basis set size, going to the next level x2c-5ZVPall basis set will most likely result in differences at the level of $\sim 0.1$ eV and we should not expect any major quantitative differences between x2c-QZVPall and x2c-5ZVPall basis sets.

Enlarging the size of GTO basis sets is similar to adding high energy local orbitals  (HLO) within the LAPW framework. A similar basis convergence behavior with respect to HLOs can be found for LAPW calculations in Ref.~\onlinecite{AgX_LAPW_HLOs}. 
In the presence of Ag $d$ orbitals, although a less severe basis set error is found in our GTO basis compared to the standard LAPW basis set, the x2c-TZVPall basis set used in the present work still suffers from the basis set error. 

\section{DFT band gaps\label{appendix:AgCl_AgBr_PBE_gaps}}
\begin{table}[bth]
\begin{ruledtabular}
\begin{tabular}{c|c|cccc}
Non-relativistic & $a_{0}$ & $L-L$ & $\Gamma-\Gamma$ & $X-X$ & $L-X$ \\
\hline
PBE & 5.692 & 5.00 & 3.46 & 5.55 & 1.72 \\
PBE~\cite{Kadek_PRB2019} & 5.692 & 4.93 & 3.47 & 5.47 & 1.68 \\ 
PBE~\cite{Rundong_JCP2016} & 5.692 & 4.72 & 3.44 & 5.29 & 1.67 \\
\hline
Scalar relativistic \\
\hline
sfX2C1e-Coulomb & 5.612 & 4.44 & 3.09 & 4.27 & 0.94 \\ 
sfX2C-Coulomb~\cite{Rundong_JCP2016} & 5.613 & 4.31 & 3.09 & 4.23 & 0.92 \\
\hline
Fully relativistic \\
\hline
X2C1e-Coulomb & 5.612 & 4.39 & 2.98 & 4.03 & 0.89 \\
DKS-Coulomb~\cite{Kadek_PRB2019} & 5.612 & 4.47 & 2.93 & 4.20 & 0.87 \\
X2C-Coulomb~\cite{Rundong_JCP2016} & 5.612 & 4.27 & 2.99 & 4.03 & 0.88 \\ 
\end{tabular}
\end{ruledtabular}
\caption{Lattice constants $a_{0}$ and energy gaps (eV) of AgCl. \label{tab:AgCl_tab}}
\end{table}

\begin{table}[bth]
\begin{ruledtabular}
\begin{tabular}{c|c|cccc}
Non-relativistic & $a_{0}$ & $L-L$ & $\Gamma-\Gamma$ & $X-X$ & $L-X$ \\
\hline
PBE &  5.937 & 4.54 & 2.92 & 4.81 & 1.60 \\
PBE~\cite{Kadek_PRB2019} & 5.937 & 4.36 & 2.96 & 4.81 & 1.59 \\ 
PBE~\cite{Rundong_JCP2016} & 5.937 & 4.31 & 2.97 & 4.81 & 1.57 \\
\hline
Scalar relativistic \\
\hline
sfX2C1e-Coulomb & 5.843 & 3.93 & 2.43 & 3.89 & 0.70 \\ 
sfX2C-Coulomb~\cite{Rundong_JCP2016} & 5.843 & 3.87 & 2.43 & 3.87 & 0.68 \\
\hline
Fully relativistic \\
\hline
X2C1e-Coulomb & 5.843  & 3.85 & 2.24 & 3.65 & 0.62 \\
DKS-Coulomb~\cite{Kadek_PRB2019} & 5.843 & 3.82 & 2.24 & 3.68 & 0.61 \\
X2C-Coulomb~\cite{Rundong_JCP2016} & 5.843 & 3.77 & 2.25 & 3.67 & 0.60 \\
\end{tabular}
\end{ruledtabular}
\caption{Lattice constants $a_{0}$ and energy gaps (eV) of AgBr. \label{tab:AgBr_tab}}
\end{table}
Here we present PBE band gaps calculated using the non-relativistic and series of relativistic Hamiltonians for a similar direct comparison as performed in Sec.~\ref{subsec:x2c1e_AgI}. 
For AgCl and AgBr, the lattice constants optimized at the PBE level~\cite{Rundong_JCP2016} were used. 
As shown in Table.~\ref{tab:AgCl_tab} and~\ref{tab:AgBr_tab}, the PBE band gaps calculated using the sfX2C1e-Coulomb and the X2C1e-Coulomb Hamiltonian show similar agreement with the more sophisticated relativistic Hamiltonians, which is consistent to what we observe for AgI in Sec.~\ref{subsec:x2c1e_AgI}.  

\section{Finite-size corrections\label{app:finite_size}}
\begin{figure}[tbh]
\includegraphics[width=0.42\textwidth]{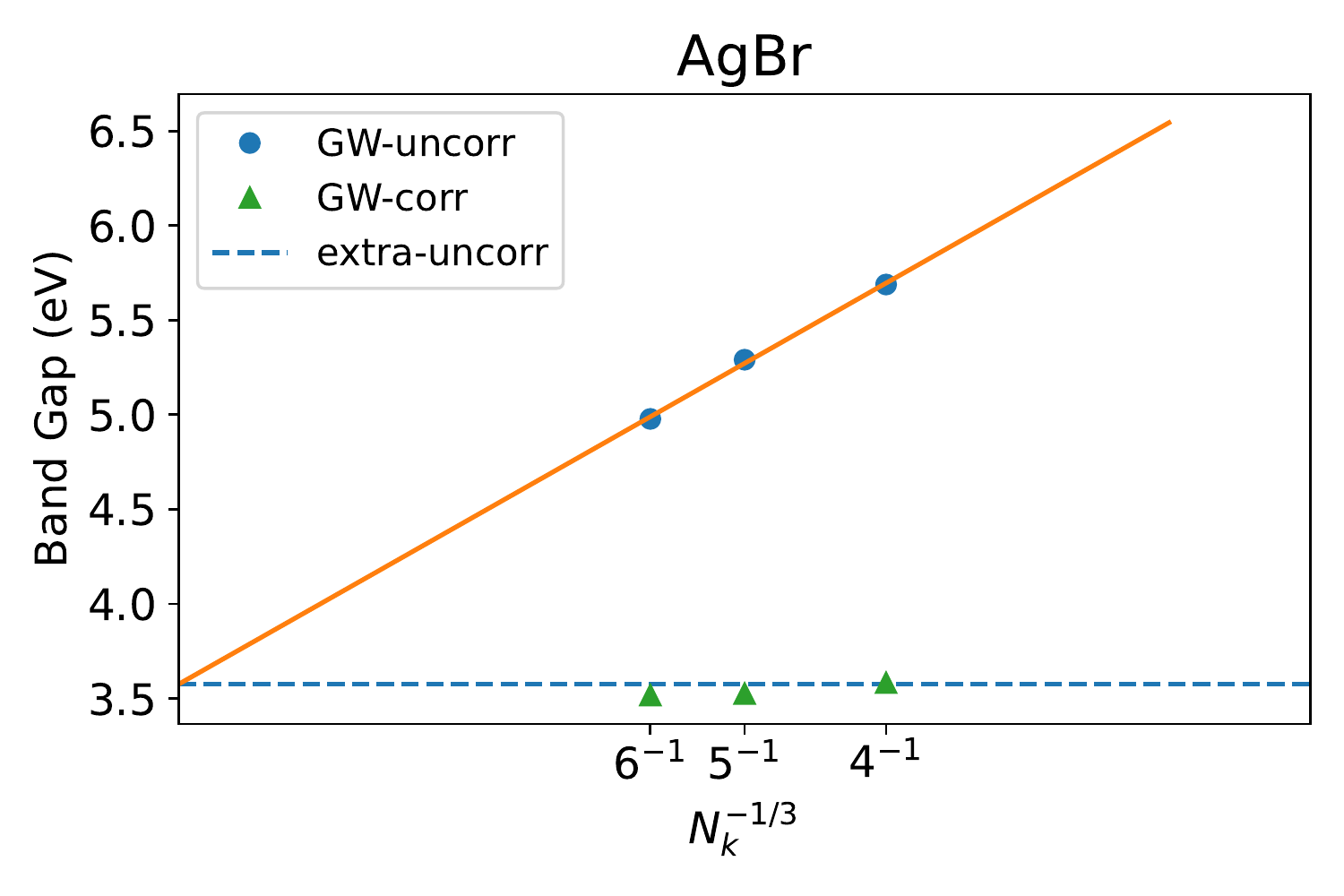}
\includegraphics[width=0.42\textwidth]{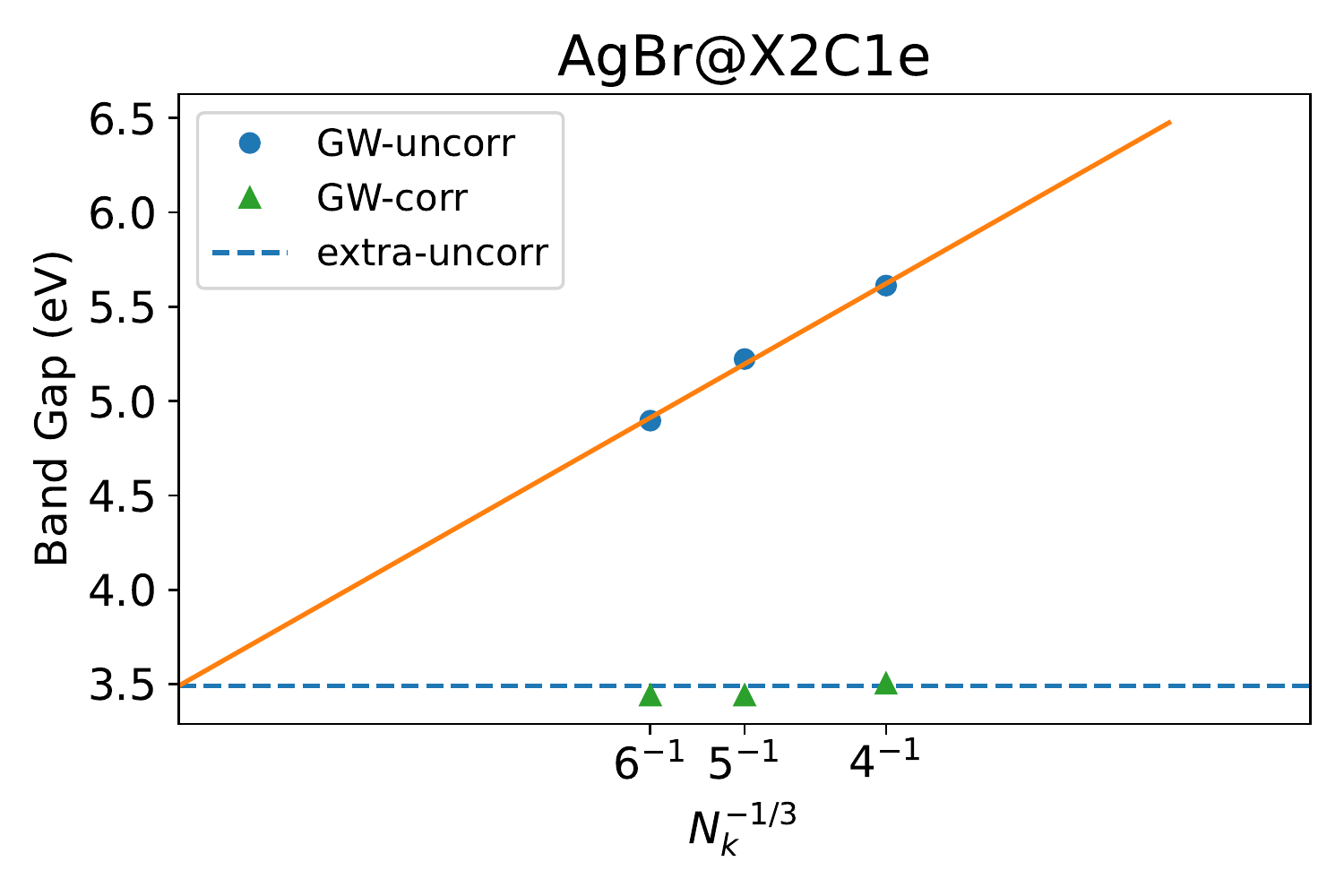}
\caption{sc$GW$ band gaps of AgBr calculated using the sfX2C1e-Coulomb (top panel) and the X2C1e-Coulomb (lower panel) Hamiltonian as a function of $N^{-1/3}_{k}$. Both the sc$GW$ band gaps with ($GW$-corr) and without ($GW$-uncorr) the head corrections to the dynamic part of the $GW$ self-energy as shown. }\label{fig:finite_size}
\end{figure}
The finite-size effects in our relativistic sc$GW$ are investigated as shown in Fig.~\ref{fig:finite_size}. The sc$GW$ band gaps of AgBr with ($GW$-corr) and without ($GW$-uncorr) the head correction to the dynamic part of the $GW$ self-energy (Eq.~\ref{eqn:GW_head_corr}) as a function of $N^{-1/3}_{k}$. Note that the finite-size corrections to the HF exchange potential is always included in both $GW$-corr and $GW$-uncorr. 

The sc$GW$ band gaps without the head correction calculated using both the sfX2C1e-Coulomb and the X2C1e-Coulomb Hamiltonians show a linear dependence with respect to $N^{-1/3}_{k}$, as expected. We then fit the sc$GW$ band gap to $\Delta(N_{k}) = \Delta_{\mathrm{TDL}} + aN_{k}^{-1/3}$ and extrapolate it to the thermodynamic limit (TDL) value $\Delta_{\mathrm{TDL}}$ which is shown as the blue dotted lines. In spite of the nice linear dependence with respect to $N^{-1/3}_{k}$, the slow convergence to TDL values makes the realistic finite size-uncorrected calculations impractical. 
The explicit inclusion of the head correction to the integrable divergence in the dynamic part of the $GW$ self-energy results in a significantly faster convergence of $GW$ band gaps with respect to the number of $k$-points. A similar convergence pattern is observed for both the sfX2C1e-Coulomb and the X2C1e-Coulomb Hamiltonians. 
The band gap is converged within 0.01 eV from $5\times5\times5$ to $6\times6\times6$ $k$-meshes. 
The differences between sc$GW$ band gaps with the head corrections at a $6\times6\times6$ $k$-mesh and the extrapolated TDL values are -0.06 and -0.05 for the sfX2C1e-Coulomb and the X2C1e-Coulomb Hamiltonian, respectively. 
The same convergence pattern is observed in AgCl and AgI as well, and the differences between the corrected values at a $6\times6\times6$ $k$-mesh and the extrapolated TDL values are all within 0.1 eV.

\end{document}